\documentclass[sts]{imsart}
\RequirePackage{amsthm,amsmath,amsfonts,amssymb}
\RequirePackage[]{natbib}
\RequirePackage[colorlinks,citecolor=blue,urlcolor=blue]{hyperref}
\usepackage{booktabs}
\usepackage{subfigure}
       
\usepackage[utf8]{inputenc}         

\usepackage{amsthm,amsmath,amsfonts,amssymb} 
\usepackage{etoolbox}

\usepackage{graphicx}                 

\usepackage{listings}                 
\usepackage{tikz}

\usepackage{algorithm2e}
\usepackage{mathtools}
\usepackage{enumerate}

\usepackage{epstopdf}
\usepackage{bbm}
\usepackage{placeins}
          
\usepackage{standalone}
\usepackage{url}
\usepackage{multirow}

\theoremstyle{plain}

\theoremstyle{definition}

\theoremstyle{remark}

\begin{document}

\begin{frontmatter}
        \title{The Two Cultures of Prevalence Mapping: Small Area Estimation and Model-Based Geostatistics}
        \runtitle{The Two Cultures for Prevalence Mapping}

\begin{aug}
\author[A]{\fnms{Jon}~\snm{Wakefield}\ead[label=e1]{jonno@uw.edu}},
\author[B]{\fnms{Peter A.}~\snm{Gao}\ead[label=e2]{peter.gao@sjsu.edu}},
\author[C]{\fnms{Geir-Arne}~\snm{Fuglstad}\ead[label=e3]{geir-arne.fuglstad@ntnu.no}}
\and
\author[D]{\fnms{Zehang Richard}~\snm{Li}\ead[label=e4]{lizehang@ucsc.edu}}

\address[A]{Jon Wakefield is Professor,
                 Departments of Statistics and  Biostatistics,
                 University of Washington, 
                 USA\printead[presep={\ }]{e1}.}
\address[B]{Peter A. Gao is Assistant Professor,
                 Department of Mathematics and Statistics,
                 San Jos{\'e} State University, 
                 USA\printead[presep={\ }]{e2}.}
\address[C]{Geir-Arne Fuglstad is Professor,
                Department of Mathematical Sciences, 
                 Norwegian University of Science and Technology,
                 Norway\printead[presep={\ }]{e3}.}
\address[D]{Zehang Richard Li is Assistant Professor,
                Department of Statistics, 
                 University of California Santa Cruz,
                 USA\printead[presep={\ }]{e4}.}

\end{aug}

\begin{abstract}
In low- and middle-income countries (LMICs), accurate estimates of subnational health and demographic indicators are critical for guiding policy and identifying disparities. Many indicators of interest are proportions of binary outcomes and the task of estimating these fractions is often called prevalence mapping. In LMICs, health and vital records data are limited, so prevalence mapping relies on data from household surveys with complex sampling designs. However, estimates are often desired at spatial resolutions at which data are insufficient for reliable  weighted estimation. We review two families of approaches to prevalence mapping: small area estimation (SAE) methods (from the survey statistics literature) and model-based geostatistics (MBG) methods (from the spatial statistics literature). SAE models can be ``area-level" or ``unit-level" and commonly use area-specific random effects and rely upon high-quality covariate data, often obtained from administrative sources. 
Unit-level models for binary responses are relatively underdeveloped. 
MBG approaches explicitly specify binary response models, incorporate continuous spatial random effects, and leverage alternative sources of data such as those arising from satellite imagery. These models are usually studied under a Bayesian framework. SAE methods often address the design by incorporating sampling weights or modeling the sampling mechanism. Two delicate issues arise when using MBG methods for prevalence mapping. First, aggregating unit level predictions to create area-level summaries requires population-level information that is rarely directly available.
Second, MBG approaches typically assume the sampling design is ignorable. We review both SAE and MBG approaches to prevalence mapping, and argue that binary response models can be improved using insights from both the survey sampling and the spatial statistics literature. We highlight these issues using household survey data from the Zambia 2018 Demographic Health Survey to estimate subnational HIV prevalence for woman aged 15--49.
\end{abstract}



\end{frontmatter}

\section{Introduction}\label{sec:introduction}

In  low- and middle-income countries (LMICs), subnational estimates of key health and demographic variables  are often used for determining progress toward targets, designing effective policy interventions and assessing disparities. For example, the United Nations General Assembly's 2030 Agenda established a set of Sustainable Development Goals for global development  \citep{GA2015}, each associated with specific aims such as eliminating preventable deaths under five years of age and reducing maternal mortality \citep{un2017equality}. Accurate tracking of the goals requires subnational estimates of indicators such as the neonatal mortality rate (proportion of children who die within the first month of life), vaccination coverage, poverty measures, contraceptive use and female attainment of secondary education. We focus on indicators that can be expressed as a \textit{prevalence}, meaning the proportion of individuals in a group that satisfy a specific condition. Producing subnational estimates of the prevalence of an outcome is often called \textit{prevalence mapping}.

When the data are sparse, this problem is an example of small area estimation (SAE), itself defined as the task of ``producing reliable estimates of parameters of interest...for subpopulations (areas or domains) of a finite population for which samples of inadequate sizes or no samples are available'' \citep[p.~xxiii]{rao:molina:15}. Much of the work on SAE is based on finite population estimation from the survey sampling literature. Historically, this research has focused on applications in high-income countries with readily available auxiliary data, often derived from administrative sources. However, in LMICs, health and vital records data are limited and although the most reliable sources of data are from household surveys, traditional SAE methods are not universally used. As an alternative, a body of research has emerged that uses so-called model-based geostatistics (MBG) approaches, drawn from the spatial statistics literature, for prevalence mapping \citep{diggle:giorgi:16}. 


We draw a distinction between two distinct families of approaches to prevalence mapping in LMICs. The first route, which we call the SAE approach, comprises methods drawn primarily from the survey sampling literature. All major surveys in LMICs collect data via a complex survey design, which involves stratification and cluster sampling, which we detail shortly. SAE methods often account for the complex survey design explicitly and, when sparsity of data prevents the use of design-based weighted estimators, frequently rely upon linear mixed effects models (LMEMs), leveraging auxiliary administrative data. The second route, which we call the MBG approach, encompasses a broader collection of spatial statistical modeling methods that have been adapted specifically for estimation in low-data settings. These strategies generally model the risk of experiencing the outcome of interest using a spatially continuous surface and use alternative sources of data such as satellite imagery. However, MBG approaches often pay less attention to the sampling design of the surveys, or to the nuances of aggregating point level predictions, which is required to produce area-level estimates. 

Inferentially, MBG approaches often take a Bayesian approach and use modern computational techniques such as the integrated nested Laplace approximation (INLA) of \cite{rue:etal:09} which provides a computationally efficient alternative to Markov chain Monte Carlo (MCMC) approaches. Subjective priors may also be specified. The SAE literature has seen an adoption of Bayesian methods, but computation is generally carried out with MCMC and ``uninformative priors'' are often favored \citep[Chapter 10]{rao:molina:15}.

In this article, we review both SAE and MBG approaches and examine their differences. In some respects, these differences result from a basic divergence with respect to the target estimand of interest. SAE approaches estimate \textit{prevalence}, and focus on a finite population of inference. The approach begins with a fixed set of individuals in a sampling frame. By contrast, MBG approaches assume the existence of a continuous spatial \textit{risk} surface, representing the probability of experiencing the outcome of interest for a hypothetical individual located at any particular location, and the average risk over an area is typically used as a proxy for the prevalence of that area.
This review follows in the footsteps of previous reviews of SAE \citep{ghosh:rao:94, pfefferman:13} but focuses on prevalence mapping in LMICs and gives an overview of MBG methods, which are less familiar to those in the design-based SAE community. Similarly, those in the spatial community may be less familiar with design-based finite population approaches.

\begin{figure}
        \centering
        \includegraphics[width=10cm]{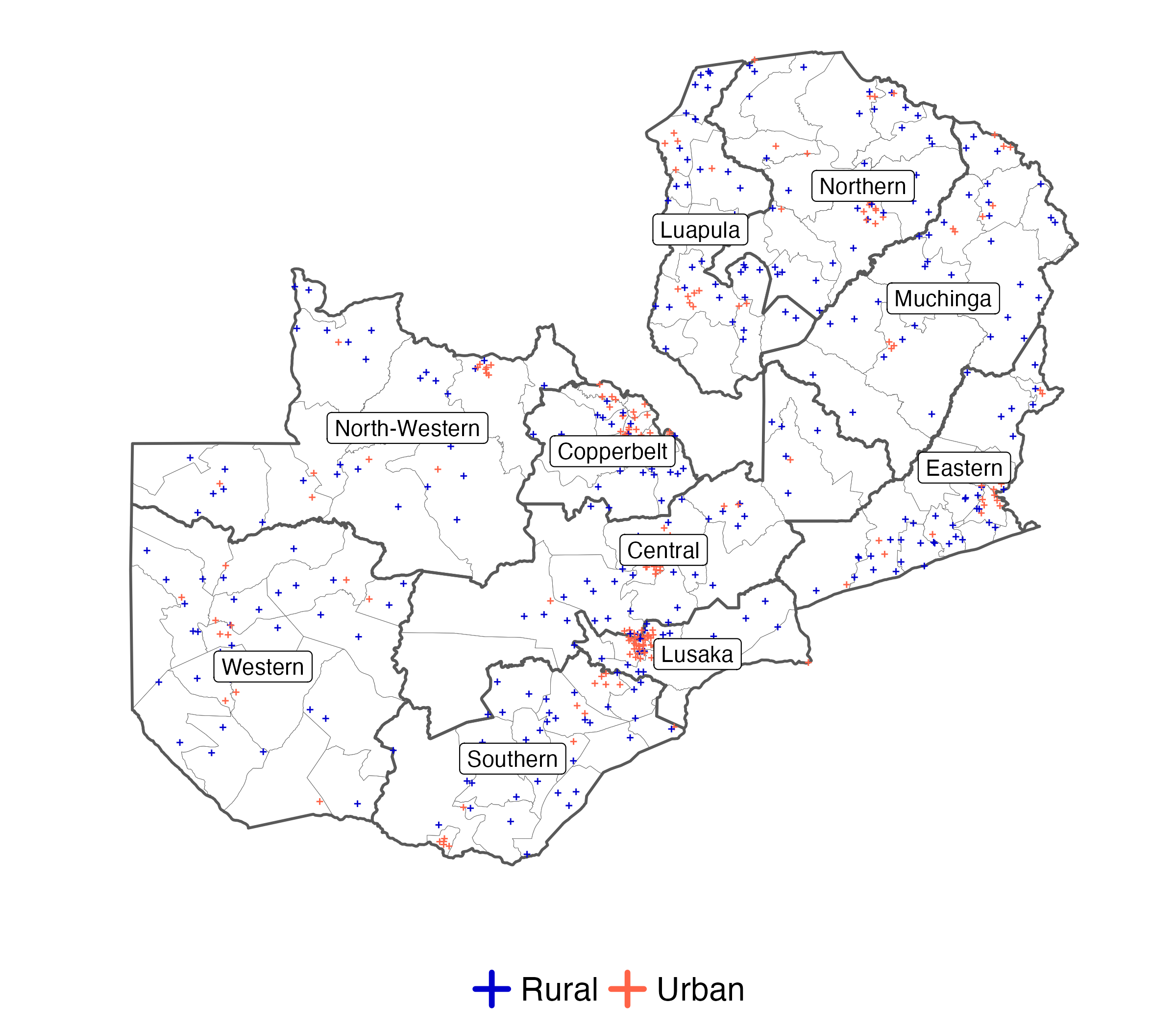}
        \caption{Locations of sampled urban and rural clusters in the 2018 Zambia DHS (jittered to preserve privacy) with Admin-1 and Admin-2 boundaries indicated and Admin-1 labels.\label{fig:Zambia_map}}
\end{figure}

To motivate the discussion, we briefly introduce an application in which we examine spatial variation in female HIV prevalence. We begin with key definitions:~all countries are divided into principal administrative divisions, called Admin-1 regions, which are further subdivided into secondary administrative regions, called Admin-2 regions.  In our example, we wish to characterize variation in HIV prevalence in women across 10 provinces (Admin-1 areas) and 115 districts (Admin-2 areas) of Zambia, based on data from the 2018 Demographic Health Survey (DHS), which used stratified  two-stage unequal probability cluster sampling. The clusters correspond to census enumeration areas (EAs) and the strata are the  urban/rural status of the cluster crossed with Admin-1 areas. 
The two stages of sampling are clusters within strata and households within clusters. For DHS (and more recent MICS) surveys, responses are reported with their cluster location, so that all individuals in the cluster are reported to be located at the geographical location of the cluster. Figure \ref{fig:Zambia_map} maps the approximate locations of the 545 sampled clusters, indicating which were urban/rural in the original sampling frame based on the 2010 census. HIV infection and geographic data were recorded for 12,893 women aged 15--49, from the 13,625 sampled households. The sampling units are households, while the observation units are women. The Admin-1 and Admin-2 boundaries are also shown in Figure \ref{fig:Zambia_map}, and we see a relatively large number of both urban and rural clusters in each Admin-1 area (as expected since these are planned domains), but sparser sampling in Admin-2 areas (unplanned domains). Government agencies often require high precision when reporting domain estimates. For example, Statistics Canada, has guidelines \citep[Table 5]{cloutier2014aboriginal} for area-level estimates: for an area with a coefficient of variation (CV) of less than 16.7\% the estimate can be used without restriction, when the CV is above this but less than 33.3\%, it should be used with caution, and an estimate with a CV greater than 33.3\% is deemed too unreliable to be published. For this example, all Admin-1 (weighted) estimates have CVs less than 16.7\%. However, at the Admin-2 level, out of 115 areas, three areas have no clusters, and fourteen have insufficient data to yield usable variance estimates for weighted estimates; only 27 of the areas have a CV
smaller than 16.7\% while 32 of the areas have a CV larger than 33.3\%. Hence, at the Admin-2 level, there is a need for modeling. In Section \ref{sec:cultures}, we fit a variety of SAE and MBG models that include spatial smoothing terms and auxiliary data.

The structure of the paper is as follows. The journey begins in Section \ref{sec:prevmap}, with a description of household surveys and an outline of the particular characteristics of prevalence mapping in LMICs that complicate the use of traditional SAE methods and have led to the emergence of MBG approaches. 
 In Section \ref{sec:sae}, we review traditional SAE methods, and discuss the finite population estimation perspective and design-based and model-based estimators. Section \ref{sec:MBG} considers MBG approaches, focusing on a number of issues including the aggregation of a continuous surface based on limited population information. In Section \ref{sec:additional}, we discuss a number of miscellaneous topics related to prevalence mapping. There is a huge literature on the SAE and MBG approaches and what we review, in the context of LMICs, is by necessity selective and not comprehensive. Section \ref{sec:cultures} compares SAE and MBG methods in the context of mapping HIV prevalence among women aged 15--49 in Zambia, and provides suggestions for modeling, taking elements from both of the ``cultures''. The paper ends with a discussion in Section \ref{sec:discussion}. Model details, additional results and code can be found in the Supplementary Materials \citep{supplement}. The code is also available at \url{https://github.com/peteragao/two-cultures-prev-map}.

\section{Prevalence Mapping using Household Survey Data}\label{sec:prevmap}

\subsection{Issues with Household Survey Data in LMICs}

 We begin by introducing some of the challenges that complicate prevalence mapping in LMICs, before contrasting, at a high level, traditional SAE and MBG approaches.

\vspace{.1in}
\noindent
{\it Complex sampling design}
\vspace{.05in}

\noindent
In many LMICs, household surveys including the DHS\footnote{\url{https://dhsprogram.com/}} and the Multiple Indicator Cluster Surveys (MICS)\footnote{\url{https://mics.unicef.org/}} collect data on health outcomes. Both survey programs implement a stratified multistage design. We will describe the DHS sampling design in detail, since it is relevant to our HIV prevalence mapping example. It is representative of the designs used by other large household surveys, including MICS, that are carried out in LMICs.

In a given country, DHS aims to conduct household surveys approximately every 5 years, with the \emph{sampling units} being households. The sampling frame is typically constructed from the most recent national census and aims to enumerate the households in the country at the time of the census and groups households into EAs (clusters). In the survey sampling literature the clusters are referred to as primary sampling units (PSUs). Clusters are classified as urban or rural by national authorities at the time of the census, based on a country-specific definition \citep{alkema2013levels}, but in some cases the classifications are updated between the time of the census and the survey. For example, the last census in Nigeria was 2006, but a new definition was taken to label each cluster as urban or rural, in 2017.

For each survey, households are generally sampled using a stratified two-stage cluster sampling design, as detailed
by \citet{DHSsampling2012}. The strata are usually defined by splitting each Admin-1 area into urban and rural parts. In the first stage of sampling, a specific number of urban clusters and rural clusters  are sampled from each Admin-1 area using probability-proportional-to-size (PPS) sampling, where a cluster's size is often defined as the number of households in the cluster. Urban clusters are often over-sampled. Since the list of households based on the census sampling frame will be outdated, the surveyor re-enumerates the households in each selected cluster. In the second stage of sampling, a fixed number of households (25 in the 2018 Zambia DHS) are sampled with equal probability from the updated list of households in each cluster. In the language of survey sampling, the
households are termed secondary sampling units (SSUs). Household members are interviewed to provide information on health and demographic variables.  As a result of the sampling procedure, each household has an inclusion probability that describes the \emph{a priori} probability that it will be included in the sample. 
The response rate of DHS surveys are high, but the reported weights do contain a non-response adjustment. Only the final weights are reported, and not the constituent parts, and the weight is scaled, so the weights alone cannot be used to estimate totals, only ratios, such as the prevalence. 

\vspace{.1in}
\noindent
{\it Sparse response data at the resolution of interest} 
\vspace{.05in}

\noindent
Often, estimates are desired not only at the Admin-1 level, but also at the Admin-2 level, as many health policy decisions are made at this level \citep{hosseinpoor:etal:16}. At this resolution, many areas may have severely limited data on an outcome of interest (and some may have no response data at all), since the surveys are typically powered to produce reliable estimates for Admin-1 regions. As a result, weighted estimation may not be feasible and model-based approaches are needed.

\vspace{.1in}
\noindent
{\it Acknowledging the Design and Aggregation} 
\vspace{.05in}

\noindent
Failing to account for stratification, clustering, and unequal probability sampling in a model-based approach can produce biased small area estimates (because of informative sampling in which the selection probabilities depend on the outcome values) and poorly calibrated estimates of uncertainty. Urban households are often overrepresented in in DHS  and if the outcome is associated with urban/rural location, then estimates of prevalence will be biased if the oversampling is not acknowledged. Unfortunately, relevant design variables, such as the number of households in a cluster, which is used in the PPS sampling, are unavailable. In addition, model-based methods that treat observations as independently and identically distributed (iid) are inappropriate for cluster sampling.

When moving from points to areas, aggregation of a risk surface is required and, formally, the locations of all of the target population, along with any associated auxiliary information that is used in a non-linear (binary data) model, is required, but this is never available. Hence, by necessity, the aggregation will be an approximate operation.


\vspace{.1in}
\noindent
{\it Limited availability of high-quality auxiliary data} 
\vspace{.05in}

\noindent
Traditional SAE model-based approaches often rely on covariate information from administrative sources such as national censuses. If recent and reliable census data is unavailable, as is often the case in LMICs, then alternative data  is needed to explain differences between areas.  



\subsection{An Overview of SAE and MBG  Approaches}

We first establish some terminology  \citep{rao:molina:15}.
\textit{Direct estimators} use response data from the area in question only when estimating the prevalence in an area. 
In areas where data are limited, \textit{indirect estimators} use statistical models that share information between areas. 
SAE approaches, which we review in depth in Section \ref{sec:sae}, typically account for features of survey design to minimize bias. The most basic SAE methods are based on H\'ajek estimators that incorporate the sampling weights. In the SAE literature, models are usually categorized as either \textit{area-level}, meaning that area-level weighted estimators are treated as response data in a smoothing model, or \textit{unit-level}, meaning that individual survey responses are modeled. The well-known Fay-Herriot model \citep{fay:herriot:79} is the most popular area-level model. In the first step of the Fay-Herriot approach, weighted estimates and associated estimates of the sampling variances are computed for each area. In the second step, these direct estimates are used as response data in a smoothing model that incorporates area-level covariates and Gaussian random effects, that are assumed (in the original formulation) to be iid across areas. The resulting estimators account for the sampling design because they take as input the direct estimates and their associated variances.

Unfortunately, if the target areas contain few samples, direct estimators (and the corresponding variance estimators) may be unreliable. For example, when using DHS data to map health outcomes at the Admin-2 level, it is common for many areas to have few or even zero sampled clusters, making direct estimation infeasible. Area-level modeling may still be possible, particularly if variance smoothing is incorporated, but for sparse data, unit-level models are the only option. These models specify a sampling model for individual responses, and are more flexible, incorporating unit-level covariates and smoothing via random effects. However, unit-level approaches often assume the sampling design is \textit{ignorable}, meaning that the same response model is appropriate for the sampled and non-sampled units.
But, as already noted, the data required to fit an appropriate model may be unavailable for the sampled responses, and so we may not be able to specify a model for which the design is ignorable. The variance model should also account for the effects of clustering. Moreover, unit-level models produce individual level predictions which must be aggregated to produce area-level estimates. For non-linear models and unit-level covariates, aggregation requires covariate information for each unobserved unit, but in the LMICs context, sampling frame information, such as the locations of all clusters and the population sizes within those clusters is not available. In Sections \ref{sec:model-based} and \ref{sec:spatAgg} we describe approximate aggregation strategies.

Like the unit-level models used in SAE, MBG approaches to prevalence mapping in LMICs treat individual responses as data and leverage recent advances in continuous spatial modeling and alternative sources of auxiliary information,  but less attention is paid to the survey design and in particular little reference is made to the sampling frame and finite populations. 

As already briefly discussed, 
SAE approaches 
estimate finite population {\it prevalences} while geostatistical approaches estimate a continuously indexed spatial surface representing {\it risk}. For example, when estimating an outcome such as the proportion of children who die within the first month of life, geostatistical models assume the existence of an unobservable spatial surface representing the hypothetical risk of death at {\it any} location. Each point-referenced observation is viewed as the result of an individual being exposed to the risk at that given location. Continuous spatial models are widely used in the environmental sciences to represent quantities such as pollutant levels or climate-related variables over a continuously-indexed study area and this is certainly reasonable in many contexts where the outcome variable (e.g.,~airpollution or temperature) does truly exist at every location.
SAE approaches thus perform \emph{prevalence mapping}, estimating the empirical proportions for areas, while MBG approaches approximate prevalences through \emph{risk mapping}, extracting areal estimates by averaging over a spatially continuous risk surface. 

Practically, geostatistical approaches borrow strength across space using Gaussian random field (GRF) models and spatial covariates, which may help to address some of the shortcomings of SAE methods in low-data settings. Using continuous spatial random effects enables smoothing even when response data is sparse or unavailable.  MBG approaches often use covariates derived from data sources such as satellite imagery. Such data, which may include variables such as intensity of night time lights or vegetation indices, are usually provided as raster data with measurements on a high-resolution grid across the domain of interest. Risk predictions can be made for any location covered by the covariate rasters. Of course, traditional SAE unit-level models could utilize satellite imagery data, but following the rationale of SAE approaches, for non-linear models, one would require the set of locations of all units in each area, which is unavailable. In practice, geostatistical unit-level models overcome this issue by generating predictions on high-resolution (often $1\times 1\,\text{km}$ or $5\times 5\,\text{km}$) pixel maps across the spatial domain of interest, and weighting by population density. 
Rasters of population density are commonly used for weighting, but the population density values are themselves modeled quantities with (often, unspecified) uncertainty and it is not clear how to propagate this uncertainty into the areal estimates. In addition, predictions that are optimal at the pixel level are not guaranteed to result in optimal predictions at coarser levels such as Admin-2 when aggregated using imprecise and inaccurate knowledge about the population. Benchmarking, to ensure consistency across nested geographical levels, is not carried out as part of MBG analyses, though there is a large SAE literature on this issue, see \citet[Section 6.4.6]{rao:molina:15}.

Interpreting and quantifying the uncertainty of small area estimates produced by MBG approaches is difficult. As noted, geostatistical models often assume implicitly that the sampling procedure is ignorable under the specified model and may not adjust for relevant design variables. 


Traditional SAE approaches have been applied to a number of indicators in LMICs by international organizations.  These include the World Bank, in the context of poverty mapping \citep{molina2019small,corral2021map,
edochie2025small} and 
the United Nations Inter-Agency Group on Mortality Estimation (UN-IGME), who produced subnational estimates of child mortality indicators using SAE methods \citep{wakefield:etal:19,wu:etal:21}, leaning heavily on discrete spatial models. HIV prevalence mapping is also common using SAE methods, see for example, \cite{gutreuter2019improving}. 

Similarly, a number of prominent research organizations produce subnational risk maps using MBG approaches, including the Institute for Health Metrics and Evaluation (IHME) \citep{golding:etal:17,burstein:etal:18,osgood:etal:18,graetz:etal:18,mosser:etal:19,local2021mapping}, WorldPop \citep{utazi:etal:18,utazi2020geospatial,ferreira2022geographic} and the Incorporated City Fund (ICF, who historically implemented DHS) \citep{mayala:etal:19,janocha2021guidance}. Other examples include \cite{giorgi:etal:15,diggle:giorgi:16} and \cite{giorgi:etal:18}.

As evidence of the cultural divide between the SAE and MBG approaches, prominent SAE texts \citep{rao:molina:15, pratesi_analysis_2016, morales_course_2021} and the SAE guidelines report of \cite{corral2022guidelines} cover spatial modeling only briefly and do not mention the MBG approach. From the other side of the divide, SAE is not mentioned in the index of the book-length treatment on MBG approaches of \cite{diggle:giorgi:19}, and the classic text of \cite{rao:molina:15} is not referenced anywhere in the book. 

\section{Small area estimation} \label{sec:sae}

In this section, we review area-level and unit-level SAE approaches more formally. We also outline direct estimation since it is an important ingredient for area-level modeling, though it is not strictly an SAE approach, since no between-area modeling is carried out. We begin by establishing notation and briefly reviewing the principles of finite population inference. 

\subsection{Finite Population Estimation and Inference}
\label{sec:design:finTarget}

We consider a finite target population in a study area that consists of $N$ \emph{observation units}. 
The target population can be represented as the set of indices $U = \{1, 2, \dots, N\}$ where observation unit $k$ has an associated response value $y_k \in\{0,1\}$ for $k \in U$. When a country is divided into $m$ administrative areas, these areas partition the population $U$ into $m$ disjoint subpopulations, $U = U_1 \cup U_2 \cup \cdots \cup U_m$, where $U_i$ is the set of units that belong to area $i$. 

This notation is standard for direct estimates and area-level models. For unit-level models and MBG approaches, it is easier to use nested indexing. Let $N_i$ denote the number of units in $U_i$ for $i=1,\dots,m$. Then, in anticipation of our SAE unit-level development, we can index population responses as $y_{ij}$ for unit $j=1,\dots,N_i$ in area $i = 1,\dots,m$. There is a one-to-one correspondence between unit $j$ in area $i$ and a unit $k[i,j]\in U$, and we switch between notations depending on the type of model considered.

The target parameters are the area-specific prevalences. With the former notation, 
\begin{equation}
        p_{i} =\frac{1}{N_i} \sum_{k \in U_i} y_k, \quad i = 1, \dots, m,
        \label{eq:ST_target}
\end{equation}
and in the latter notation
\[
    p_{i} = \bar{y}_i=\frac{1}{N_i}\sum_{j=1}^{N_i}y_{ij}, \quad i=1,\dots,m.
\]

We let $S=\{k_1,\dots, k_n\}\subset U$ denote the set of $n$ sampled indices, where $S$ is partitioned into $m$ areas, $S=S_1\cup\cdots\cup S_m$.
We assume a probability sampling design, and for all $k\in U$, let $\pi_k$ denote the probability that unit $k$ is sampled and $w_k= 1 / \pi_k$ denote the design weight for unit $k$.  In general, weights may contain non-response  and  post-stratification adjustments. In the DHS the weights include the former but not the latter. This is common for household surveys in LMICs, where observed population data (from, for example, a census) is generally neither recent nor reliable enough for the purpose of post-stratification. 

When conducting inference based on survey data, we can distinguish between design-based inference and model-based inference. The design-based approach assumes a fixed finite population of responses, and inference is based on the randomization distribution over the space of possible samples defined by the sampling design.
Asymptotic analysis for sample survey estimators usually considers a sequence of finite populations of growing size and an associated sequence of sampling designs. The sample size is typically assumed to grow at the same rate as the finite population and various other assumptions can be made about the sampling design as well as the finite population responses. Under this design-based framework, estimators that are unbiased, or at least design consistent, are preferred. For an overview of sample survey asymptotics, see \cite{breidt:opsomer:17}. The model-based approach assumes that the finite population responses are drawn from a data-generating model and inference is based on this model, though design consistency of model-based estimates is desirable.
The sampling design is assumed to be ignorable with respect to the model. SAE methods adopt both design-based and model-based perspectives, but generally design-based inference is used when there is sufficient data for using direct weighted estimators while model-based inference is more commonly used when data is more limited.

\subsection{Direct Estimates}
\label{sec:Design:direct}

Direct weighted estimators acknowledge the sampling design by incorporating sampling weights. For example, the   H\'{a}jek estimator \citep{hajek:71} replaces the numerator and denominator in Equation \eqref{eq:ST_target} with survey weighted estimates,
\begin{equation}
  \hat{p}^{\tiny{\text{w}}}_{i} = \frac{\sum_{k \in S_i} w_k y_k}{\sum_{k\in S_i} w_k}, \quad i = 1, 2, \dots, m.\label{eq:HT}
\end{equation}
Under the design-based perspective, the random quantity in (\ref{eq:HT}) is
$S_i$. The estimators of the numerator and the denominator are unbiased, and
$\hat{p}^{\tiny{\text{w}}}_i$ is consistent for $p_{i}$. Design-based estimates of the
variance can be constructed with linearization-based approximations or resampling methods such as jackknife or bootstrap estimators \citep{lohr2021sampling,pedersenandliu:2012}. 

Achieving reliable weighted estimates with low variances requires sufficient sample sizes for each target area, making direct estimators inadequate for SAE with small domains \cite[Chapter 2]{rao:molina:15}. 
For example, the $m$ administrative areas may be unplanned areas \citep{lehtonen2009design} which do not match the strata used for sampling, so that some administrative areas have sparse, if any, data. 

\subsection{Area-Level Models}
\label{sec:design:sDirect}

Simultaneous modeling of the data from all areas can 
 increase the precision of estimates, relative to the direct alternatives, effectively increasing the sample size.  The most commonly used approach is the two-stage Fay-Herriot model \citep{fay:herriot:79} that uses auxiliary information in the form of area-level covariates and introduces area-specific random effects. 
In the original paper, the random effects were assumed to be iid normal and a frequentist approach to inference was taken. Let $\hat{\theta}^{\tiny{\text{w}}}_i$ represent a direct estimate of an area-level parameter and 
$V_i$ be the sampling variance of $\hat{\theta}^{\tiny{\text{w}}}_i$, which is estimated from data but often treated as known. 
The Fay-Herriot model is:
\begin{eqnarray}
\hat{\theta}_i \mid \theta_i &\sim_{\mathrm{iid}}& \mbox{N} ( \theta_i, V_{i}),\label{eq:fayherriot1}\\
 \theta_i&=&\alpha+\boldsymbol{x}_i^\top \boldsymbol{\beta} + u_i , \label{eq:fayherriot2}\\
 u_i \mid \sigma^2_u&\sim_{\mathrm{iid}}&\mbox{N}(0, \sigma^2_u),\quad i = 1, \dots, m,\label{eq:fayherriot3}
\end{eqnarray}
where $\alpha$ is the intercept, $\boldsymbol{x}_i $ are area-level covariates with associated coefficients contained in the vector $\boldsymbol{\beta}$ and $u_i $ represent between-area differences not explained by covariates. Equation (\ref{eq:fayherriot1}) is referred to as the {\it sampling model} and equation (\ref{eq:fayherriot2}) the {\it linking model}. The Fay-Herriot model implicitly acknowledges the sampling design through the use of sampling weights when computing the direct estimate and its standard error, $\sqrt{V_i}$. 
The direct estimates may be transformed to make the normal approximation to the sampling distribution more accurate and to ensure that the resultant prevalence estimates are on $[0,1]$.  In particular, for each area $i$, we can define $\hat{\theta}^{\tiny{\text{w}}}_i=h(\hat{p}_i^{\tiny{\text{w}}})$ where
$h(\cdot)$ represents a transformation. The Fay-Herriot model can then be applied to model the transformed $\hat{\theta}^{\tiny{\text{w}}}_i$ parameters with the resulting modeled estimates being transformed back to the original range. The sampling variance of the transformed $\hat{\theta}^{\tiny{\text{w}}}_i$ parameters must be estimated or approximated. Transformations used with the Fay-Herriot model include the log \citep{fay:herriot:79, slud_mean-squared_2006}, logit \citep{jiang_best_2011, mercer:etal:15},  dual power \citep{sugasawa_parametric_2015}, and arcsine \citep{hirose_arc-sin_2023}.
We emphasize that $\theta_i$ represents a finite population parameter, such as the logit prevalence, logit$(p_i)$, for area $i$ and not a super-population characteristic, such as the logit risk. 

This model can be used to generate small area estimates using either frequentist or Bayesian methods. The Fay-Herriot model is an example of a LMEM and so the empirical best linear unbiased predictor (EBLUP), a standard summary for such models, can be used \citep[Chapter 6]{rao:molina:15}.
Bayesian approaches are also available, but in the SAE literature, ``diffuse" (rather than subjective/informative) priors are often sought 
\citep[Chapter 10]{rao:molina:15}. For SAE, under frequentist inference, a large emphasis is placed on estimating the mean squared error (MSE) associated with an areal estimate, with the bootstrap being a common tool. A good discussion of design-based and model-based approaches to MSE estimation is provided by \cite{datta:etal:11}. Under a fully Bayesian approach, as described in Section \ref{sec:model-based}, the natural analogue is the posterior variance, which is a model-based uncertainty measure. Another common uncertainty summary is the coefficient of variation, which is the square root MSE (or posterior standard deviation), divided by a point estimate.

Area-level models are popular as they are generally computationally easy to fit and they can improve upon the precision of direct estimators by explaining between-area differences using auxiliary data and random effects. The resulting estimates exhibit {\it shrinkage}, so that they are (conditional on the true value) biased, but their variance is reduced, and in general this results in a lower MSE for the complete collection of estimates, when compared to the MSE of direct estimates. Another major advantage of area-level models is that they account for the sampling design and achieve design consistency, in the sense that the sampling bias and variance of the estimated prevalence can be shown to converge to zero asymptotically under reasonable assumptions on the finite population and sampling design. Intuitively, since the H{\' a}jek estimator is design consistent,
Fay-Herriot estimators, though based on a model, are design consistent also, as the within-area sample size increases.

In a high-income setting, the Fay-Herriot model has seen a vast number of applications, as summarized in \cite{rao:molina:15} and \cite{ghosh2020small}. A well-known example is the production of county-level estimates of poverty for the Small Area Income and Poverty Estimation (SAIPE) program in the United States \citep{bell:etal:16}. 

We continue by discussing possible extensions to the basic Fay-Herriot model that can be used to address particular challenges of prevalence mapping in LMICs.
The basic Fay-Herriot model is called a matched model in the sense that both the sampling and linking models are on the same scale (for example, logit), each with normal models. 
For a prevalence $p_i$ and a nonlinear transformation $h(\cdot)$,  $$\mbox{E}[~h(\hat{p}_i^{\tiny{\text{w}}}) \mid h(p_i)~] \neq h(p_i),$$ even if $\hat{p}_i^{\tiny{\text{w}}}$ is design unbiased \citep[Section 4.2]{rao:molina:15}. This motivates an unmatched model in which it may be assumed that,
\begin{equation}\label{eq:unmatched}
\hat{p}_i^{\tiny{\text{w}}} \mid p_i \sim_{\mathrm{iid}} \mbox{N} ( p_i, V_{i}),\quad i = 1, 2, \dots, m,
\end{equation}
\noindent
but with a linking model that assumes that $\theta_i=h(p_i)$, and not $p_i$, is normally distributed.
 \cite{you:rao:02} assume a normal sampling model but a log-normal linking model. For estimating small area proportions, sampling model (\ref{eq:unmatched}) along with a logistic linking model has been used \citep{mohadjer_hierarchical_2012}. Other unmatched models for proportions are explored by \cite{liu_hierarchical_2014} and \cite{franco_applying_2013}. A general empirical Bayes (EB) approach for SAE with unmatched sampling and linking area-level models is outlined by \cite{sugasawa_small_2018}.

The basic Fay-Herriot model assumes normally distributed iid area-level random effects, but the model may be easily extended to allow for random effects with other correlation structures. In particular, spatial and spatiotemporal covariance matrices may be used to smooth estimates across space and space-time. \cite{chung:datta:20} describe a range of spatial models including a conditionally autoregressive (CAR) model; they provide a comparison of the traditional Fay-Herriot model with spatial alternatives, finding that a spatial area-level model can improve estimation when good covariates are not available.  \cite{ghosh_generalized_1998} applied an intrinsic CAR (ICAR) prior \citep{besag:kooperberg:95} to the random effects, while other methods have focused on the use of simultaneous autoregressive (SAR) spatial models \citep{soton8165, petrucci_small_2006, pratesi_small_2008, marhuenda_small_2013}.  One approach that we have used extensively is a model that decomposes the random effect into the sum of an unstructured iid normal random effect and a spatial ICAR random effect, which is known as the BYM model \citep{besag:york:mollie:91}. In our analyses in Section \ref{sec:cultures}, we adopt the reparameterization known as the BYM2 model \citep{riebler:etal:16}, in which the vector of random area effects has structure,
$$
\boldsymbol {u} = \sigma_u \left(\sqrt{1-\phi}\boldsymbol{e} + \sqrt{\phi} \boldsymbol{S}\right),
$$
where $\sigma_u$ is the total standard deviation, $\phi$ is the proportion of the variance that is spatial, $\boldsymbol{e}$ is a vector of iid standard normal random variables and $\boldsymbol{S}$ follows a scaled ICAR prior so that the geometric mean of the marginal variances of $S_i$ is equal to $1$, under a sum-to-zero constraint that is imposed to ensure identifiability when there is an intercept in the model \citep{rue:knorrheld:05}. More details on this model are contained in the Supplementary Materials.

Spatial models have been extensively used in disease mapping \citep{wakefield:etal:00}.
Disease mapping is distinct from SAE in that it is based on a full enumeration of events (up to accurate case enumeration). Hence, the prevalence is observed, but one is interested in risk. Consequently, model-based, as opposed to design-based, inference is carried out. Generalized linear mixed models (GLMMs) are a common approach, and the BYM model has been extensively used in disease mapping applications. 

While the vast majority of applications of the Fay-Herriot model have assumed normal iid random effects or spatial random effects with CAR, ICAR or SAR forms, a number of authors have considered different shrinkage forms. For example, \cite{datta2015small} suggested spike and slab mixture priors and \cite{tang2018modeling} consider horseshoe priors and \cite{tang2023global} extend such priors to include spatial effects.

The inputs to the Fay-Herriot model are the weighted estimates and their variances. When the data are sparse in an area, the usual variance estimation methods may produce
 estimates that are zero, undefined, or unstable.
 For example, when the logit transform is used, problems arise for areas in which $\hat p_i^{\tiny{\text{w}}}=0/1$.
To alleviate difficulties with the sampling variances, they may be modeled using generalized variance functions \citep[Chapter 7]{wolter:07}, which leverage the mean-variance relationship between $\theta_i$ and $V_i$, and/or incorporate  covariates \citep{otto:bell:95, mohadjer_hierarchical_2012, franco_applying_2013, liu_hierarchical_2014}. Uncertainty in the sampling variances may also be incorporated into the model by considering a joint model for the direct estimates and the associated sampling variance estimates \citep{you2006small, maiti_prediction_2014, sugasawa_bayesian_2017,gao2023spatial}. 
If there are just a small number of areas with 
no data then one may still fit Fay-Herriot models, treating these areas as having missing data. Spatial random effects models are particularly appealing in this regard, and the situation brightens considerably if there are strong associations with covariates. It is, however, very difficult to give guidelines on when the proportion of missing areas becomes too large to follow such a strategy.

Synthetic estimators are developed under the assumption that small areas have similar characteristics to larger areas \citep[Section 3.2]{rao:molina:15}. In simple models, synthetic estimators correspond to Fay-Herriot models with no random effects, which therefore lean on the regression part of the model. 
In general, the bias from such approaches can be considerable when between-area residual variation is significant. However, if one requires estimates for a set of geographic areas for which few were sampled, a spatial model is not useful, and strongly predictive covariates are essential. In this paper, we focus on the situation in which the majority of the target areas provide samples.

  When data are indexed by both areas and sub-areas (such as Admin-1 and Admin-2 geographies), one may introduce two levels of random effects, to provide more efficient estimation by borrowing information at each level.  \cite{torabi2014small} propose a  Fay-Herriot model of this type, see also \cite{erciulescu2019model}. Such a model is very appealing, but in LMICs the data will often be too sparse to produce reliable direct estimates and variances at Admin-2, and these are required as inputs.

There is a strand of research that models area-level responses on [0,1] using a variety of beta sampling models, see for example, \cite{janicki2020properties} and \cite{de2024small}.
 
 Surveys with temporal information are quite commonplace, and various models have been proposed; see for example, \citet[Section 4.4.3]{rao:molina:15}.
A more recent review of combining surveys over time, is provided by \cite{pfeffermann2022time}. In the LMICs setting,
  \cite{li:etal:19} use a spatiotemporal  Fay-Herriot model to estimate Admin-1 under-5 mortality in 35 African countries, using BYM2 spatial priors, random walk temporal priors \citep{rue:knorrheld:05} and space-time interactions, as described in \cite{knorrheld:00}.  





\subsection{Unit-Level Models}\label{sec:model-based}

When it is not possible to obtain reliable area-level direct estimates and standard errors, one must consider alternative approaches. 
Unlike area-level models, unit-level models treat individual responses as random variables drawn from some superpopulation model \citep{royall1970finite,valliant2000finite}, making the finite population small area means and totals random as well.

For continuous responses without a clustering component, \cite{battese1988error} proposed the nested error regression model, also called the basic unit-level model. Again define $N_i$ as the population size in area $i$, so that $N_i=|U_i|$ with $j=1,\dots,N_i$, indexing units in the area $i$ population. The unit-level nested error model is defined for all members of the population as, 
\begin{equation}\label{e:bhf}
Y_{ij} =\alpha+\boldsymbol{x}_{ij}^\top \boldsymbol\beta+u_{i}+\varepsilon_{ij},\quad 
\end{equation}
for  $j=1,\dots,N_i$, $i=1,\dots,m$, and 
where $\alpha$ is the intercept, $\boldsymbol{x}_{ij}$ are unit-level covariates, and $\boldsymbol\beta$ are regression coefficients.  We have area-level random effects $u_i \mid \sigma_u^2\sim_{\mathrm{iid}}\mbox N(0,\sigma_u^2)$ and
$\varepsilon_{ij} \mid \sigma_\varepsilon^2\sim_{\mathrm{iid}} \mbox N(0,\sigma_\varepsilon^2)$ represent iid unit-level effects, which are a combination of true unit-level variation, and measurement error. We use $Y_{ij}$ to avoid ambiguity between the stochastic variable (in the model) and the realized value $y_{ij}$ in the fixed finite population. A key requirement is that (\ref{e:bhf}) holds for all members of the population, sampled and unsampled. We define $S_i^{\tiny{\mbox{u}}}$ as the set of sampled units in area $i$, with $n_i=|S_i^{\tiny{\mbox{u}}}|$ so that the observed data are $\{ y_{ij}, j \in S_i^{\tiny{\mbox{u}}},i=1,\dots,m\}$.

For any area $i$, the finite population mean of the continuous response according to the model is, under the assumption that $\varepsilon_{ij}$ represents true signal, $$\overline{Y}_i=\alpha+\overline{\boldsymbol{x}}_i^\top\boldsymbol\beta +u_i+\overline{\varepsilon}_i,$$ where 
$$\overline{Y}_i = \frac{1}{N_i} \sum_{j=1}^{N_i}Y_{ij},\quad \overline{\boldsymbol{x}}_{i} =\frac{1}{N_i} \sum_{j=1}^{N_i}\boldsymbol{x}_{ij},\quad \overline{\varepsilon}_i= \frac{1}{N_i} \sum_{j=1}^{N_i}\epsilon_{ij},$$ denote the area means, for $i = 1, \dots, m$. This illustrates that for a linear model, only area-level population covariate averages are needed. Since $\overline{\varepsilon}_i \mid \sigma_\varepsilon^2 \sim  \mbox N(0,\sigma_\varepsilon^2/N_i),$ we can set $\overline{\varepsilon}_i = 0$, for large finite population sizes.
Hence, inference for $\overline{Y}_i$, is close to that for,
\begin{eqnarray*}
    \mu_i&=& \mbox{E}\left[ ~\overline{Y}_i \mid \alpha, \boldsymbol\beta,u_i ~\right]\\&=&\alpha+\overline{\boldsymbol{x}}_i^\top\boldsymbol\beta +u_i,\qquad i = 1,  \dots, m.
    \end{eqnarray*}
Further, $\overline{Y}_i$ is a description of our beliefs about the unknown population average $\overline{y}_i$.
Hence, model-based estimation of $\mu_i$ is asymptotically equivalent to estimation of the finite population mean. The area-level summary is the product of an {\it aggregation step} which is more difficult when attempted with nonlinear prevalence models, as we see shortly. 

There have been many applications of unit-level models \cite{rao:molina:15} and extensions to the original model. For example, \cite{molina2010small} proposed a linear nested error model to estimate nonlinear poverty measures.

As with area-level models, unit-level models can be analyzed using frequentist or Bayesian approaches. Model-based approaches are more common than design-based, since inference is based on a super-population model that is responsible for the randomness. However, design consistency for the unit-level linear model, via a pseudo-EBLUP approach that explicitly uses the weights, was considered by \cite{you2002pseudo}, see also \cite{jiang:lahiri:06,pfeffermann2007small} and \cite{guadarrama2018small}. 

Our interest is primarily in binary outcomes such as the prevalence of health and demographic indicators. For such a response, a logistic unit-level model may be specified. For binary variables, multilevel logistic regression models have been used previously \citep{malec:etal:97,jiang2001empirical,congdon:2010,zhang2014multilevel,hobza2016empirical}. 
  
  A unit-level binary response model for all members of the population is, 
  \begin{eqnarray}
      Y_{ij} \mid r_{ij} &\sim& \mbox{Bernoulli}(r_{ij})\label{eq:binunit1}\\
      \mbox{logit}(r_{ij}) &=& \alpha + \boldsymbol{x}_{ij}^\top\boldsymbol\beta + u_i
  \label{eq:binunit2}
  \end{eqnarray}
  for $j=1,\dots,N_i$, $i=1,\dots,m$, and with individual-level covariates $\boldsymbol{x}_{ij}$ with associated log odds ratios $\boldsymbol\beta $. This is an example of a GLMM and inference can again proceed from either frequentist or Bayes perspectives. The EB approach extends the EBLUP method beyond the LMEM to the GLMM 
\citep[Chapter 9]{rao:molina:15}. Sometimes this approach is referred to as empirical best prediction (EBP).

Aggregation under nonlinear models is not straightforward. Under the population model (\ref{eq:binunit1}) and (\ref{eq:binunit2}),
  the area-level risk is
  \begin{equation}
  r_i = \frac{1}{N_i} \sum_{j=1}^{N_i} \mbox{expit}( \alpha + \boldsymbol{x}_{ij}^\top\boldsymbol\beta + u_i),\label{eq:unitrisk}
  \end{equation}
  for $i = 1, \dots, m$.
Notice that we require covariates, $\boldsymbol{x}_{ij}$, on all the individuals, which is difficult to achieve in LMICs. Specifically, censuses are often infrequently carried out and may be inaccurate, and if we wish to use satellite imagery or meteorological variables, we need the locations of all units, which is never achievable. These challenges are further discussed in Sections \ref{sec:spatAgg} and \ref{sec:auxiliary}.

The above description adopts notation that is typical of that used in the SAE literature, with units $j$ in areas $i$. 
We now define notation and provide a development for data available from household surveys in LMICs, in which units $j$ in the same cluster are recorded with the same location, and so we combine together the data within each cluster. Note that, for a given indicator, all  units in the same cluster have the same design weight and the same non-response adjustment (these adjustments are constant within sampling strata, since strata are used as the 
response groups), which provides further justification for combining data within the same cluster.

Let $S_i^{\tiny{\mbox{c}}}$ index the set of sampled clusters $c$ in area $i$, with $n_i=|S_{i}^{\tiny{\mbox{c}}} |$.
An added complication is that not all households (and therefore individuals)  are sampled in the selected clusters. We let
 $S_{ic}^{\tiny{\mbox{c}}}$ index the set of sampled individuals (units) in cluster $c$ of area $i$, with $n_{ic}=|S_{ic}^{\tiny{\mbox{c}}} |$. In the population, the total number of clusters in area $i$ is $C_i$ and the total number of individuals in cluster $c$ is $N_{ic}$, $c=1,\dots,C_i$, $i=1,\dots,m$. In the sampled clusters, the observed totals for the sampled units are,
$$y_{ic}^{\tiny{\mbox{CL}}} =\sum_{
j\in 
S_{ic}^{\tiny{\mbox{c}}}}
y_{ij},\qquad c \in S_{ic}^{\tiny{\mbox{c}}},\qquad i=1,  \dots, m.$$
These totals should be distinguished from the unobserved population totals, for all clusters,
$$y^{\tiny{\mbox{POP}}}_{ic} =\sum_{j=1}^{N_{ic}}y_{ij},\qquad c=1,\dots,C_i,\qquad i=1,\dots,m.$$


We denote the cluster-specific {\it risk parameter}
by $r_{ic}$. This will not cause ambiguity with individual-level risks $r_{ij}$ since the models will only involve cluster-level risks. We assume that all individuals in the cluster experience the same risk so that $\mbox{E}[ Y^{\tiny{\mbox{POP}}}_{ic}/N_{ic}  \mid r_{ic}] = r_{ic}$, 
and assume, for all clusters in the population, the observation model is,
\begin{equation}
                \label{e:logit-bhf-obs-pop}
Y^{\tiny{\mbox{POP}}}_{ic} \mid r_{ic} \sim \text{Binomial}(N_{ic},r_{ic}),
\end{equation}
for $c=1,\dots,C_i, i=1,\dots,m$, and 
where $Y^{\tiny{\mbox{POP}}}_{ic}$ and $y^{\tiny{\mbox{POP}}}_{ic}$ again differentiate between stochastic variables in the model and realized values in the fixed finite population.
This model is assumed to also hold for selected individuals, so that the sampling model is,
\begin{equation}
                \label{e:logit-bhf-obs}
Y_{ic}^{\text{\tiny{CL}}} \mid r_{ic} \sim \text{Binomial}(n_{ic},r_{ic}),
\end{equation}
for $c \in S_i^{\tiny{\mbox{c}}}$, $i=1,\dots,m$.

 We model the risk parameters using a logistic model, with cluster-level covariates $\boldsymbol{z}_{ic}$ (where we assume that the covariates are common to all units in the cluster) and area-level random effects $u_i$. As we have described, DHS and MICS data are collected via stratified two-stage cluster sampling, with clusters selected within strata, and households sampled within clusters. To account for between-area variability and within-cluster variation, two levels of random effects can be used, for areas and for clusters \citep{stukel_estimation_1997, marhuenda_poverty_2017}. This gives the logistic regression model, assumed to hold for all clusters in the population, as 
\begin{equation}
        \mbox{logit}(r_{ic}) = \alpha+\boldsymbol{z}_{ic}^\top\boldsymbol\beta + u_i+ \delta_{ic}\label{eq:unit:SAErisk}
\end{equation}
where   $\delta_{ic} \mid \sigma_\delta^2
\sim_{\mathrm{iid}} \mbox{N}(0, \sigma_\delta^2)$ denote cluster level random effects for $c=1,\dots, C_i$, $i=1,\dots,m$. The area-level random effects may be iid or assigned a spatial model, such as the BYM2.
All units (households) have the same geographical (recorded) cluster location but there is within-cluster dependence of units, which is why we specify an overdispersion model for the aggregated counts in each cluster. 
 The cluster-specific random effects, $\delta_{ic}$, account for the dependence that two individuals within the same cluster will display \citep{scott:smith:69},  which will induce {\it overdispersion} in the sampling model. If we take a beta model for the cluster random effects (as opposed to the normal model for $\delta_{ic}$, assumed above), we obtain an alternative model for  overdispersion. This leads to a betabinomial marginal sampling model, and this model is discussed more fully in Section \ref{sec:cultures} and in the Supplementary Materials. \cite{marino2019semiparametric} consider a logistic model with the distribution of the area-level random effects being unspecified, and illustrate its use with data on unemployment incidence.

Now now consider aggregation for the SAE unit-level model (\ref{eq:unit:SAErisk}) 
and suppose $\delta_{ic}$ represents within-area variation. Then, under the superpopulation model, $r_i = \mbox{E}[P_i]$
and from (\ref{eq:aggintpop}), we obtain,
\begin{eqnarray*}
  r_i   &=&  \sum_{c=1}^{C_i}r_{ic} \times \frac{N_{ic}}{N_i}  \\&=&  \sum_{c=1}^{C_i} \mbox{expit} 
( \alpha + \boldsymbol{z}_{ic}^\top \boldsymbol{\beta}+ u_i + \delta_{ic}) \times \frac{N_{ic}}{N_i} .
\end{eqnarray*}
In practice, we may sometimes know the number of clusters in the sampling frame, but not their locations or sizes. \cite{dong:wakefield:21} estimate subnational vaccination prevalence in Nigeria using the 2018 DHS.  The sampling frame is approximated by simulating locations $\boldsymbol{s}_{ic}$ from a population density surface, and then taking information available in the DHS survey manual to simulate cluster-level (child) population sizes, $N_{ic}$. Clearly this approach contains a number of approximations, but is more consistent with an SAE, rather than an MBG, approach, since it attempts to recover the final population.
Samples are obtained from the posterior via
\begin{eqnarray*}
r^{(b)}_i &=&  \sum_{c=1}^{C_i} \mbox{expit} ( \alpha^{(b)} + \boldsymbol{z}(\boldsymbol{s}_{ic})^\top \boldsymbol{\beta}^{(b)} + u^{(b)}_i + \delta^{(b)}_{ic}) \times \frac{N_{ic}}{N_i}
\end{eqnarray*}
with $\delta^{(b)}_{ic}\mid \sigma^{2 (b)}_\delta \sim \mbox{N}(0,\sigma^{2 (b)}_\delta)$ and with samples from the posterior, $\{ \alpha^{(b)}, \boldsymbol{\beta}^{(b)},\sigma^{2 (b)}_\delta\}$ for $b=1,\dots,B$. This provides a method of accounting for within-area variation in risk, when a discrete spatial model is used. Here we have assumed that the $\delta$ terms should be included in the aggregation, a point we now discuss in more detail.

Model (\ref{eq:unit:SAErisk}) describes between-cluster variation in the target population through covariates and captures unexplained between-area variation through the area-specific random effects. Models such as this were considered by \cite{wakefield:okonek:pedersen:20}.
The $\delta_{ic}$ terms may also represent ``real'' between-cluster signal, since the spatial random effect term $u_i$ is constant within areas.  Further discussion of $\delta_{ic}$ is postponed to Section \ref{sec:interp}.

The above cluster-level model may be described as a two-fold nested error regression unit-level model \citep[Section 4.5.2]{rao:molina:15} for binary data, with a binomial sampling model. 

When estimating poverty, income or inequality measures, linear models like the basic unit-level model have been used and there is great interest in such approaches. \cite{elbers2003micro}  use a nested error regression model with random effects at the cluster level instead of at the area-level. Their method, commonly called the ELL approach (after the authors of the paper,  Elbers, Lanjouw, and Lanjouw), has been widely used to estimate small area means of poverty-related quantities, especially by the World Bank.
However, as pointed out by \cite{das_comparison_2019}, the ELL model assumes that all between-area differences are explained by the covariates, which may not be reasonable when covariates are unavailable or weakly related to the outcome, see \cite{diallo:rao:18} for an extended ELL model, see also \cite{molina2010small}. \cite{molina2019small} review further improvements and discuss other aspects of poverty mapping. 

There are two major challenges to the use of unit-level prevalence models:~1)~appropriately accounting for the design and 2)~aggregating from clusters to areas. In the context of the household surveys that are common in LMICs, both aspects are even trickier, because of the lack of available information on the sampling frame. We next discuss 1), with 2) being delayed to Section \ref{sec:spatAgg}.

\subsection{Accounting for the Complex Design}\label{sec:accounting}

 Unlike Fay-Herriot area-level models, which incorporate sampling weights when computing direct estimates and their variances (and hence account for the design), unit-level models do not generally include sampling weights in the estimation procedure. Hence, it is crucial to include design  information in the unit-level model specification. 

One might naively include the weights in the regression model as a covariate, but aggregation would require the weights for the whole frame, and one is then still faced with the problem of getting an appropriate area-level variance estimate. There are more complex procedures that include incorporating functions of the selection probabilities in the model  \citep{verret_model-based_2014}, or model the sampling weights to account for informative sampling, see \cite{pfefferman:07} for a frequentist approach and \cite{si:etal:15} for a Bayesian slant. \cite{parker2023comprehensive} review various approaches to account for informative sampling in the unit-level model.

 It is (often implicitly) assumed that the sampling design is ignorable with respect to the model used for estimation. We need both (i) the specified risk model to be correct for all sampled and non-sampled individuals in both observed and unobserved clusters, and (ii) outcomes for any set of individuals from the target population to be independent, conditional on their risk parameters. Requirement (i) is often justified by including relevant design variables as covariates in the model, for example, including strata variables as covariates \citep{little:12}. However, when some of the relevant design variables are unavailable or their functional relationship to the response is unknown, then this approach may not be successful.

For household surveys in LMICs the stratification is usually Admin-1 area crossed with urban/rural (with each cluster being classified as urban or rural). To account for this, one could include fixed effects for Admin-1 areas, urban/rural status, and their interaction. A more pragmatic and parsimonious approach is to include random effects (spatial or otherwise) at Admin-1 or Admin-2, and the main effect of urban/rural only (thereby assuming that the association with urban/rural is approximately constant across Admin-1 areas).

For many indicators, non-negligible associations between prevalence and urban/rural will be present. Including an urban/rural indicator in the model is straightforward, but to then aggregate, the urban and rural population sizes must be known for each area of interest. Typically these sizes may be available at the Admin-1 level (in the survey reports) but are usually unavailable at the Admin-2 level. Note also that the urbanicity status of a cluster may change over time due to urbanization, but when unit-level models procedures adjust for urbanicity, it is not the current urban/rural classification of the cluster that is relevant, but rather, the classification when the sampling frame was constructed. The recorded stratification variable should thus be viewed as a fixed partition over time. A method for constructing urban/rural fractions has recently been developed, by producing an urban/rural surface over the study region, based on a classification algorithm \citep{wu2024modelling}. In Section \ref{sec:cultures} we describe an approach to estimating urban/rural fractions at Admin-2, in the context of subnational HIV prevalence estimation.

If forming an urban/rural surface is deemed too difficult, one may include readily available variables such as population density and/or nighttime lights in the regression model, as surrogates for urban/rural. The success of this strategy is difficult to anticipate. As a starting point, it is prudent to first estimate the association between prevalence and urban/rural. If the association is negligible, one might sleep a little better. One can also examine a sequence of cluster-level models that include urban/rural alone and also with population density.

Beyond stratification, the DHS typically implements PPS sampling of clusters, with size corresponding to the number of households in the cluster. However, the size of sampled clusters is not usually made public for privacy reasons and the size of non-sampled clusters is typically not known, meaning that cluster size may not be used as a covariate. As a result, if cluster size is associated with the outcome of interest, then the design is not ignorable. Flexible models to account for PPS sampling have been proposed \citep{zheng:little:05}.

Since accounting for stratification and clustering via model specification is delicate, other methods have been proposed for acknowledging the sampling design when using unit-level models. Many of these methods leverage the sampling weights during parameter estimation, including pseudo-EBLUP estimation \citep{you:rao:02}.
Pseudo-Bayesian approaches that produce weight-adjusted posteriors are also under development, but are tricky to extend to random effects models \citep{leon2019fully,williams2021uncertainty,gao2023pseudo}.


Binary data unit-level models in the SAE literature are often used in a high income countries context in which there are high quality census data available. In this case, the logistic regression model may include demographic groups (e.g.,~by age, sex, race) as covariates. In this case, aggregation consists of averaging over the group-specific risks in each area, weighting by the population in each cross-classification cell (with these subpopulation sizes being available from the census). Both \cite{malec:etal:97} and \cite{congdon:2010} follow this approach, which has strong connections with the multilevel regression and poststratification approach described in Section \ref{sec:MRP}. Aggregation for cluster-level models will be considered in Section \ref{sec:spatAgg}.

In Section \ref{sec:MBG} we describe further MBG cluster-level models which are particular unit-level models that have seen widespread use for analyzing household survey data in LMICs. 

\subsection{Computation}

Fay-Herriot models are simply LMEMs and computation is straightforward. 
In general, maximum likelihood (ML), restricted maximum likelihood (REML) and method of moment estimators fall under the frequentist umbrella for estimating variance components in the model \citep{rao:molina:15}. Linear unit-level models are also similarly relatively easy to implement, since thay are also LMEMs. Non-linear (e.g.,~logistic) models, are examples of GLMMs and frequentist inference is  more involved, particularly with spatial random effects, since unlike their linear counterparts, the integration over random effects cannot be carried out analytically. However, likelihood methods are still available.

Bayesian inference can be carried out for LMEMs and GLMMs with iid or spatial random effects using MCMC or INLA. There are many MCMC algorithms that could be used but spatial models, and the dependent posteriors they produce, can be problematic in the sense of displaying slow mixing \citep{rue:knorrheld:05}. While MCMC algorithms tailored for specific models can overcome some issues \citep[Chapter 4]{rue:knorrheld:05}, INLA offers an accessible alternative. INLA is extremely fast and LMEMs and GLMMs with iid or spatial random effects are exactly the kinds of models that INLA was designed for. The accuracy of the INLA method has been demonstrated on many occasions, see for example \cite{osgood2023statistical}.

Specific packages to implement area-level and unit-level models are described in Section \ref{sec:Rpackages}.


\section{Model-Based Geostatistics}\label{sec:MBG}

Like unit-level models from the SAE literature, MBG approaches treat individual outcomes as response data. Almost all unit-level SAE approaches assume iid or discrete spatial random effects, while in contrast MBG approaches assume continuous spatial surfaces that represent the risk of experiencing an outcome at any given location. 
These risk surfaces may be used to compute spatially averaged risks for areas of interest, and can be viewed as proxies for finite population prevalences. 
\subsection{Geostatistical Models}

For prevalence mapping with MBG, the individual response data are typically aggregated to the cluster level, with each cluster representing a distinct spatial location. 

The following model for mapping health outcomes at the Admin-2 level is typical (see for example \cite{mayala:etal:19}), and uses a binomial response model with a logit link and latent spatial random effects. The sampling model is again given by (\ref{e:logit-bhf-obs})
with,
\begin{equation}
        \mbox{logit}(r_{ic})=\alpha+\boldsymbol{z}(\boldsymbol{s}_{ic})^\top\boldsymbol\beta+\omega(\boldsymbol{s}_{ic})+\delta_{ic}\label{eq:unit:Geo}
\end{equation}
where $\boldsymbol{s}_{ic}$ denotes the cluster location, the logit baseline odds is $\alpha$, $\boldsymbol{z}(\boldsymbol{s}_{ic})$ is a vector of cluster-specific covariates  and the associated log odds ratios are  $\boldsymbol\beta$. The spatially correlated random effects are denoted $\omega_{ic}=\omega(\boldsymbol{s}_{ic})$ and $\delta_{ic} \mid \sigma_\delta^2 \sim_{\mathrm{iid}}\mbox{N}(0, \sigma_\delta^2)$ are independent cluster random effects.  The area-specific random effects used in the previously described SAE unit-level model (\ref{eq:unit:SAErisk}) have been replaced by 
$\omega_{ic}$, which in most MBG approaches is assumed to arise from a GRF model, often with a Matérn correlation structure \citep{stein:99}.  The discussion on acknowledging the design in Section \ref{sec:accounting} carries over to MBG models, so that when samplng is informative, one is required to include design variables in the covariate model.

This model describes risk as a function of location (where the GRF is continuous, but covariates can be discontinuous), with $\delta_{ic}$ modeling ``random deviations'' from the risk surface at an observed cluster. In summary, the MBG form \eqref{eq:unit:Geo} aims to describe as much of the residual spatial variation as possible, while the SAE version \eqref{eq:unit:SAErisk} emphasizes the prevalences in the discrete \emph{target areas} and the area-specific random effect does not try to explain residual spatial variation at a finer scale than the target areas. The former can lead to improved unit-level predictions compared to the latter, but for area-level summaries, it is not clear which model will be beneficial.

\subsection{Interpretation of the Cluster-Specific  Error Term}\label{sec:interp}

In Equation \eqref{eq:unit:Geo}, as discussed in detail by \cite{dong:wakefield:21}, $\delta_{ic}$ may represent one of three distinct types of variation. It may be a surrogate for ``measurement error'', that is, the misrecording of responses. Given the binary nature of the responses, this would not be an ideal model, which would instead contain misclassification probabilities. If this is the interpretation given, then  $\delta$ contributions would not be included in predictions. Some authors have not included the cluster-specific effect in prediction \citep{wakefield:etal:19,diggle:giorgi:19}. 

Another, very plausible interpretation is that the terms are a mechanism to induce {\it overdispersion}, i.e.,~excess-binomial variation, due to within-cluster dependence in responses. In this case, the predicted risk at location $\boldsymbol{s}_{ic}$ is the marginal risk, averaged over $\delta_{ic}$, so that for a generic cluster,
 \begin{eqnarray}   
    \mbox{E}[~ r_{ic}&\mid&\alpha,\boldsymbol\beta,\omega(\boldsymbol{s}_{ic})~] \nonumber \\
&=& \mbox{E}[ ~\mbox{expit}(\alpha+\boldsymbol{z}(\boldsymbol{s}_{ic})^\top\boldsymbol\beta+\omega(\boldsymbol{s}_{ic})+\delta_{ic})~], \label{eq:deltmarginal}
\end{eqnarray}
where the expectation is over $\delta_{ic} \mid \sigma_{\delta}^2\sim \mbox{N}(0,\sigma_{\delta}^2)$.

The third explanation is that $\delta_{ic}$ represents ``true'' variation in the risk surface, i.e.,~unstructured between-cluster variation. In this case, when one aggregates across an area, $\delta$ contributions must be included. Hence, for an unobserved cluster, operationally the risk is (\ref{eq:deltmarginal}), regardless of whether we assume that $\delta$ is included to accommodate overdispersion or true signal, because in both cases we need to average over $\delta \mid \sigma_\delta^2$.



\subsection{Estimating Prevalence across Administrative Areas: Aggregation}
\label{sec:spatAgg}

In the MBG spatial statistics formulation with a GRF model, an aggregated area-level risk target is conceived as,
\begin{equation}\label{eq:aggint}
r_i = \int_{A_i} r(\boldsymbol{s}) d_i(\boldsymbol{s}) ~\mbox{d}\boldsymbol{s},
\end{equation}
where $A_i$ is the geographical extent of area $i$ and $d_i(\boldsymbol{s})$ is a normalized density surface which represents the locations of the relevant population.
In contrast, the finite population prevalence  target in the SAE world is,
\begin{equation}\label{eq:aggintpop}
P_i = \frac{Y^{\text{\tiny{POP}}}_i}{N_i} = \sum_{c = 1}^{C_i} \frac{Y^{\text{\tiny{POP}}}_{ic}}{N_{ic}} \times  \frac{N_{ic}}{N_i},
\end{equation}
where (recall) $C_i$ is the total number of clusters in area $i$, and $N_{ic}$ and $Y^{\text{\tiny{POP}}}_{ic}$ are the population denominators and numerators, with $$N_i =\sum_{c=1}^{C_i} N_{ic} ,\qquad Y^{\text{\tiny{POP}}}_i = \sum_{c=1}^{C_i} Y^{\text{\tiny{POP}}}_{ic},$$
being the totals in area $i$, $i=1,\dots,m$. 

The links  between (\ref{eq:aggintpop}) and (\ref{eq:aggint}) are clear with the prevalence  $Y^{\text{\tiny{POP}}}_{ic}/N_{ic}$ being replaced by the   risk $r(\boldsymbol{s})$  and the population fraction $N_{ic}/N_i$ being replaced by the density $d_i(\boldsymbol{s})$. The numerator  and denominator  of the weighted estimator  $ \hat{p}^{\tiny{\text{w}}}_{i} $, defined in \eqref{eq:HT}, directly estimate the numerator ($Y^{\text{\tiny{POP}}}_{ic}$) and denominator ($N_i$) in \eqref{eq:aggintpop}. We also contrast with the unit-level model that lead to the averaged risk (\ref{eq:unitrisk}) in which  a superpopulation model provides estimates for the set of $N_i$ finite individuals in area $i$. In a perfect world, master sampling frame information would be available for each cluster, including the relevant population sizes and locations, along with all covariates used in the model. Unfortunately such information is never available.



Under the MBG approach, risk may be modeled via  \eqref{eq:unit:Geo}, yielding a surface representing risk at any location $\boldsymbol{s}$ in the study region, regardless of whether a cluster exists at the location.
To approximate \eqref{eq:aggint}  in area $i$, a grid consisting of $G_i$ cells with centroids $\boldsymbol{s}^{\mbox{\tiny{G}}}_{ig}$, $g = 1, \dots, G_i$, is created, for example, at resolution  $1\, \text{km}\times 1\, \text{km}$ or $5\, \text{km}\times 5\, \text{km}$ \citep{utazi2020geospatial,local2021mapping}. 
The risk for area $i$ is then estimated by the population-weighted estimated risk across the grid,
\begin{equation}\label{eq:rsmooth}
 r_i = A \sum_{g = 1}^{G_i} r(\boldsymbol{s}_{ig}^{\mbox{\tiny{G}}} ) d_i(\boldsymbol{s}_{ig}^{\mbox{\tiny{G}}} ) 
\end{equation}
where the factor $A$ is defined as the area of the cells and $d_i(\boldsymbol{s}_{ig}^{\mbox{\tiny{G}}} )$ is the normalized population density at grid location $\boldsymbol{s}_{ig}^{\mbox{\tiny{G}}} $.
Current MBG approaches implicitly assume that the population
in the administrative area is so large that the difference between prevalence and
risk is minor.  With reference to \eqref{eq:aggintpop}, this implies changing the target of inference to
$r_i = \mathrm{E}[P_i 
] $.

Following the discussion in Section \ref{sec:interp}, under either overdispersion or true signal interpretations, we should include $\delta_{ic}$  contributions to the area-level averages, i.e.,~with $\delta_{ig}^{\mbox{\tiny{G}}}$ terms in the
predictions at each grid cell \citep{burstein2019mapping,utazi2020geospatial,utazi2021vacc}. 
For unobserved clusters, we use (\ref{eq:deltmarginal}) 
    and it is this expectation which is being implicitly used when aggregation is carried out.
    However, since grid cells are not matched to clusters, there is  an arbitrary element here, since the results will change when different gird sizes are used.
    This argues for taking $G_i=C_i$, though the latter may not be known and it still leaves the problem of choosing locations, $\boldsymbol{s}_{ig}^{\mbox{\tiny{G}}} $, $g=1,\dots,G_i$.
    

    It is often not explicitly stated whether the  cluster random effects $\delta$ are included in the aggregation. 
\citet{paige2021spatAgg} discuss population-weighted spatial aggregation in detail. Issues related to interpreting the cluster-specific non-spatial variation have received little discussion in recent efforts to use MBG for prevalence mapping of demographic and health indicators \citep{JSSv078i08,burstein2019mapping}.  



\cite{chan2019bayesian} take a somewhat different geostatistical view, explicitly acknowledging the complex design. They develop theory for a continuous response, with available covariates assumed to acknowledge any informative sampling. Follow up work includes \cite{finley2024models} and \cite{chan2022bayesian} who take a preferential sampling viewpoint with a selection model that depends on location through covariates. Complex survey designs lead (implicitly) to spatially varying sampling efforts. This resembles geostatistical preferential sampling \citep{diggle:etal:10}, but
in survey sampling, the inclusion probabilities are not directly linked to the level of risk in the clusters. The inclusion probabilities are instead determined by a survey design that aims to reach the desired statistical power for areas of interest in a cost efficient way. This does not mean that the survey design should be haphazardly ignored, however, as we have emphasized. 

In many applications that we have looked at using MBG approaches (and the SPDE approximation) we have found that the reported prediction intervals are narrower than the corresponding area-level intervals.
The MBG approaches make more assumptions and so, if those assumptions are met, can result in more precise estimation. However, the reported uncertainty 
 could be artificially low because of missing sources of error, errors in the aggregation process, or other model failures.

\subsection{Computation}\label{sec:computation}


%

MBG analyses usually adopt a Bayesian approach to inference, in contrast to the frequentist approaches which are dominant in the SAE literature. Samples may be obtained from the posterior distribution of $r(\boldsymbol{s})$ for any location $\boldsymbol{s}$, or for the spatially averaged area-level risks $r_i$, via (\ref{eq:rsmooth}). The posterior uncertainty for any variable may be quantified using credible intervals or by reporting the posterior variance or posterior coefficient of variation.

When using a continuous spatial model, the number of observed clusters $n=n_1+\dots+n_m$, is large, and computation is an issue because we
need to manipulate $n \times n$ matrices, which involves $O(n^3)$ operations \citep{rue:knorrheld:05}.
Many approximations have been proposed to overcome this problem, see \cite{heaton:etal:18}. The stochastic partial differential equation (SPDE) approach pioneered by \cite{lindgren:etal:11} is particularly popular for MBG approaches in LMICs. SPDE models, implemented using INLA, have been used in LMICs to produce high resolution maps of health and demographic outcomes, including vaccination coverage \citep{utazi2020geospatial,utazi2021vacc}, neonatal, infant, and child mortality \citep{golding:etal:17,burstein2019mapping}, childhood malnutrition \citep{kinyoki_mapping_2020}, contraceptive usage rates \citep{bosco_exploring_2017}, and malaria prevalence rates \citep{bhatt:etal:15}. 
The SPDE method is still seeing active development \citep{lindgren2022spde}.
The INLA software implementation has restrictions on the models that can be fitted, and so Template Model Builder (TMB) \citep{kristensen:14} has been used an alternative for GRF models \citep{burstein2019mapping}, when more flexibility in modeling is required. TMB and INLA are compared in \cite{osgood2023statistical}.

\section{Additional Issues}\label{sec:additional}

In this section, we discuss a number of other issues that are pertinent to prevalence mapping.

\subsection{Geographical Issues}\label{sec:geogissues}

Although the clusters in DHS and MICS surveys are typically census enumeration areas, GPS coordinates are supplied for the center of each cluster only, and we treat these as points in our analyses, following convention. For confidentiality, urban/rural cluster locations are displaced by up to $2\, \mathrm{km}$/$5\, \mathrm{km}$ with the locations of a random sample of a further 1\% of rural clusters being jittered by up to $10\, \mathrm{km}$.  In many cases, the jittering algorithm is constrained to keep the location in the same sampling stratum (i.e.,~Admin-1 crossed with urban/rural). As a result, some clusters may be randomly relocated to an Admin-2 area that does not correspond to the true area. The displacement has implications for modeling and has been investigated methodologically, as we detail shortly, and from an applied perspective, see for example \cite{tonye2024influence}.

For continuous spatial models, the jittering distances are short compared to the typical effective range parameter for GRFs in geostatistical models and the displacement has been demonstrated to have only a minor impact on parameter estimation and predictive accuracy \citep{wilson2021estimation,altay2022accounting}. However, rasters can show large variation in values within typical jittering distances and failure to account for this local variability reduces predictive power \citep{altay2024impact}.  With respect to covariate modeling, \cite{perez-heydrich:13} and \cite{warren:etal:16} have suggested using a buffer zone of $5\, \mathrm{km}$ within which to average the covariates. However, these methods still exhibited reduced predictive power compared to methods fully accounting for covariate uncertainty and complicate the interpretation of covariate models \citep{altay2024impact}.  

In older surveys (especially MICS surveys), the geographical coordinates of the clusters may not be available, with only the areas within which the cluster lies being accessible. \cite{marquez2021harmonizing} develop a  model for data in which polygon information only is available, by averaging over potential locations, in contrast to previous approaches that proposed ad hoc algorithms \citep{golding:etal:17,reiner2018variation,utazi:etal:18}. 

In this paper, we argue that in many situations we may include
discrete spatial random effects at a level that corresponds to  the level at which inference is required, particularly at coarser geographical levels.  However, one situation in which continuous spatial models are more appealing is when multiple datasets are available over areas with different boundaries. Continuous spatial models provide a natural way to model risk at each spatial location, 
particularly when the boundaries shift over time, or the different geographies are not nested.

\subsection{Auxiliary Variables}\label{sec:auxiliary}

To produce reliable subnational estimates, at fine geographical resolution in particular, requires auxiliary variables with good predictive power. One simple viewpoint is that SAE is a predictive exercise and so one can be flexible in the strategy used for constructing a covariate model. But auxiliary information may be available from  a variety of sources, and some are more preferable than others. 
The use of census variables is common in high income countries, but in LMICs censuses are less frequently carried out and the data they produce are less reliable. Administrative source data such as from health facilities offer great potential
but are currently not routinely available. Satellite data, such as night time lights and vegetation indices, are becoming widely available,  and are being extensively used, see Section \ref{sec:sausage}. Meteorological and climate variables may be useful, though are perhaps less obviously related to demographic and health variables, and in some cases will show little variation over the study region. Economic, household, cultural and lifestyle variables (for example, dietary variables) may show strong associations in regression settings. However, to obtain these variables at finer  geographical resolutions requires modeling, based on survey data \citep{gething:etal:15}, which opens up a whole can of worms.  The utilization of mobile phone and social media data are exciting prospects but their use is currently in its infancy.

One should also be careful in selecting variables to include in the model.
In an extreme case, suppose we are modeling child mortality and there is an available covariate that is a composite measure of poverty. We would clearly not want to use such a variable if it were constructed using a measure of child mortality.

When modeling at a fine pixel-level scale there are a number of issues that require consideration.
The displacement of cluster locations is relevant to the use of pixel-level covariates, as discussed in Section \ref{sec:geogissues}. For a fine-scale pixel surface, the associated covariates on the pixel grid may take a large range of values, but the cluster-level data from which the covariate model is estimated may not adequately cover this range, particularly when there are a large number of covariates. This may result in aberrant predictions or pixels with sets of covariates that are not adequately represented in the observed clusters.

Model choice is in its infancy in SAE, but is important, particularly if there are a large number of covariates. Machine learning methods can deal with large numbers of variables, but have their own challenges, as discussed in Section \ref{sec:sausage}.
 
\subsection{Uncertainty Measures}

The presence of random effects in SAE models complicates the interpretation of uncertainty intervals. 
Under the frequentist paradigm, bias and variance are calculated with respect to repeated sampling of both the data and the random effects. 
Consider a linear model with area-level random effects, $u_1,\dots, u_m$. Then the BLUP $\hat{u}_i$, for a generic area $i$, is unbiased in the sense that $\mbox{E}[\hat{u}_i-u_i]=0$, where the expectation is with respect to the distributions of both the distributions of $y_{ij}$, and $u_{i}$. However, in a real population, $\hat{u}_i$ is biased in the sense that  $\mbox{E}[\hat{u}_i-u_i\mid u_i] \neq 0$, $i = 1,\dots, m$,  due to shrinkage. Hence, the ``Unbiased'' in BLUP is deceptive and is not relevant for making inference about {\it any particular area} because the area-level predictor $\hat{u}_{i}$ is clearly biased.

When frequentist analyses are carried out, the mean squared error (MSE) is often used to quantify the uncertainty for each area, and is defined as E$[(\hat u_{i} - u_{i})^2] = \mbox{var}(\hat u_{i} - u_{i})$ where, again, the expectation is taken over both data and random effects distributions. As with the first moment discussion above, $\mbox{var}(\hat{u}_i-u_i)$ differs from $\mbox{var}(\hat{u}_i-u_i\mid u_i)$. 
Conditional MSE has been considered, \citet[Section 6.2.7]{rao:molina:15}.

The posterior variance $\mbox{var}(u_{i} \mid \boldsymbol{y})$ is a Bayesian measure of uncertainty which represents the variance of the random variable $u_{i}$ given the data and chosen priors, and conditional on the model being correct. Credible intervals can be constructed from the posterior for any parameter of interest when samples from the posterior are available. For prevalences close to zero and with small samples in particular, the posterior will be skewed and the posterior intervals will reflect this, which is desirable. Such intervals are also exact, up to the numerical approximation that is used for computation. If the posterior approximations can be improved, these intervals can be made arbitrarily close to exact. In contrast, frequentist intervals are based on asymptotic normality, which is somewhat ironic given the ``small'' in SAE. 

Direct estimates have asymptotically correct coverage for every area since they use data from the relevant area only, and do not employ shrinkage.
Depending on one's inferential persuasions, one may be comforted when Bayes intervals coincide with frequentist intervals in large samples. However, for random effects models, the interpretation of both frequentist confidence intervals and Bayesian credible intervals is fraught with difficulties. 
The discussion that follows is true for more complex random effects models (including unit-level models), but for simplicity, consider a simple Fay-Herriot model with E$[\hat \theta_{i} \mid u_{i}] = \alpha + u_{i}$ and to keep the discussion simple, assume $\alpha$ is known.
The 95\% prediction interval is, $(\alpha + \hat u_{i}) \pm 1.96 \times \sqrt{\mbox{MSE}(\hat u_{i} )}$.
For this area, and assuming a fixed $u_{i}$, if we repeatedly sample data and produce intervals, those intervals will not contain the true $u_i$ 95\% of the time. What we can say is that the coverage will be approximately 0.95 if we average over {\it all} possible areas, that is, we need to also average over the random effects distribution.  Due to the shrinkage, areas which have extreme, ${u}_i$, in the tails (away from zero) will have coverage below the nominal 0.95 while areas in the center of the distribution, will have coverage above 0.95. This realization is deeply disturbing, since it will often be areas with relatively low or high prevalence that will have the lowest coverage.

The silence on these issues in both the SAE and MBG literatures has been deafening.
A notable exception is \cite{burris:hoff:19}, who discuss this issue in detail, and introduce a new method for constructing intervals that borrow strength across areas but retain the correct frequentist coverage for each area.
The average width of the intervals is narrower than the width of direct estimates, but not as narrow as from the random effects model. 

\subsection{Model Checking}

We briefly discuss model checking for each of weighted, area-level, unit-level, and MBG approaches.
One of the beauties of the weighted (direct) estimator is the minimal assumptions under which it is based. However, achieving nominal coverage requires a sufficiently large sample size to ensure the accuracy of the normal sampling model for the estimator. Instability of the variance estimator can reduce the accuracy. As mentioned in Section \ref{sec:introduction}, agencies produce guidelines for reporting weighted estimates (often in terms of coefficients of variation), which is presumably at least in part implicitly based on considerations of accuracy of the normality assumption and variance stability.

Beyond the appropriateness of the sampling distribution, Fay-Herriot models assume a particular random effects model on a particular linking scale. 
In our experience, the particular form of discrete spatial model (e.g.,~BYM2, CAR, SAR), will not be critically important in many examples, as all perform some form of local smoothing. Plotting the smooth small areas estimates versus the direct estimates (when available) allows the shrinkage to be examined. Deciding on when there is ``too much shrinkage" is a critical question, with currently no clear guidelines. 

SAE unit-level models are far more dependent on model assumptions. Unfortunately, the sparsity of data within each cluster (since there are usually a small number of units within each) makes critiquing the sampling model difficult. 

MBG approaches are the most dependent on assumptions, with all difficulties associated with unit-level models being present, along with a more delicate spatial model. 
Non-parametric exploration of the correlation function through, for example, semi-variograms, is challenging since we have sparse observations and count data. In general, in the context of the survey sample sizes in LMICS, spatial parameters (such as the total variance and proportion spatial parameter in the BYM2 model, and the spatial variance and range parameter in the GRF model) can be poorly estimated, because of the sparsity of the survey data and paucity of information in spatially dependent data \citep{zhang2004inconsistent}. Comparing the posteriors to the priors, and examining sensitivity of predictions (point and interval estimates) to spatial hyperpriors is recommended.

Cross-validation provides an obvious approach for model assessment, though there is no consensus on how it should be performed in a spatial setting with complex survey data. The former aspect is discussed in \cite{roberts:etal:17}, particularly in the context of blocking strategies where care should be  taken with the manner in which the data are split. We routinely carry out leave-one Admin-1 area out cross-validation, and then predict the direct estimate, using the remainder of the areas. This can be useful, but when there are a small number of Admin-1 areas (such as in our Zambia example) model assessment is difficult. This procedure is also not directly answering the question of interest, ``Is the model adequate at Admin-2?''. Clearly, it will be more difficult to assess model performance when the data are sparse. There are many choices to make when cross-validation is carried out, beyond the splitting strategy, including which metrics to use to compare  left out data with predictive distributions. The procedure for leaving out data should also be consistent with the survey design.
In Section \ref{sec:cultures} we examine a number of different options.

\subsection{Multilevel Regression and Poststratification}\label{sec:MRP}

Multilevel Regression and Poststratification (MRP)  is a technique that was introduced by \cite{gelman1997poststratification}.  Under one version, a multilevel regression model is used to model a binary outcome at the unit level, as a function of a potentially large set of categorical variables. The population is partitioned into cells representing each possible combination of categorical variables and units within the same cell are aggregated. In the poststratification step, the population prevalence in each area is computed as a weighted average of the cell estimates with weights corresponding to the population fractions in the area. In this weighting step, there are close links with Generalized Regression (GREG) estimators; see \citet[Section 2.3.2]{rao:molina:15}.

A key component of the approach is the random effects prior on the parameters corresponding to each categorical variable. The existence of a large number of cells means that the use of hierarchical smoothing is essential to overcome data sparsity. Many of the variables, including space, time, age and education levels, are candidates for smoothing priors. To ensure that no bias results from informative sampling, the relevant design variables must be included in the set of  variables that are included in the model. MRP is particularly popular among political scientists, see \cite{park2004bayesian} for an early example and \cite{wang2015forecasting} for a successful use of the method when the original data (which were based on a sample of Xbox users) were not representative. \cite{valliant2020comparing} reports a simulation study in which MRP does not perform particularly well in a situation in which a limited number of covariates were available. 
MRP can be viewed as a unit-level SAE model in which individual responses within the same cross-classification cell of (say) demographic subgroups are combined and then modeled across areas, with averaging over these subgroups corresponding to the aggregation step.

\subsection{Spatial Machine Learning}\label{sec:sausage}

Given the abundance of geospatial covariates now available, in the context of prevalence mapping in LMICs (as discussed in Section \ref{sec:auxiliary}), it is natural that 
machine learning (ML) methods are growing in popularity. ML approaches are appealing in principle, as they can provide flexible modeling (to capture nonlinearities and interactions, for example) when there are a large number of auxiliary variables, while potentially avoiding a difficult model selection process. ML approaches have been used both with SAE area-level and unit-level models, and with MBG models. 

While it is relatively straightforward to use generic ML algorithms (once the covariate data have been massaged into a usable form), there are a number of issues that require careful thought, including accounting for the survey design, choosing tuning parameters in the ML algorithm, and, most importantly, obtaining valid interval estimates. There are serious complications with the latter since great care is required to find a justifiable approach  with a vast literature available on resampling methods such as the bootstrap, jackknife, infinitesimal jackknife and split conformal prediction methods. For many ML algorithms it is not clear which, if any, of these methods will give a valid prediction interval in any given context. \cite{dezeure2015high} provide a discussion of various aspects, including describing ML algorithms for which the vanilla bootstrap does not work, because the limiting distribution of the predictor
is a complicated object which may not be continuous. The validity of a particular approach in a mapping context may also depend on an assumption of 
iid outcomes which will not hold for data from surveys, in which there are dependencies due to both the sampling design and the spatial nature of the data. In addition, a recurring theme is that the binary nature of the data is rarely acknowledged when ML techniques are applied.


\cite{jean2016combining} were early proponents of the use of convolutional neural network (CNN) approaches for poverty mapping, but notably there was little discussion of the validity of uncertainty estimates.
\cite{burke2021using} provide a review of ML techniques in the context of the SDGs, but again avoid the thorny question of uncertainty quantification. 

We highlight a number of applications of different classes of ML algorithms, beginning with random forests (RFs). 
\cite{georganos2021geographical} describe a ``geographical random forest'' in which a local RF is constructed and illustrated by modeling population density in Dakar, using satellite data.
\cite{krennmair2022flexible} examine the use of a RF covariate model embedded within a linear mixed effects model and describe a bootstrap procedure to produce measures of uncertainty (while noting that more theory is required on the procedure). \cite{bilton2017classification} apply classification trees to poverty mapping at a low geographic level using survey data from Nepal. The authors use weights both in the classification algorithm and in calculating measures of uncertainty using a bootstrap.

\cite{stevens2015disaggregating} use RFs to produce high-resolution gridded maps of population, which is an extremely important endeavor. 
\cite{ratledge2022using} apply a CNN with satellite data to attempt to assess the causal impact of electrification. \cite{far2023small}, in a wide-ranging review critiques and compare various approaches (including CNNs and tree-based methods) for wealth mapping using geostatistical data.



 \cite{bosco_exploring_2017} compare artificial neural networks (NNs) with a MBG for estimating a number of development indicators in Bangladesh, Kenya, Nigeria and Tanzania, using a range of geospatial coordinates. They report mixed success in prediction across indicators and countries.

 A quite different use of NNs is described by \cite{semenova_priorvae_2022} who use variational autoencoders to create a class of Gaussian process priors, which can then be used to model spatial random effects such as the $\omega$ process used in \eqref{eq:unit:Geo}. \cite{wikle2023statistical} provide a review of deep NNs for spatial and spatiotemporal data, describing many approaches to constructing flexible spatial models. The review highlights difficulties with the use of deep learning:~large datasets and high computational burden; the black box nature of the fitting that makes interpretation hazardous; and the difficulties in obtaining measures of uncertainty for the predictions.
The second and third of these flaws is common to many spatial ML approaches.

\cite{chi2022microestimates} link DHS data on household wealth variables to high-dimensional data from a variety of sources including satellite imagery, mobile phone networks, topographic maps and Facebook. A data reduction exercise is carried out and then a gradient-boosted regression tree is used for prediction on a $2.4\, \mathrm{km} \times 2.4\, \mathrm{km}$  grid for 135 LMICs. They acknowledge that providing a measure of the uncertainty of the map is important, but their approach to computing this uncertainty is ad hoc. \cite{yeh2020using} model a wealth index using satellite data  and DHS data, with a CNN. As is typical the complex survey design is ignored and uncertainty estimates are not reported. However, an interesting aspect of this paper, is attempting to look at changes in wealth over time. \cite{corral2025poverty} examine the performance of ML approaches to poverty mapping. They compare validation techniques that compare predictions with direct estimates, rather than census-based measures, showing that the former can be misleading while the latter can provide a more realistic measure of the performance of an algorithm. \cite{newhouse2025small} examine poverty mapping using poverty data from a survey and compare CNNs using satellite covariate data with traditional census-based small area estimates. The model they develop uses the (scaled) design weights and is a unit-level model with two levels of random effects.


\cite{bhatt:etal:17} use an ensemble method known as stacked generalization to provide a more flexible mean function within an MBG model. Unfortunately, there is no theory to give valid inference with such an approach but it has still been used by IHME on numerous occasions, see for example, \cite{golding:etal:17,osgood2018mapping,browne2021global}. \cite{lloyd_using_2020} outline a stacked generalization approach to predicting the residential status of buildings in urban areas.


As pointed out above, a large impediment to the routine use of ML techniques in a spatial context is constructing valid interval estimates on predictions. A modified split conformal procedure is used by 
\cite{michal2024model} to obtain intervals for lasso and RF models. \cite{wieczorek2024design} describes the general use of conformal prediction in a survey sampling setting. \cite{mao2024valid}, in the context of continuous response data, develop methods for applying conformal prediction algorithms to spatial data by exploiting approximate local exchangeability.
\cite{mcconville2017model} describe a model-assisted regression model with a lasso, and derive the variance of the estimator (to give an uncertainty estimate), and illustrate its use in the context of tree canopy cover estimation. In general, asymptotic arguments are used to derive variance estimates, which again should be taken in the context of sparse data situations. Conformal prediction approaches (at least in their current guise) also have sample size limitations, which may lead to interval estimates that are not practically useful. The same drawback is also true for prediction-powered inference (PPI) approaches to forming intervals \citep{angelopoulos2023prediction}, which have strong links with a number of literature strands including model-assisted estimation \citep{sarndal:etal:92} and semi-parametric inference \citep{robins1995semiparametric}.

 \cite{saha2023random} and \cite{zhan2024neural} consider GRF models with RF and NNs regression models, respectively. Rigorous frequentist analysis of point predictions is carried out, but predictive intervals are more difficult to calculate, as they point out, for example in the discussion of \cite{zhan2024neural}. More recent related work on modeling binary data with RFs is reported in \cite{saha2025randombinary}. 

There are Bayesian implementations of ML models, which offer the possibility of obtaining valid uncertainty intervals, under correct model specification. \cite{macbride2025spatial} embed a RF in a Bayesian hierarchical discrete spatial smoothing algorithm, but the frequentist RF implementation does not give a procedure that fully acknowledges the total uncertainty. 
\cite{jiang2023bart} combine a GRF with a Bayesian adaptive regression tree component, so the uncertainty quantification is fully Bayesian, but computational complexity is a serious deterrent to this approach and the simulations show that achieving nominal frequentist coverage with such an approach is challenging.

 Though ML approaches for prevalence mapping are very much in their infancy, they provide an exciting prospect for fully exploiting auxiliary information and providing estimates at granular levels, though assessing the accuracy and calibration of predictions and intervals is likely to remain a challenge.

\subsection{SAE {\tt R} Packages}\label{sec:Rpackages}

There are many R packages implementing variations of Fay-Herriot area-level models and unit-level linear models and here we mention only a subset.  The {\tt sae} package \citep{molina2015r} uses frequentist inference, and allows for spatial random effects, via a SAR model and space-time estimation with an autoregressive regression model of order 1 (AR1) model. The {\tt emdi} package \citep{kreutzmann2019r} also includes methods for area-level and unit-level modeling, allowing for SAR spatial random effects with linear unit-level models. The possibility of a Box-Cox transformation on the response may be integrated within the modeling procedure.
The package {\tt PovMap} \citep{povmappkg} adds features to {\tt emdi}. The {\tt tipsae} package \citep{de2024r} allows the fitting of area-level beta models.

Many Bayesian implementations of SAE models use the
\texttt{R} package \texttt{INLA} \citep{lindgren:rue:15,Rue2017review,Bakka2018review}, which 
implements approximate full Bayesian inference using the INLA method \citep{rue:etal:09}, and can be used for both discrete and continuous spatial models. The package is a popular choice for those using BYM2 or SPDE models. 
The \texttt{TMB} package \citep{JSSv070i05} provides empirical Bayesian
inference that allows fast inference for spatial models \citep{osgood2023statistical}.

Implementations of unit-level logistic models are less common.  
The {\tt RiskMap} package \citep{riskmappkg} allows unit-level logistic modeling, using GRF spatial models with Matérn correlation functions, and is the successor to the {\tt PrevMap} package  \citep{JSSv078i08}. 
An SPDE implementation of geostatistical models is available in the {\tt mbg} package \citep{mbgpkg}. \cite{altay2023geoadjust} provide an {\tt R} package for MBG models, accounting for DHS jittering.

The {\tt SUMMER} package \citep{li2025space} allows Bayesian spatial and space-time modeling for area-level and unit-level models, with the latter offering linear and logistic choices. In the space-time formulation, the BYM2 spatial model is combined with AR1 or random walk models of either order 1 or 2 (RW1 or RW2) temporal models, along with a variety of space-time interaction models. Implementation is based on INLA. 
The \texttt{SUMMER} package was used in \citet{li:etal:19} and \citet{UNbrochure} for subnational child mortality modeling over time using area-level and unit-level betabinomial models. A number of the SAE models that are available in the {\tt SUMMER} package have also been incorporated in the {\tt survey} package \citep{ lumley:04,surveypkg}. The \texttt{surveyPrev} package \citep{surveyprevpkg} further streamlines the full workflow of spatial prevalence estimation, from data preparation to modeling and visualization, and focuses on DHS and MICS data. It has an accompanying shinyApp\footnote{\url{https://sae4health.stat.uw.edu}}.

\section{Two Cultures in Practice: HIV prevalence mapping for adult women in Zambia}\label{sec:cultures}

\subsection{SAE and MBG Approaches}\label{sec:Zmodels}

We return to the 2018 Zambia DHS data introduced in Section \ref{sec:introduction}, in which the aim is to estimate the HIV prevalence/risk, among women aged 15--49, across Admin-1 and Admin-2 areas. Zambia has one of the highest HIV burdens in the world, as highlighted by \cite{dwyer2019mapping} and the subnational distribution of prevalence has been the subject of a number of studies, see for example, \cite{mweemba2022estimating} and \cite{cuadros2023geospatial}. Our analyses are not intended to be definitive or comprehensive, but rather to compare and contrast SAE and MBG approaches using various models.

Below we distinguish between methods that estimate the prevalence, $p_i$, versus those that estimate the risk, $r_i$, in areas $i=1,\dots,m$. 
We now summarize the seven models we use for comparison. Additional details on each model are provided in the Supplementary Materials, where we also report on a number of other models.

\begin{enumerate}

        \item[M1] \textbf{Direct:}  We use the H\'ajek estimator  $  \hat{p}^{\tiny{\text{w}}}_{i}$, as defined in \eqref{eq:HT}, for weighted (direct) estimation. 
    
        \item[M2] \textbf{Fay-Herriot BYM2 without covariates:} Given direct estimates $\hat{p}^{\tiny{\text{w}}}_{i}$, we compute $\hat{\theta}^{\tiny{\text{w}}}_i=\mathrm{logit}(  \hat{p}^{\tiny{\text{w}}}_{i})$ and estimate the associated sampling variances, $V_i$. These are used as inputs to an area-level model following \eqref{eq:fayherriot1} and \eqref{eq:fayherriot2} with a linking model that includes a BYM2 spatial prior on the area-level random effects, but no covariates. 
        \item[M3] \textbf{Fay-Herriot BYM2 with covariates:} As the previous model, except with area-level covariates.
        \item[M4] \textbf{Betabinomial BYM2 without covariates:} A betabinomial unit-level sampling model that allows for overdispersion at the cluster level, with a linking model that includes BYM2 spatial prior on the area-level random effects, but no covariates apart from an urban/rural indicator, which we discuss more fully below.
    
        \item[M5] \textbf{Betabinomial BYM2 with covariates:} As the previous model, except with unit-level covariates.
    
        \item[M6] \textbf{Betabinomial GRF without covariates:} A betabinomial unit-level sampling model that allows for overdispersion at the cluster level, with a GRF spatial prior on the area-level random effects and without covariates, apart from an urban/rural indicator.

        \item[M7] \textbf{Betabinomial GRF with covariates:} As the previous model, except with unit-level covariates.

\end{enumerate}
Models M1--M3 estimate prevalence, while models M4--M7 estimate risk.
Models M2--M5 are more in the spirit of an SAE approach, while M6 and M7 are examples of an MBG formulation. 


For the unit-level models (M4--M7) we need to account for the design.
Since the stratification is urban/rural crossed with Admin-1 regions, there is a justification for  Admin-1 crossed with urban/rural interactions. However, for reasons of parsimony, we assume a single fixed effect term for urban/rural combined with spatial random effects.
Under all approaches, we wish to produce estimates at both Admin-1 and Admin-2 levels. The GRF continuous spatial models give both levels under a single model, suitably aggregated. We could fit Admin-2 spatial BYM2 models only, and then aggregate up to Admin-1, but we prefer to pick as simple a model as possible to achieve estimates at the desired Admin level.
Hence, for each of the BYM2 approaches (models M2--M5), we fit two models one with BYM2 spatial effects at Admin-1 and one with these effects at Admin-2. 

At the Admin-2 level, out of the 115 areas, there are 3 areas with no data, and 14 for which the variance estimates are either zero or (inaccurately) close to zero because of sparse data. For all 17 areas, when carrying out the Fay-Herriot analyses at Admin-2, we include these areas as missing data.
We comment further on this aspect in Section \ref{sec:results}.

  A model that we have found useful for estimation at the Admin-2 level in other contexts is a nested BYM2 model with fixed effects at the Admin-1 level and {\it nested} BYM2 models within each Admin-1 area (with a sum-to-zero effect on the random effects within each Admin-1 area). We fitted this model also, but the results were very similar to the non-nested versions (as can be seen in the Supplementary Materials), and so  we do not include the results in the main paper. The similarity of results in this example is due to the relative abundance of data at Admin-1 (10 areas only) and the non-rarity of the outcome. 
  
  The betabinomial model  with overdispersion parameter $d$, for a generic cluster with sample size $n$ and risk $r$, has variance, $$n r(1-r)[1+(n-1)/( d+1)].$$ Hence, the excess binomial variation is $1+(n-1)/(d+1)$, which helps in the interpretation of $d$.
 
We use three covariates:~{\it access}, based on estimated travel times to cities in 2015 \citep{weiss2018global}; {\it malaria incidence}, as estimated for 2018 by the Malaria Atlas Project 
\citep{hay2006malaria} and {\it night time light intensity}, as observed via satellite imagery in 2016 \citep{roman2018nasa}.
 We transform the  night time lights variable to be $\log(1+x)$ where $x$ is a measure of the intensity of night time lights, so that the distribution of these variables are  not too skewed which could result in a small number of points influencing the predictions heavily.
All covariates are standardized to have zero mean and unit variance. We extract all covariate values on a $1\, \mathrm{km} \times 1\, \mathrm{km}$ grid of locations based on the population raster used by WorldPop. 
For the unit-level models, we assign covariate values to each sampled cluster based on the centroid of the cluster location. 

In order to perform aggregation with the unit-level models we need to evaluate the area-level estimate \eqref{eq:rsmooth}. The values of the three covariates and the population density values are all routinely available at the grid locations. In addition, we require  aggregation over urban/rural. The complexity of this aggregation depends on the model fitted, as detailed in the Supplementary Materials. In the most complex case, we require the urban/rural indicator at each grid point, which is not routinely available. We use the following approach to produce pixel-level maps of urban/rural status. 

We assume that we know the correct fraction of urban to total population in each Admin-1 area, which will be used to ensure the pixel urban/rural map is consistent with the area fraction. When no changes have been made to the sampling frame, the latter may be
found in the survey reports; for this study we were able to obtain the fraction of the urban population at Admin-1, directly from the census.  
When the sampling frame has been changed, an updated list must be acquired. The urban/rural pixel map is produced using the following steps:
\begin{itemize}
        \item[1.] 
        From WorldPop, download 
        $100\, \mathrm{m} \times 100\, \mathrm{m}$
        pixel maps of population density for the year matching the list of known urban proportions (this is 2010 for Zambia, the year of the last census),
        and age-specific population density maps for the year for which estimates are desired (this is 2018 for Zambia). 
        \item[2.] For each Admin-1 area, select a threshold and set pixels with population above the threshold as urban and values below the 
        threshold as rural. Using the all-age population density map, the thresholds are set to the level for which the resulting urban proportions are equal to the known urban proportion for each Admin-1 area.
        \item[3.] This Admin-1 consistent threshold is used across the whole map to obtain a grid-level set of urban/rural indicators.
\end{itemize}
In general, step 1. can introduce error, as the populations may neither be accurate, nor match exactly (in terms of age) the target population. Quantifying this uncertainty is difficult, however.

All computations were implemented in {\tt R} \citep{r_core_team_r_2024}. We conduct approximate Bayesian inference via INLA using  the {\tt SUMMER} and {\tt surveyPrev} packages for models M2--M5, and our own INLA code with an SPDE representation of the GRF for models M6 and M7. For all covariate models, we adopt default priors on the intercept  and slope parameters (independent Gaussian with zero mean and precision 0.001). For models with BYM2 random effects, we use penalized complexity (PC) priors for the variance parameters,  as suggested by \cite{riebler:etal:16}, with a scaled precision matrix. For the GRF models with SPDE random effects, we also use PC priors, as suggested by \cite{fuglstad:etal:19a}. See the Supplementary Materials for specific hyperparameter choices for these priors. 

\subsection{Results}\label{sec:results}

Ninety-three percent of women who were eligible for HIV testing and were
interviewed, consented and  provided a blood specimen for HIV testing.
Removing observations without geographic locations, the national weighted estimate of HIV prevalence for women 15--49 is 14.3\% (95\% interval: 13.2--15.4\%) and is more than twice as high in urban areas, 20.4\% (18.6--22.2\%) as in rural areas, 8.9\% (8.0--9.9\%). Based only on direct point estimates, $\hat p^{\tiny{\text{w}}}_i$, (see Figure \ref{fig:Zambia_map} for area names), Copperbelt has the highest female HIV prevalence among Admin-1 areas, with 19.9\% (16.5--23.4\%) and is more than three times greater than Muchinga, the lowest, with prevalence 6.3\% (4.6--8.0\%).

\begin{figure}
        \centering
        \subfigure[][Admin-1 prevalence.\label{fig:geography:Admin-1}]{
\includegraphics[width=9.5cm]{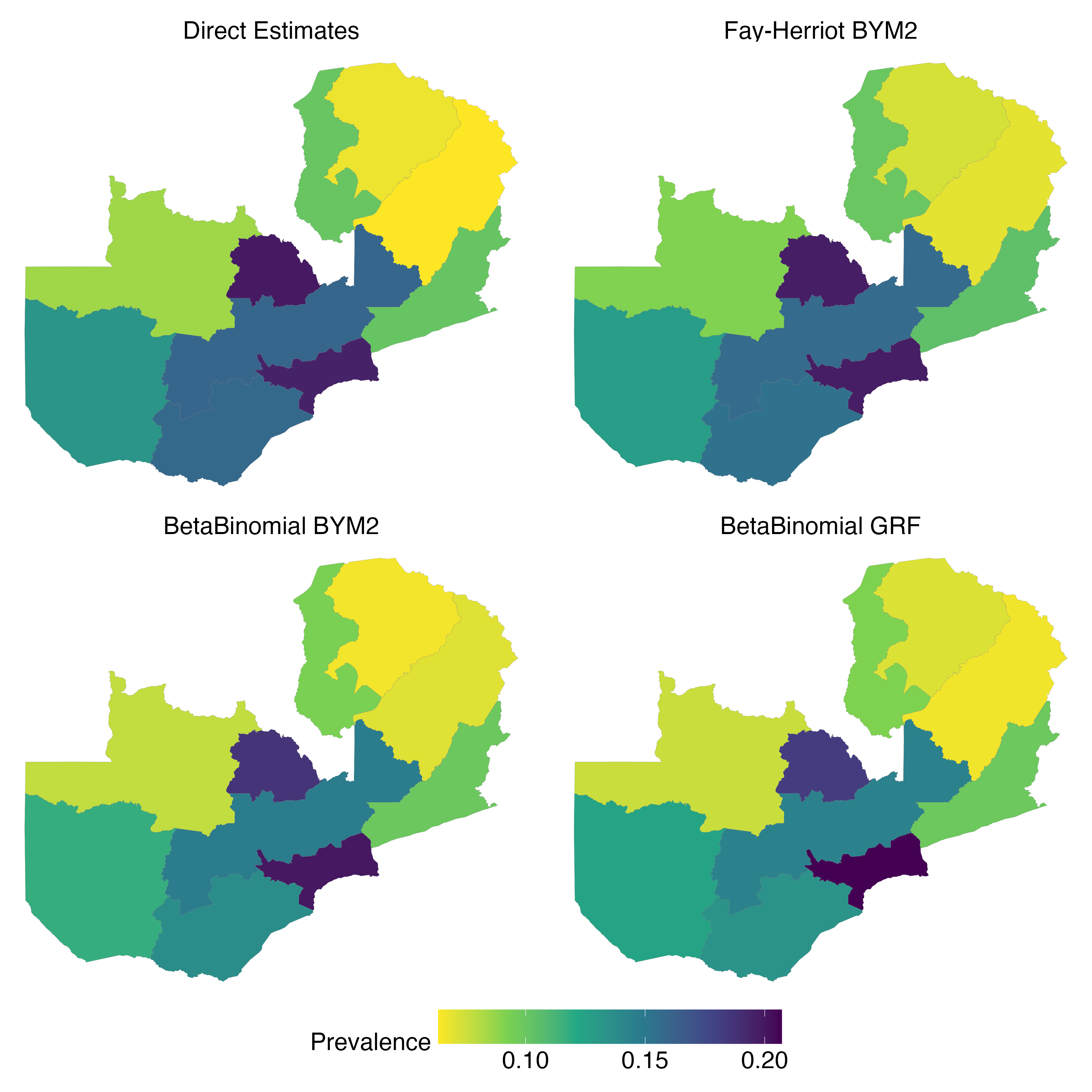}
                }
\subfigure[][Admin-2 prevalence.\label{fig:geography:Admin-2}]{
\includegraphics[width=9.5cm]{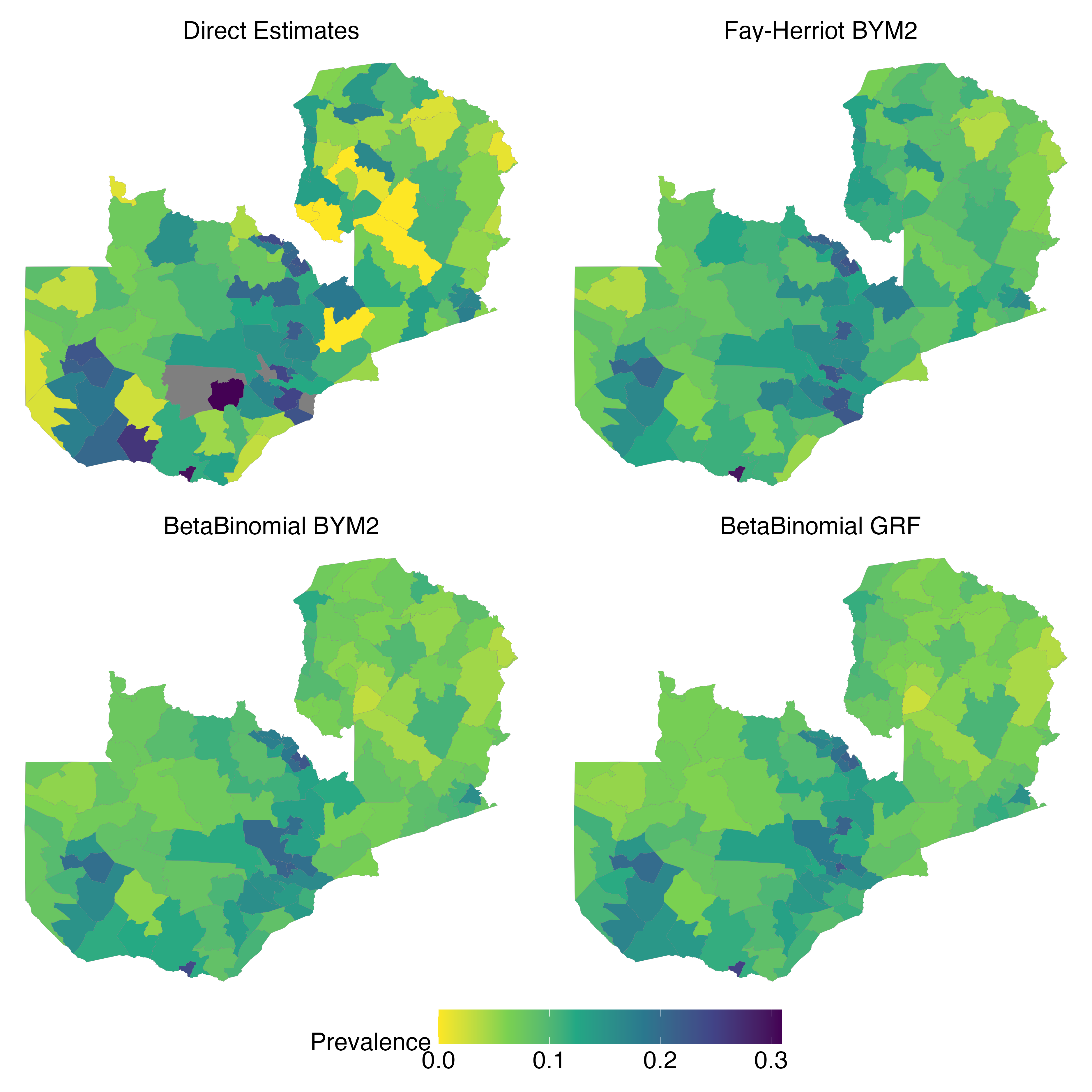}
                }
    \caption{HIV prevalence/risk estimates for women aged 15--49 in Zambia in 2018: \subref{fig:geography:Admin-1} Admin-1 areas, and
            \subref{fig:geography:Admin-2}  Admin-2 areas. The BYM2 and GRF models contain covariates.}\label{fig:geography}
\end{figure}
\begin{figure}
        \centering
        \subfigure[][Admin-1 CV.\label{fig:geography:Admin-1:cv}]{
                \includegraphics[width=9.5cm]{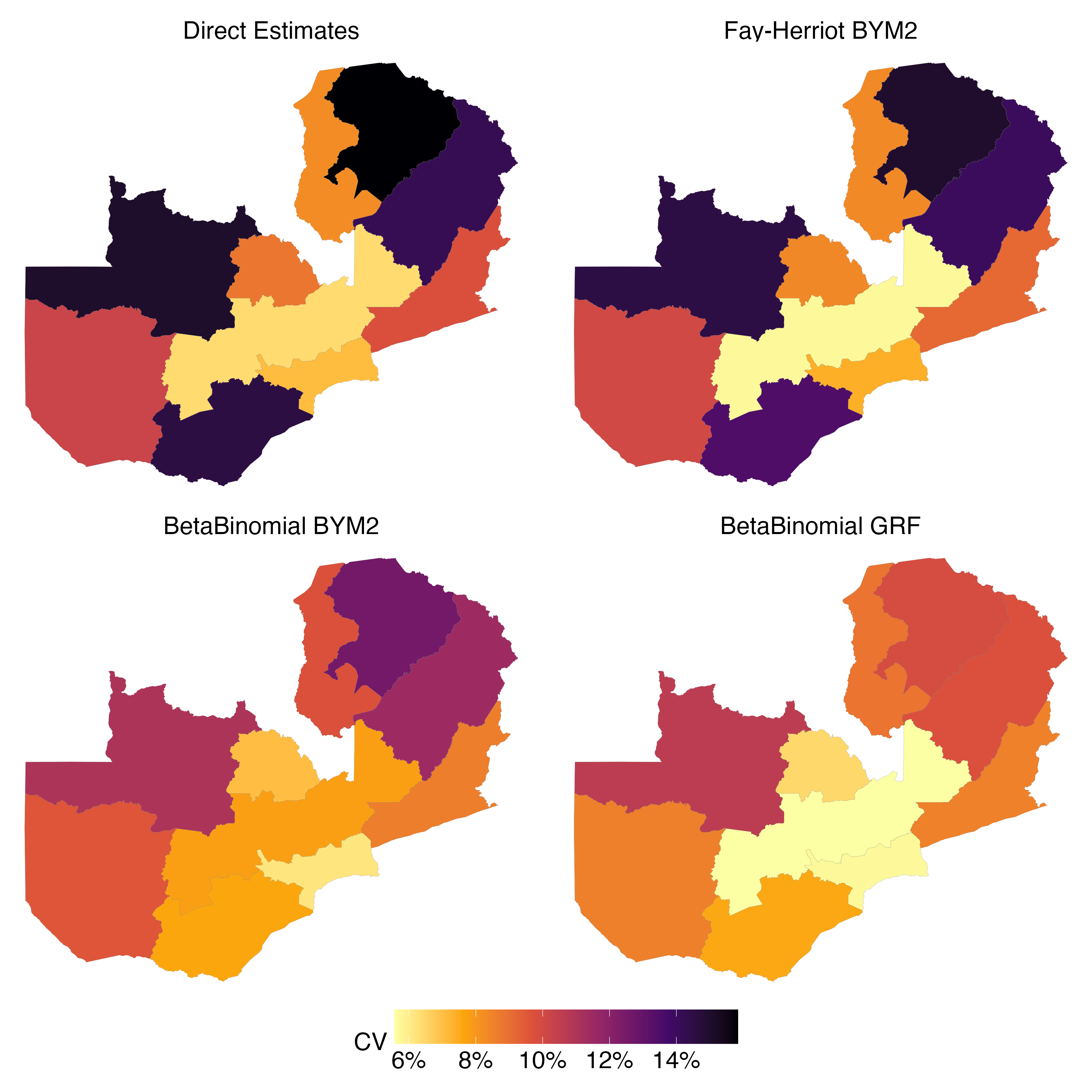}
        }
        \subfigure[][Admin-2 CV.\label{fig:geography:Admin-2:cv}]{
                \includegraphics[width=9.5cm]{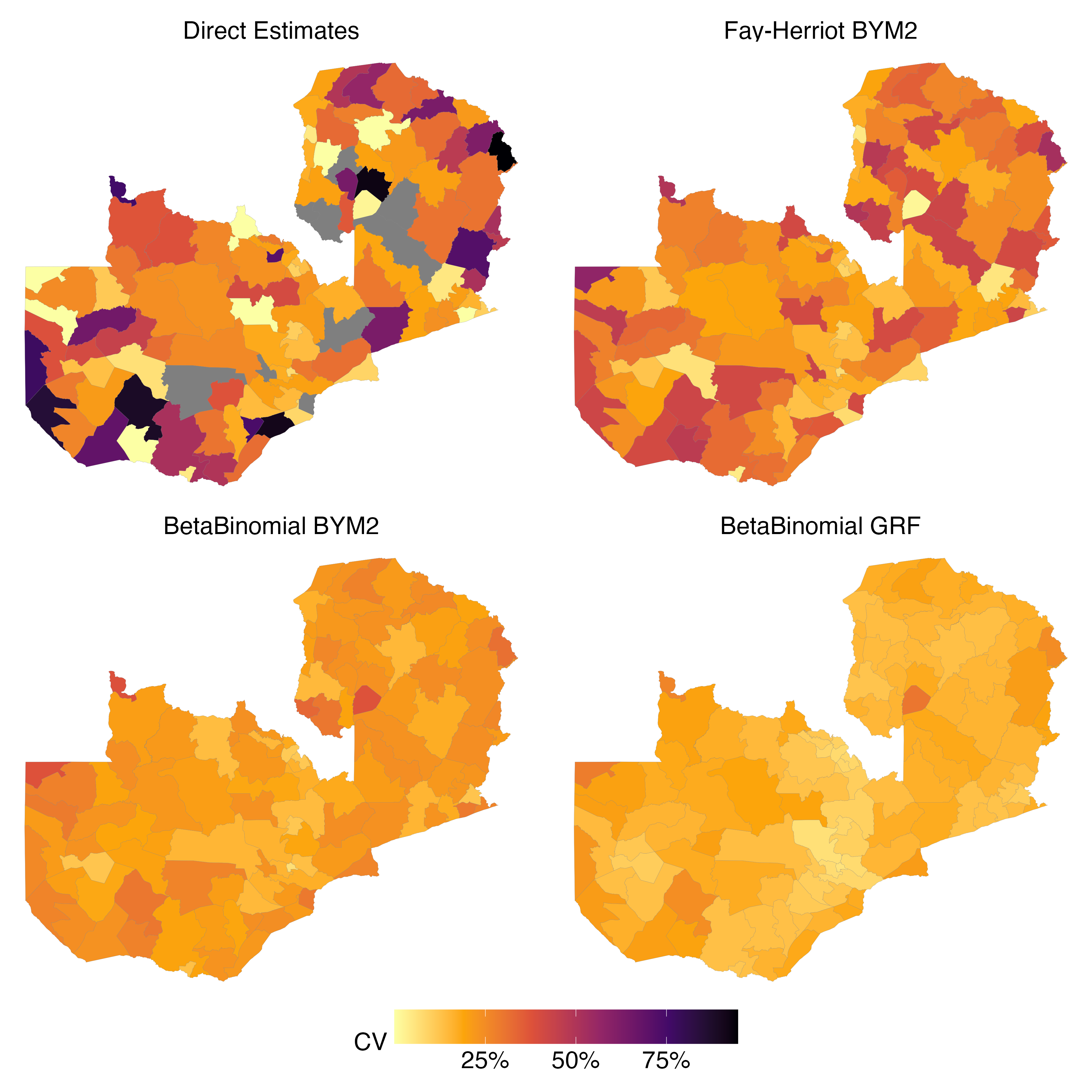}
        }
        \caption{Coefficients of variation, expressed as a percentage, for prevalence/risk estimates for women aged 15--49 in Zambia in 2018: \subref{fig:geography:Admin-1:cv} Admin-1 areas, and
                \subref{fig:geography:Admin-2:cv}  Admin-2 areas. The BYM2 and GRF models contain covariates.}\label{fig:geography:cv}
\end{figure}

Figures \ref{fig:geography} and \ref{fig:geography:cv} provide Admin-1 and Admin-2 level HIV prevalence point estimates and coefficients of variation (CV). To keep the figures at a readable size, we report on the direct estimation model and the covariate models only, as described in Section \ref{sec:Zmodels}.

We first discuss the Admin-1 level results. At this level, all point estimates produce visually similar maps. However, a more detailed comparison included in the Supplementary Materials indicates that the four unit-level models produce nationally aggregated estimates that show slight downward bias when compared with the weighted estimate, indicating  that the design is not being fully accounted for in the unit-level models, and/or there is an aggregation issue. Interestingly, the covariate model results show slightly more bias than the no covariate results, which hints at aggregation being at least part of the issue.

The top half of Table \ref{tab:results-space} contains a range of inferential summaries for the Admin-1 analyses. The unit-level betabinomial BYM2 and GRF models shows noticeable excess-binomial variation. The overdispersion factor, for the median cluster sample size of 24, gives increases of 46--60\% increase over the nominal binomial variation. The overdispersion decreases a little when covariates are added to the models, but there is little change in the spatial standard deviation parameters. The spatial standard deviation drops a little when covariates are added to the model, but in general, the changes are small.

There are marked differences in the CVs at the Admin-1 in the top half of Figure \ref{fig:geography:cv} and in the interval widths in Table \ref{tab:results-space}. 
As expected, the direct estimates have the greatest uncertainty, which is reduced a little in the Fay-Herriot BYM2 models. At Admin-1, adding covariates reduces the uncertainty in general, though not by much, and the GRF models produce the narrowest intervals. 

For Admin-1,  we would use the weighted or Fay-Herriot BYM2 results, given the fewer assumptions, as compared to the unit-level models. The GRF models have potential aggregation issues and the narrow intervals may not correctly reflect uncertainty, since there is evidence that these intervals have low frequentist coverage, at least at Admin-2 (as we discuss shortly, see Table \ref{tab:results}).



\begin{table}[htp]
\caption{Model summaries, with Cov = No/Yes being a label for presence of covariates in the model. Average interval width (expressed as a percentage) and variance components, at Admin-1 (top half) and Admin-2 (bottom half). Interval Width is for a 95\% interval, and is expressed as a percentage. The excess binomial variation (Excess Binom) is the increase over the binomial model for a cluster of median size. In the BYM2 models the spatial Std Dev is the standard deviation of the BYM2 random effects, and in the GRF models it represents the Mat\'ern standard deviation (so these are not comparable). Note that the same GRF model is used at both Admin-1 and Admin-2 levels, so that the spatial Std Dev estimates are the same; the Admin-1 and Admin-2 results are obtained through aggregation from the continuous surface. The Interval Width summaries for the direct model in the bottom half are calculated over the Admin-2 areas with data and valid variance estimates (98 districts versus the full set of 115).}\label{tab:results-space}
\begin{tabular}{llrrr}
\toprule
\multicolumn{1}{l}{\multirow{2}{*}{Model}}&\multicolumn{1}{l}{\multirow{2}{*}{Cov}}& Interval& Excess&Spatial\\
&&Width&Binom&Std Dev\\
\midrule
Direct & $-$ & 5.07 & $-$ & $-$ \\
Fay-Herriot BYM2& No& 4.80 & $-$ & 0.37\\
Fay-Herriot BYM2& Yes& 4.82 & $-$ & 0.33\\
Betabinomial BYM2 &No & 4.00 & 1.60 & 0.26\\
Betabinomial BYM2& Yes& 3.97 & 1.53 & 0.26\\
Betabinomial GRF &No  & 3.64 & 1.48 & 0.40\\
Betabinomial GRF &Yes & 3.51 & 1.46 & 0.41\\
\midrule
Direct & $-$& 12.75 & $-$ & $-$ \\
Fay-Herriot BYM2 &No & 11.72 & $-$ & 0.64\\
Fay-Herriot BYM2 &Yes & 10.61 & $-$ & 0.53\\
Betabinomial BYM2 &No  & 8.69 & 1.48 & 0.34\\
Betabinomial BYM2 &Yes& 7.93 & 1.44 & 0.33\\
Betabinomial GRF &No  & 6.38 & 1.48 & 0.40\\
Betabinomial GRF &Yes& 5.59 & 1.46 & 0.41\\
\bottomrule
\end{tabular} 
\end{table}%

Turning to the Admin-2 point estimates, and referring to Figures \ref{fig:geography} and \ref{fig:geography:cv}, we see greater differences between methods, due to the sparsity of data at this level. The greatest variation across areas occurs with direct estimation (recall that three areas have no clusters, and therefore no estimates, and 14 additional areas have insufficient data to compute variance estimates). The least variation across areas (i.e.,~the greatest shrinkage) is seen with the GRF models. In terms of the magnitude of spatial variation, the BYM2 models lie between the direct and GRF approaches. In the bottom half of Table \ref{tab:results-space} we see that, as we would expect, the covariate models have reduced interval estimates, when compared to the no covariate models. Details on the parameter estimates are in the Supplementary Materials, but here we note that the only variable that had a significant impact was {\it access}. The GRF models have the narrowest intervals but, as we discuss further in Section \ref{sec:comparison}, these intervals appear poorly calibrated.

The decision as to which model we prefer at Admin-2, is more difficult than at Admin-1. Due to the poor calibration of unit-level models, and aggregation issues, we would choose the Fay-Herriot model with BYM2 random effects and covariates. One change we would make, if we were involved in a substantive collaboration, would be to use some form of variance smoothing, as described in 
Section \ref{sec:design:sDirect}, for the 14 areas with data but problematic variance estimates. 

 In Zambia, the data are relatively abundant at the Admin-2 level, but this is frequently not the case and in other analyses in LMICs, for which the data are more sparse, we have used unit-level models, because the amount of missing data meant that the fitting of Fay-Herriot models was no longer tenable. For unit-level models (at Admin-2 or lower) we have found it useful to include fixed effects at higher levels, with spatial random effects nested within these levels.

 



The Supplementary Materials contain a variety of graphical summaries. These include ranking plots, which can be very informative and are straightforward to construct, given samples from the posterior distribution.
%

\subsection{Model Comparison}\label{sec:comparison}

To assess the different models, we carried out a leave-one-out (LOO) cross validation procedure. We do not directly observe prevalence $p_i$ (or risk $r_i$) and the input data for the direct and Fay-Herriot approaches is different to the input data for the unit-level models. Consequently, we predict the weighted estimates for Admin-2 areas, based on datasets in which we systematically leave out each Admin-2 area in turn. We then predict the logit of the direct estimate logit$(\hat p^{\tiny{\mbox{w}}}_i)$ using the remaining data (all but seventeen areas contained a viable direct estimate and standard error). A variety of metrics were considered to assess the consistency between the left out data and the model predictions. In Table \ref{tab:results} we show the results for a subset of metrics, namely the Mean Absolute Error (MAE), the coverage of an 80\% predictive intervals and the log score (which is sometimes referred to as the log of the conditional predictive ordinate). 
We emphasize that all metrics are computed after logit-transformation of the estimates, i.e.,~we predict logit$(\hat p_i)$. A full description of the metrics  are given in the Supplementary Materials. 

We emphasize that these metrics provide only part of the consideration on choosing a model and should not be taken as definitive. Cross-validation for dependent data is currently a topic of great interest, and LOO has been critiqued but as yet no consensus on the best approach has emerged \citep{burkner2020approximate,adin2024automatic}.
 
In Table  \ref{tab:results}, we see that in terms of coverage and the log score, the Fay-Herriot BYM2 model with covariates performs well. A notable feature is the poor coverage of the unit-level models, with the betabinomial GRF with covariates model having a coverage of just 56\%. 
All of these results are consistent with the average interval widths in Table \ref{tab:results-space} and we conclude that, in this example at least, the unit-level models are all producing interval estimates that are overly-optimistic, a dangerous phenomena.
The log score results favor the Fay-Herriot models over the betabinomial models but all models are close, apart from the betabinomial GRF model with covariates, which has significantly poorer performance. For the MAE, the unit-level models provide the lowest values, with the GRF models giving slighter lower values than the BYM2 models.


%

\begin{table}[htp]
\caption{Cross validation summary measures over left-out Admin-2 areas, with Cov = No/Yes being a label for presence of covariates in the model. Mean Absolute Error (MAE) is on the logit scale, nominal coverage is at 80\% level and log score is the log of the predictive density measured at the observed data, which is $\mbox{logit}(\hat p_i^{\tiny{\mbox{w}}})$.
The best scores are highlighted in {\bf bold}.}\label{tab:results}
\begin{tabular}{lrrrr}
\toprule
Model & Cov & MAE & Coverage & log score\\
\midrule
Fay-Herriot BYM2 &No& 0.601&  74 &  -1.033 \\
Fay-Herriot BYM2 & Yes& 0.578 & {\bf 82} & {\bf -1.003}  \\
 Betabinomial BYM2 &No & 0.552& 62&-1.062\\
Betabinomial BYM2&Yes & 0.555&64&-1.150 \\
Betabinomial GRF& No  & {\bf 0.545}& 60 &-1.180 \\
Betabinomial GRF&Yes  & 0.546 & 56 & -1.492 \\
\bottomrule
\end{tabular}

\end{table}%

\section{Discussion}
\label{sec:discussion}

Prevalence mapping is a crucial aid to revealing inequalities in LMICs. A beneficial approach to modeling
 is to bridge the rich literatures of survey sampling 
and spatial statistics, while avoiding a dogmatic approach in one's thinking. To elaborate, one should not expect one approach to be universally best, but rather 
the models chosen should depend on the geographical target level, the data sparsity, the informativeness of the design, and the availability of useful auxiliary information.

Weighted estimates are very appealing, since they are robust and design consistent, which is very desirable when the target is an empirical prevalence. If they produce estimates with acceptable precision, they should be used. Unfortunately, in many situations the data will be too sparse for successful use of weighted estimation. Area-level Fay-Herriot models retain design consistency properties, but use  the totality of data to increase precision. Including spatial random effects at the target level of inference will be advantageous when strong covariates are not available, though one may want to include fixed effects at higher levels, to guard against over-smoothing. For example, if the target is Admin-2, one may model with fixed effects at Admin-1 and nested spatial random effects for each collection of Admin-2 areas within each Admin-1.  This approach will also provide an approximate form of benchmarking \citep{mcgovern2024direct} and has an appealing link with the design since stratification is often carried out at Admin-1. 
For Fay-Herriot approaches, modeling the variances of the weighted estimates to provide usable estimates and robustness may be advantageous \citep{maples2009small,gao2023spatial}. 
The inferential path followed (frequentist or Bayes) and the particular type of discrete spatial model used (ICAR, CAR, SAR,...) are of lesser importance. However, when the data are sparse, a full posterior distribution has advantages over frequentist point estimates and uncertainty summaries, as it accounts for non-normality of sampling distributions and appropriately averages over nuisance parameters via marginalization. 

Benchmarking subnational estimates
to officially accepted national levels is desirable \citep{sarndal2007calibration,zhang2014multilevel}, but it is rare for accurate national prevalences to be available in LMICs. 
Generalized regression (GREG) approaches \citep{sarndal2007calibration} are appealing as a way of extending Fay-Herriot type models via the use of more complex models \citep{gao2024smoothed}.

When weighted estimates, with reliable standard errors, are not available for sufficient areas, so that Fay-Herriot models are not an option, one may turn to unit-level models. We have a preference for discrete spatial models at the target geography, under the rationale, of not wanting to model at levels any lower than needed.
As an extreme example, suppose we wanted a national estimate. If model assumptions are met and precise information required for aggregation is available then the theory tells us that 
a more precise estimate will be obtained from a spatial model. However, simply calculating a weighted estimate will be preferable, since it is based on few assumptions, avoids difficult aggregation steps, and will account for the survey design (and will often be precise enough for most purposes). A simple sanity check for any unit-level model is aggregating to the national level and comparing with the weighted estimate and its uncertainty interval.


Spatially continuous MBG models are aesthetically appealing, since they can produce estimates at any required geographical level.
However, the aggregation step may inject  bias. 
MBG approaches may  produce estimates with the lowest uncertainty (because we are putting in the most assumptions), but one must then question whether these estimates are accurate measures of the true uncertainty. The issue is that the full uncertainty in the data, including the inputs to the aggregation process, may not be being fully acknowledged. We elaborate on this point shortly. 

Aggregation can be very challenging with unit-level models. 
For example, if there is informative sampling over urban/rural strata, we need to investigate whether ignoring this aspect will lead to bias.
Estimating an urban/rural association is relatively uncomplicated, but producing reliable 
fine-scale maps of urban/rural, which are needed for spatial aggregation,
is an ongoing research problem. Treating space as continuous increases the challenge
as one needs to rely
on fine-scale population density rasters to perform aggregation to the areal level.
Both population and covariate rasters are estimated based on other data sources, and
there is a need for further investigation to understand
the consequences of not acknowledging the inherent uncertainty when used
in an aggregation scheme. The result of aggregation is typically an areal risk and
examining the distinction between
risk and prevalence at different spatial scales is worthwhile.

This all being said, MBG models may be the only viable option when data are sparse relative to the geographical target scale, particularly when strongly predictive covariates are available. In this case, one should, however, view the results more cautiously. 

There are cases in which a continuously-indexed model is necessary, for example, to give a principled
way of handling data that is available with limited geographical information (e.g.,~area only information, rather than points) or under different geographical partitions \citep{marquez2021harmonizing,wilson:wakefield:18}.

A fundamental issue with nonlinear unit-level models, including MBG approaches, is that they do not produce design consistent estimates. The only way to guarantee such consistency is to use the weights, but these are not utilized in standard unit-level models. \cite{you:rao:02,pfeffermann2007small,guadarrama2018small} and \cite{cho2024optimal} have made some progress in achieving design consistency for linear unit-level models, but it is not straightforward to extend to logistic models, let alone continuous spatial models, especially since DHS and other surveys only report the final weights and not the constituent parts, which are required for areas with no data.  Unit-level models may neglect to account for all aspects of the sampling design and frequently make the strong assumption that the design is ignorable. Often in LMICs the design is not ignorable because of urban/rural stratification and the strong association between many outcomes and this classification.

%



Section \ref{sec:cultures}
demonstrates the clear importance of model choices both in terms
of central predictions and uncertainty. Since it will always be possible
to argue for many reasonable models, there is a strong need for formal validation techniques to
compare different models in terms of their ability to estimate areal prevalences.
However, the spatial misalignment between the point-referenced observations and the desired areal quantities makes validation a challenging task.
One method for model validation is an approach that frames direct estimates as observations and scores different models based on their ability to predict direct estimates, but there are many choices (such as which data to leave out, and which metric to use for comparison of predictions and left-out data), and it is not an ideal approach since it is only possible at a large geographical scale which will often not coincide with  the inferential target.
One should be wary of using more complex models unless one can show they are better than simpler models through validation. Cross-validation can also be carried out at the unit-level, but there is little experience with this enterprise, which is clearly an important area for future research.

The whole consistency of estimates argument is best framed within a spectrum of approaches, with weighted estimation at one end and MBG models at the other, and it is natural to change the approach as the sparsity of data changes.
 A very important research question is when, as a function of sample size, to transition between approaches.

 In our experience, for estimation when the data are rich, for example (usually) when constructing Admin-1  estimates using DHS or MICS
surveys, direct (weighted) estimation is a safe choice. 
If the survey does not provide precise enough Admin-1
estimates using weighted estimation, one should 
turn to models. Our preference is to begin with simple Fay-Herriot area-level models.
For finer
levels such as Admin-2 or Admin-3 (depending on the country, survey size and rarity of outcome), one can use spatial statistical unit-level models, but ensure that the model contains terms to acknowledge the complex design. In general, one should use models that one is comfortable with, whether they be traditional SAE style unit-level models or MBG versions. We are cautious with unit-level  approaches, particularly with respect to acknowledging the design and aggregation. A critical evaluation of the latter is crucial. 
A key point is that the goal of the analysis should determine the approach, and  different goals may call for different approaches, and one should not expect a single model to provide the best choice for all questions.  \citet{corral2021map} reached a similar conclusion. When producing official statistics using SAE, 
\citet{tzavidis:etal:18} suggest that 
one consider progressively  finer geographies while at each
level assessing  the feasibility of producing SAE estimates for the current
level. This is broadly consistent with our recommendations.

\begin{acks}[Acknowledgments]
We are grateful to the Space Time Analysis Bayes (STAB) working group
for discussion and feedback on the paper. We also acknowledge the  DHS for access and use of the data, and for permission to make available the cluster aggregates and their displaced locations.
\end{acks}

%
%
\bibliographystyle{chicago}
\bibliography{spatepi2}


\clearpage

\newcommand{\beginsupplement}{
        \setcounter{page}{1}
        \setcounter{section}{0}
        \setcounter{table}{0}
        \renewcommand{\thetable}{S\arabic{table}}
        \setcounter{figure}{0}
        \renewcommand{\thefigure}{S\arabic{figure}}
        \renewcommand{\thepage}{S\arabic{page}}  
        \renewcommand{\thesection}{S\arabic{section}}
        \renewcommand{\theequation}{S\arabic{equation}}

     }

\beginsupplement

\section{Model Description}

\subsection{Direct estimation}

Weighted (direct) estimation is done using a H\'ajek estimator \citep{hajek:71}, 
\begin{equation}
  \hat{p}^{\tiny{\text{w}}}_{i} = \frac{\sum_{k \in S_i} w_k y_k}{\sum_{k\in S_i} w_k},\label{eq:HT}
\end{equation}
where $S_i$ is the set of indices of sampled individuals in area $i$,  $i = 1,  \dots, m$.

\subsection{Fay-Herriot BYM2 without covariates} 

The form of the  
Fay-Herriot model without covariates is,
\begin{eqnarray}
\mathrm{logit}(\hat{p}^{\tiny{\text{w}}}_{i}) \mid p_i &\sim_{iid}&\mbox{N} ( \mathrm{logit}(p_i), V_{i}),\label{eq:fayherriot1}\nonumber \\
 \mathrm{logit}(p_i)&=&\alpha
 + u_i , \label{eq:fayherriot2}\\
 \boldsymbol{u}\mid \phi, \sigma_u^2 &\sim& \mathrm{BYM2}(\phi,  \sigma^2_u),\qquad i=1,\dots, m ,\nonumber
\end{eqnarray}
where $\alpha$ is the intercept and the spatial random effects $u_i$ are the sum of iid and intrinsic conditional autoregressive (ICAR) random effects.
We adopt the reparameterization known as the BYM2 model \citep{besag:etal:91,riebler:etal:16}, in which the vector of random area effects has structure,
$$
\boldsymbol {u} = \sigma_u \left(\sqrt{1-\phi}\boldsymbol{e} + \sqrt{\phi} \boldsymbol{S}\right),
$$
where $\sigma_u$ is the total standard deviation, $\phi$ is the proportion of the variance that is spatial, $\boldsymbol{e}$ is a vector of iid standard normal random variables and $\boldsymbol{S}$ follows a scaled ICAR prior so that the geometric mean of the marginal variances of $S_i$ is equal to $1$, under a sum-to-zero constraint that is imposed to ensure identifiability when there is an intercept in the model \citep{rue:knorrheld:05}. The unscaled ICAR prior before the sum-to-zero constraint satisfies
$$S^\star_i \mid S^\star_j \in \mbox{ne}(i) \sim \mbox{N}(\overline{S}^\star_i,\sigma^2_s/n_i),$$
where $\mbox{ne}(i)$ is the set of neighbors of area $i$ (which we take as those areas that share a common boundary with area $i$), $n_i$ is the number of such neighbors and $$\overline{S}^\star_i = \frac{1}{n_i}\sum_{j \in \mbox{\small{ne}}(i)} S^\star_j.$$


 We take a Bayesian approach to inference, placing priors on the hyperparameters $\alpha, \boldsymbol\beta, \phi,$ and $\sigma^2_u$. We adopt the default prior on the intercept (Gaussian with zero mean and zero precision). We use penalized complexity (PC) priors for the variance parameters \citep{simpson:etal:17}. We specify the prior for the area effect variance $\sigma_u^2$
such that $\Pr(\sigma_u>1)=0.01$ and for the spatial correlation parameter $\phi$ such that $\Pr(\phi>0.5)=2/3$. 

The prevalence in area $i$ is,
\begin{equation}\label{eq:agg2}
p_i = \mathrm{expit} ( \alpha + u_i),
\end{equation}
and predictions are computed by transforming posterior samples for $\alpha,u_i$ to get samples for $p_i$, $i=1,\dots,m$.

\subsection{Fay-Herriot BYM2 with covariates} 

In this model, everything as in the no covariate version except we replace (\ref{eq:fayherriot2})  with,
$$ \mbox{logit} (p_i ) = \alpha + \boldsymbol{x}_i^\top \boldsymbol{\beta} + u_i,$$
and the prevalence in area $i$ is,
\begin{equation}
p_i = \mathrm{expit} ( \alpha + \boldsymbol{x}_i^\top \boldsymbol{\beta} 
 + u_i),\label{eq:aggFH}
\end{equation}
with covariates access, night time lights, and malaria prevalence included in $\boldsymbol{x}_i$, for $i=1,\dots,m$. For the covariate model slope parameters we take the default INLA priors (Gaussian with zero mean and precision 0.001)

\subsection{Betabinomial BYM2 without covariates} 

  We begin with the (superpopulation) sampling model for the total HIV cases, $Y_{ic}^{\text{\tiny{CL}}} $, out of $n_{ic}$ sampled in cluster $c$ in area $i$,
    \begin{equation*}
    Y_{ic}^{\text{\tiny{CL}}}  \mid r^\star_{ic} \sim \text{Binomial}(n_{ic},r^\star_{ic})
    \end{equation*}
with conditional risk,
    \begin{equation}
    r^\star_{ic}  \mid r_{ic}, d\sim \mathrm{Beta}(r_{ic}, d) ,
    \label{eq:beta}
    \end{equation}
    where the beta distribution is parameterized so that,
    \begin{align*}
        \mbox{E}[r^\star_{ic}\mid r_{ic}, d]&=r_{ic},\quad
        \mathrm{var}(r^\star_{ic}\mid r_{ic}, d)=\frac{r_{ic}(1-r_{ic})}{1+d},
    \end{align*}
    for the set of sampled clusters $c \in S_i^{\tiny{\mbox{c}}}$, $i=1,\dots,m$. The idealized rationale here is that if we envisage repeated sampling from a cluster, for each of the samples we would draw $r^\star_{ic}$ from the beta distribution (\ref{eq:beta}). The marginal risk, $r_{ic}$, is the target that we would like to estimate.
The parameter $d>0$ represents the level of overdispersion. Higher levels of overdispersion correspond to lower values of $d$, with the excess binomial variation being $1+(n-1)/(d+1)$, for a cluster with sample size $n$. 

This two-stage model produces the marginal sampling model that describes the realizations we are expecting to see, 
\begin{equation}
		\label{eq:betabin}
Y_{ic}^{\text{\tiny{CL}}}  \mid r_{ic} \sim \text{BetaBinomial}(n_{ic},r_{ic},d).
\end{equation}

    The risks are modeled as
    $$\mbox{logit}(r_{ic}) = \alpha+
    1( \boldsymbol{s}_{ic} \in \mbox{ urban })\gamma+ u_i,$$
    where $\boldsymbol{s}_{ic}$ is the reported location of cluster $c$ in areas $i$ and the only cluster-level covariate used in this model is an urban/rural indicator $1( \boldsymbol{s}_{ic} \in \mbox{ urban })$, and the $u_i$ are modeled as BYM2 random effects. The parameter $\exp(\gamma )$ is the ratio of the odds of HIV prevalence in an urban cluster  to the odds of HIV prevalence in a rural cluster, in a typical area (i.e.,~an area with $u_i=0$).
    
    We take a Bayesian approach to inference, placing priors on the hyperparameters. Priors on the intercept and regression parameter are independent zero mean Gaussian with precision 0.001. We specify the prior for the area effect variance $\sigma_u^2$ such that $\Pr(\sigma_u>1)=0.05$ and for the spatial correlation parameter $\phi$ such that $\Pr(\phi>0.5)=2/3$. In INLA, overdispersion is parameterized using the parameter $1/(d+1)$, and we use the default INLA prior for this parameter.


For this model, the area-level risk is,
\begin{equation}
r_i = q_i \times \mbox{expit}(\alpha +u_i) + (1-q_i) \times \mbox{expit}(\alpha + \gamma + u_i)
\\  
\label{eq:agg3}
\end{equation}
where $q_i$ is the fraction of the 15--49 female population in  area $i$ that reside in rural clusters, $i=1,\dots,m$.

\subsection{Betabinomial BYM2 with covariates}

 We extend the previous sampling model (\ref{eq:betabin}) to add further covariates via,
\begin{equation}\label{eq:covariates}
\mbox{logit}(r_{ic}) = \alpha+\mathbf{z}^\top_{ic} \boldsymbol{\beta}+1( \boldsymbol{s}_{ic} \in \mbox{ urban })\gamma + u_i,
 \end{equation}
 with the cluster-level covariates incorporated in $\mathbf{z}_{ic}=\mathbf{z}(\boldsymbol{s}_{ic})$ being access, night time lights, and malaria prevalence. Priors are the same as in the previous model.

The risk is continuous over space because of the continuous covariate surfaces. We have,
\begin{eqnarray}
r_i &=&\int_{A_i } r(\boldsymbol{s}) d(\boldsymbol{s})~\mbox{d}
\boldsymbol{s}, \label{eq:cnstarget}
\end{eqnarray}
and the grid approximation is,
\begin{eqnarray}
\tilde{r}_i &=&
 A \sum_{g = 1}^{G_i} r(\boldsymbol{s}_{ig}^{\mbox{\tiny{G}}} ) d_i(
 \boldsymbol{s}_{ig}^{\mbox{\tiny{G}}} ) \label{eq:unitagg}
\end{eqnarray}
For this model,
 \begin{eqnarray}
r(\boldsymbol{s}_{ig}^{\mbox{\tiny{G}}} )=
\mbox{expit}(
\alpha+\mathbf{z}^\top_{ig} \boldsymbol{\beta}+1( \boldsymbol{s}^{\mbox{\tiny{G}}}_{ig} \in \mbox{ urban })\gamma + u_i
)
\nonumber \\  \label{eq:agg4}
\end{eqnarray}
where the factor $A$ is defined as the area of the grid cells and $d_i(\boldsymbol{s}_{ig}^{\mbox{\tiny{G}}} )$ is the normalized female 15--49 population density at grid location $\boldsymbol{s}_{ig}^{\mbox{\tiny{G}}} $.
For implementation, we therefore require gridded covariates for access, night time lights and malaria prevalence, in addition to the urban/rural status of each grid cell. We use $1\, \text{km}\times 1\, \text{km}$ grid cells.

\subsection{Betabinomial GRF without covariates} 
    
    We use the superpopulation sampling model  (\ref{eq:betabin}),
    with latent model,
    $$\mbox{logit}(r_{ic}) = \alpha+1( \boldsymbol{s}_{ic} \in \mbox{ urban })\gamma+ \omega(\boldsymbol{s}_{ic}),$$
    where $\omega$ is a GRF with approximate Mat\'ern covariance structure, as specified using the INLA-SPDE approach. The approaches approximates a continuously indexed random field using a Gaussian Markov random field (GMRF) following \cite{Lindgren:etal:11}, using a set of piecewise linear basis functions on a mesh of locations over the domain. These functions have random weights modeled using a GMRF with precision matrix specified so that the resulting random field has an approximate Mat\'ern covariance function. The only covariate used in this model is a cluster-level indicator of urban/rural. 
     
    We take a Bayesian approach to inference, placing priors on the hyperparameters with priors for $d$, $\alpha$ and $\gamma$ as previously stated. For the GMRF $\omega$, we have two parameters:~the marginal variance $\sigma_\omega^2$ and the practical range $r$. We place a PC prior on the practical range $r$ such that $\Pr(r < 3)=0.5$ and $\Pr(\sigma_\omega > 2) = 0.5$, where $r$ is measured in degrees. 
    
    For prediction, we construct a grid across the spatial domain and make predictions for all pixels. These grid-cell predictions are then weighted by estimated female 15--49 populations  and aggregated to area-level predictions. 
    Viewing the risk as a continuous function in space gives the target as (\ref{eq:cnstarget}) and 
the grid approximation is (\ref{eq:unitagg}) 
where
\begin{eqnarray}
r(\boldsymbol{s}_{ig}^{\mbox{\tiny{G}}} ) 
&=& 
\mbox{expit}(
\alpha+1( \boldsymbol{s}^{\mbox{\tiny{G}}}_{ig} \in \mbox{ urban })\gamma + \omega(\boldsymbol{s}_{ig}^{\mbox{\tiny{G}}})).
\nonumber \\ \label{eq:agg5}
\end{eqnarray}
    
    \subsection{Betabinomial GRF with covariates} 
    
  We use the superpopulation sampling model  (\ref{eq:betabin}) along with latent risk model,
   $$\mbox{logit}(r_{ic}) = \alpha+\mathbf{z}^\top_{ic} \boldsymbol{\beta}+1( \boldsymbol{s}_{ic} \in \mbox{ urban })\gamma+ \omega(\boldsymbol{s}_{ic}),$$
   with all terms and priors defined as in previous models.
   Aggregation proceeds as before with (\ref{eq:cnstarget}) being approximated using (\ref{eq:unitagg}) and 
      \begin{eqnarray}
      r(\boldsymbol{s}_{ig}^{\mbox{\tiny{G}}} ) = 
%
\mbox{expit}(
\alpha+\mathbf{z}^\top_{ig} \boldsymbol{\beta}+1( \boldsymbol{s}^{\mbox{\tiny{G}}}_{ig} \in \mbox{ urban })\gamma + \omega(\boldsymbol{s}_{ig}^{\mbox{\tiny{G}}}) ).
\nonumber \\ 
\label{eq:agg6}
    \end{eqnarray}
   
\section{Model Comparison Metrics}

Leave-one-out (LOO) cross validation is used to compare models in Section 6 of the paper. In particular, suppose we leave out each of $m$ areas at some level (for example, Admin-2) in turn, and predict the (logit) direct estimate using the remaining data. We let $  \hat{p}^{\tiny{\text{w}}}_{i} $ be the weighted estimate  in area $i$, $i=1,\dots,m$. The asymptotic distribution of the estimator $\hat{\theta}^{\tiny{\text{w}}}_i = \mbox{logit}(\hat{p}^{\tiny{\text{w}}}_{i})$, is $\mbox{N}( \theta_i,V_i)$ where in the design-based formulation, $\theta_i=\mbox{logit}(p_i)$ is the logit of the prevalence and $V_i$ is the design-based variance estimate.

The LOO approach is aimed at assessing a model's ability to predict left out data. The predictive distribution for the observable (direct estimate) $\hat{\theta}^{\tiny{\text{w}}}_i$ should acknowledge the uncertainty in the target, which is $V_i$. 
For each model we obtain the posterior distribution of the area-level prevalence (area-level models) or the area-level risk (unit-level models) in each of the $m$ LOO cases.

Let $\theta_i$ represent  the logit of the area-level prevalence (area-level models) or logit area-level risk (unit-level models), which is itself a function of some subset of the parameter set.
The predictive distribution is for the weighted estimate, which we label $ \tilde{\theta}^{\tiny{\text{w}}}_i$. This distribution is not analytically tractable, and so there is no closed form for the predictive distribution:
$$p( \tilde{\theta}^{\tiny{\text{w}}}_i \mid \boldsymbol{y}_{-i} ) = \int_{\theta_i} p( \tilde{\theta}_i^{\tiny{\text{w}}} \mid \theta_i) \pi( \theta_i \mid \boldsymbol{y}_{-i} ) ~\mbox{d}\theta_i,$$
where $\tilde{\theta}^{\tiny{\text{w}}}_i \mid \theta_i \sim \mbox{N}( \theta_i , V_i )$.

A variety of metrics are considered to assess the consistency between the left out data and the model predictions.
We obtain a measure of the closeness of the mean of the predictive distribution to the weighted estimate (the mean absolute error), and then three measures that use the complete distribution, the coverage of a predictive interval, the log score and the interval score.
 \\


\noindent
{\bf Mean Absolute Error (MAE):} A measure of the closeness of the means of the predictive distribution to the weighted estimates is,
$$ \mbox{MAE} = \frac{1}{m} \sum_{i=1}^{m} \mid
\mbox{E}[ \tilde\theta^{\tiny{\text{w}}}_i \mid \boldsymbol{y}_{-i}] - \hat{\theta}^{\tiny{\text{w}}}_i
\mid,$$
where $ \boldsymbol{y}_{-i}$ represents the totality of data with area $i$ left out. Note that $$\mbox{E}[ \tilde\theta^{\tiny{\text{w}}}_i \mid \boldsymbol{y}_{-i}] = \mbox{E}[ \theta_i \mid \boldsymbol{y}_{-i}].$$

This measure assesses {\it point} estimates, meaning the location summary of the predictive distribution only, and not the spread of the distribution.\\

\noindent
{\bf Coverage estimates:}  To evaluate the coverage we need to obtain a predictive interval for the left out ``data'', which here corresponds to the logit of the weighted estimate, $\hat{\theta}^{\tiny{\text{w}}}_i$. 
For each of the $m$ Admin-2 areas, we calculate the fraction of these intervals that contain $\hat{\theta}^{\tiny{\text{w}}}_i = \mbox{logit}(\hat{p}^{\tiny{\text{w}}}_{i})$. 
We generate samples, $\tilde{\theta}^{\tiny{\text{w}}}_i \mid \boldsymbol{y}_{-i}$:
\begin{eqnarray*}
\theta_i^{(b)} &\sim &\pi( \theta_i \mid \boldsymbol{y}_{-i} )\\
\tilde{\theta}_i^{\,\tiny{\text{w}}(b)} &\sim& \mbox{N} ( \theta_i^{(b)}, V_i),
\end{eqnarray*}
for $b=1,\dots,B$. In the first step, we generate samples from the INLA approximation to the posterior.
In the area-level model the prevalence is directly modeled, while in the unit-level models we assume the prevalence is well-approximated by the risk. Based on the samples $\{ \tilde{\theta}_i^{\,\tiny{\text{w}}(b)},b=1,\dots,B\}$, we take the relevant quantiles to form the predictive interval.\\


\noindent
{\bf The log score:} The log score is the average of the predictive density of the observed data, based on the left out data: 
$$\mbox{log score} =  \frac{1}{m} \sum_{i=1}^m \log 
p( \hat{\theta}^{\tiny{\text{w}}}_i \mid \boldsymbol{y}_{-i})  
.$$
Since the predictive density is not of closed form we approximate via
$$ p( \hat{\theta}^{\tiny{\text{w}}}_i \mid \boldsymbol{y}_{-i})   \approx \frac{1}{B} \sum_{b=1}^B \phi ( \hat{\theta}^{\tiny{\text{w}}}_i  \mid \theta_i^{(b)}, V_i ),$$
where $\phi( \cdot \mid \theta_i^{(b)}, V_i  )$ is the density of a normal with mean $\theta_i^{(b)}$ and variance $ V_i$. \\

\noindent
{\bf The interval score:} To assess a predictive interval, one may calculate the interval score, with small values being favored:
$$\mbox{IS}_\alpha(l_i,u_i;\hat{\theta}^{\tiny{\text{w}}}_i) = (u_i-l_i) + \frac{2}{\alpha}1(\hat{\theta}^{\tiny{\text{w}}}_i<l_i) + \frac{2}{\alpha}1(\hat{\theta}^{\tiny{\text{w}}}_i>u_i),$$
where $1(\cdot)$ is 1 if the argument is true and 0 otherwise.
In practice, one produces the predictive interval for each area $(l_i,u_i)$, as described in the coverage section above and then averages the $m$ scores.
This score is intuitive since it rewards narrow intervals that capture the observations 
\citep{gneiting2014probabilistic}.

To emphasize, all metrics are computed after logit-transformation of the estimates.
Table \ref{tab:results_cv1} provides all scores from the LOO cross-validation.


\begin{table}[htp]
\caption{Leave-one-out performance metrics for Admin-2 models, with Cov = No/Yes being a label for presence of covariates in the model. The nominal coverage is 80\% and the best scores are highlighted in {\bf bold}. }\label{tab:results_cv1}
\begin{tabular}{lrrrrr}
\toprule
Model & Cov & MAE 
& Coverage & log score & int score
\\
\midrule
Fay-Herriot BYM2 & No  & 0.601 
&   74  & -1.033 &  2.523 \\
 Fay-Herriot BYM2& Yes  & 0.578 
 &
 {\bf 82} & {\bf -1.003} & 2.451 \\
BetaBinomial BYM2& No & 0.552 
& 62 & -1.062 & 2.489 \\
BetaBinomial BYM2&Yes & 0.555 
& 64 & -1.150 & {\bf 2.485} \\
BetaBinomial GRF& No  & {\bf 0.545} 
& 60 & -1.180 & 2.595 \\
BetaBinomial GRF& Yes & 0.546 
& 56 & -1.492 & 2.620 \\
\bottomrule
\end{tabular}
\end{table}


\section{Covariate Surfaces}

\begin{figure}[htbp]
	\centering
	\includegraphics[width=\columnwidth]{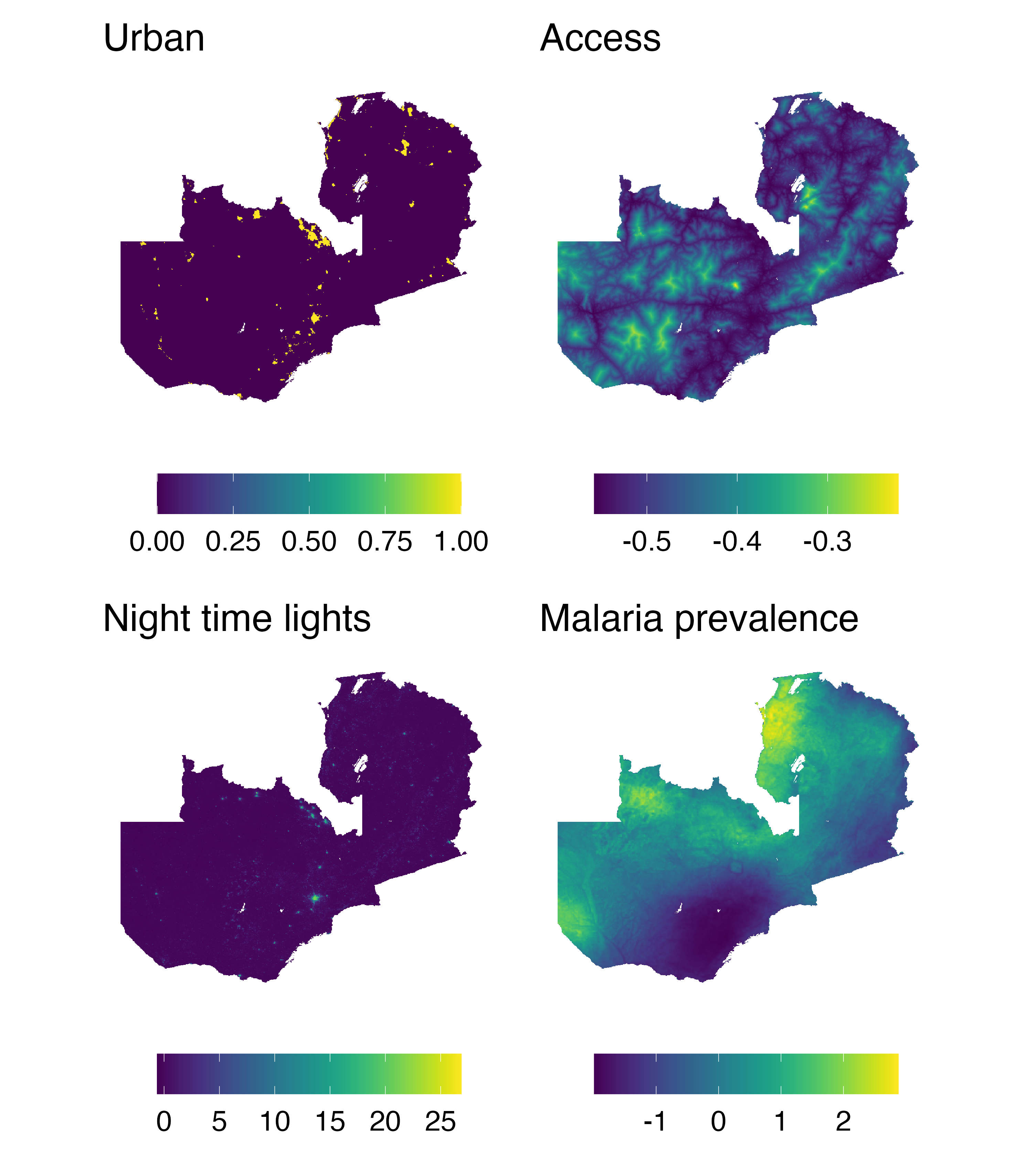}
    \caption{Pixel-level map of (untransformed) covariates used for modeling.\label{fig:zambia:cov}}
\end{figure}

Figure \ref{fig:zambia:cov} shows the urban/rural surface, along with the covariate surfaces. We transform  the night time lights variable to be $\log(1+x)$ where $x$ is a measure of the intensity of night time lights, so that the distribution of these variables are  not too skewed which could result in a small number of points influencing the predictions heavily.

\section{Triangulation used for SPDE approach}
\label{sec:triangulation}

\begin{figure}[htbp]
	\centering
	\includegraphics[width=\columnwidth]{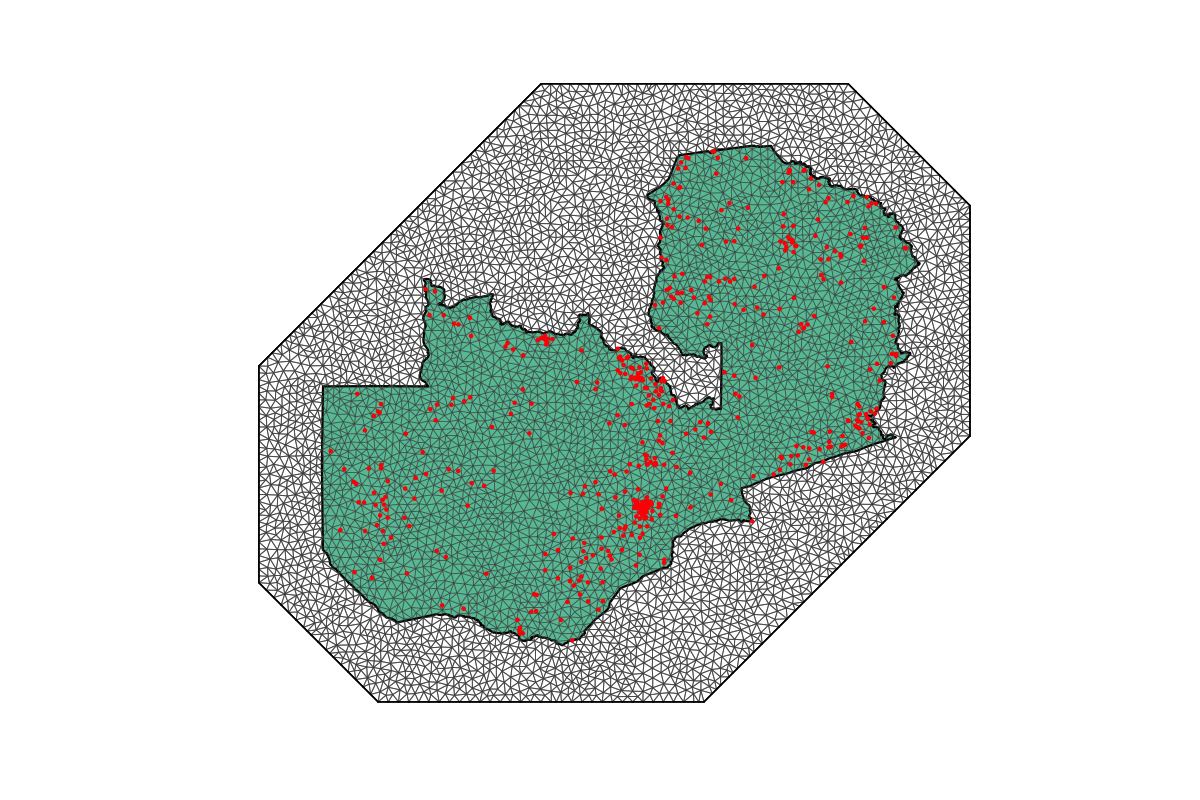}
	\caption{SPDE mesh for Zambia, with cluster locations (jittered for privacy).\label{fig:zambia:mesh}}
\end{figure}

We implement an approximate Matérn GRF using the SPDE approach
\citep{Lindgren:etal:11}. This is a common way to achieve computational
efficient GRF approximation in MBG  approaches \citep{utazi2020geospatial,local2021mapping}. We used the mesh shown in Figure 
\ref{fig:zambia:mesh}. The mesh extends beyond Zambia
to avoid boundary effects. 

\section{Comparison of Aggregated Estimates}

Given the approximations contained in aggregation strategies, when going from urban/rural partitions or pixels to areas, it is important to compare higher-level summaries as an additional method for comparing and validating models. Ideally, the aggregated model-based estimates will agree well with the weighted estimates.
In this Section we examine aggregated national and Admin-1 estimates from all models. We describe in detail how we move from Admin-2 to Admin-1, with the Admin-1 to national following analogously.

We obtain samples from area-level estimates at Admin-2, $\theta^{(b)}_{2ij}$,  
 through (\ref{eq:HT}), (\ref{eq:agg2}), (\ref{eq:aggFH}), (\ref{eq:agg3}) for the models with no continuous over spatial components, 
and via the grid approximation (\ref{eq:unitagg}) with risks (\ref{eq:agg4}), (\ref{eq:agg5}), (\ref{eq:agg6}) for the models with continuous over space components. We can then obtain Admin-1 level estimates via,
$$\theta^{(b)}_{1i} = \sum_{j \in A_i} f_{ij} \theta^{(b)}_{2j},\qquad b=1,\dots,B,$$
where $A_i$ is the set of Admin-2 areas in the target Admin-1 area and the population fractions are $f_{ij} = M_{ij}/M_i$, where $M_{ij}$ is the number of women aged 15--49 in Admin-2 area $j$ in Admin-1 area $i$ and $M_i=\sum_{j \in A_i} M_{ij}$. These totals are obtained from WorldPop.

\begin{figure}[htbp]
\includegraphics[width=\columnwidth]{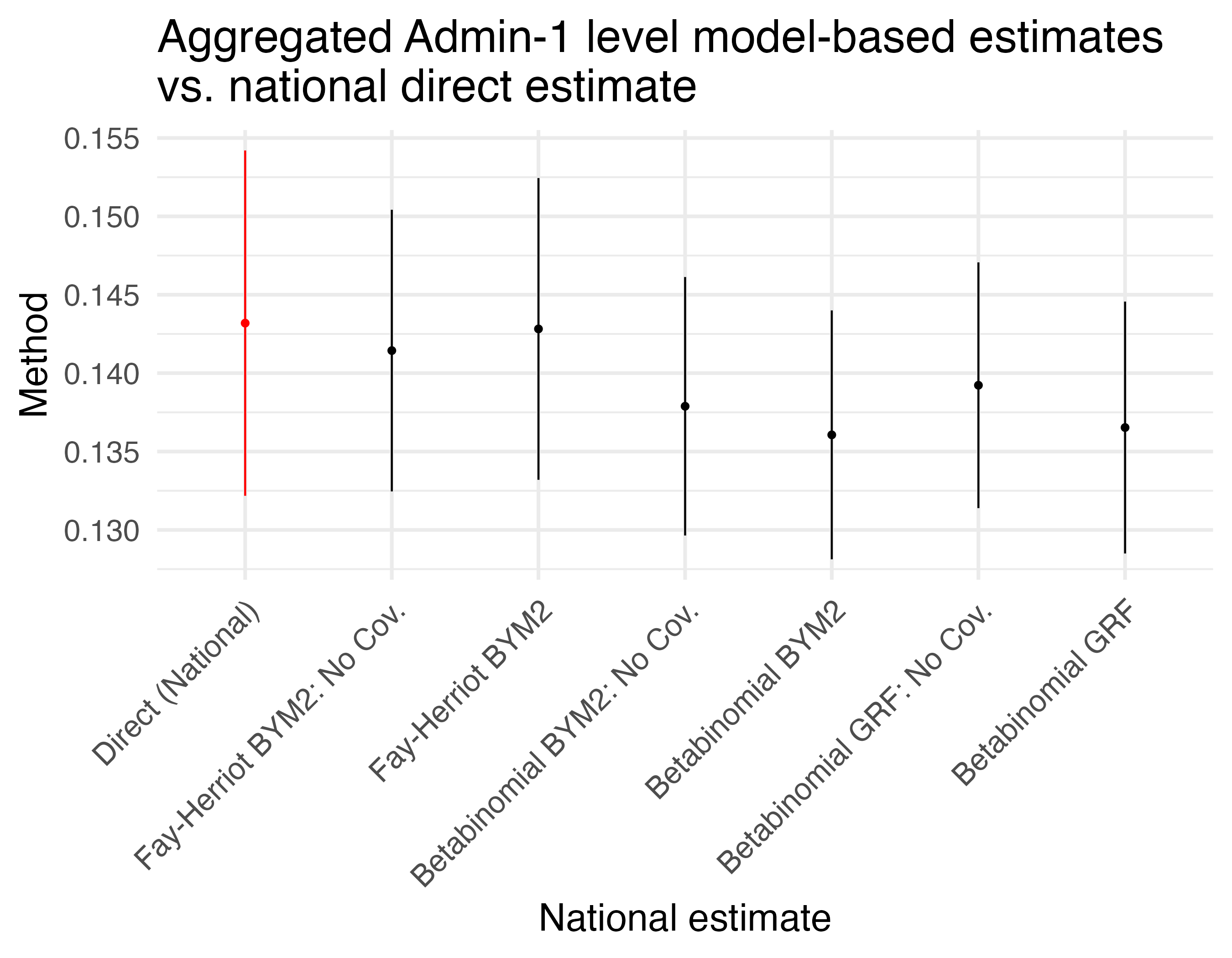}
\caption{Comparison of aggregated model-based  estimates and direct weighted national estimates. The BYM2 models are specified at Admin-1.}
\label{fig:Zambia_aggregated_adm1_natl}
\end{figure}
\begin{figure}[htbp]
\includegraphics[width=\columnwidth]{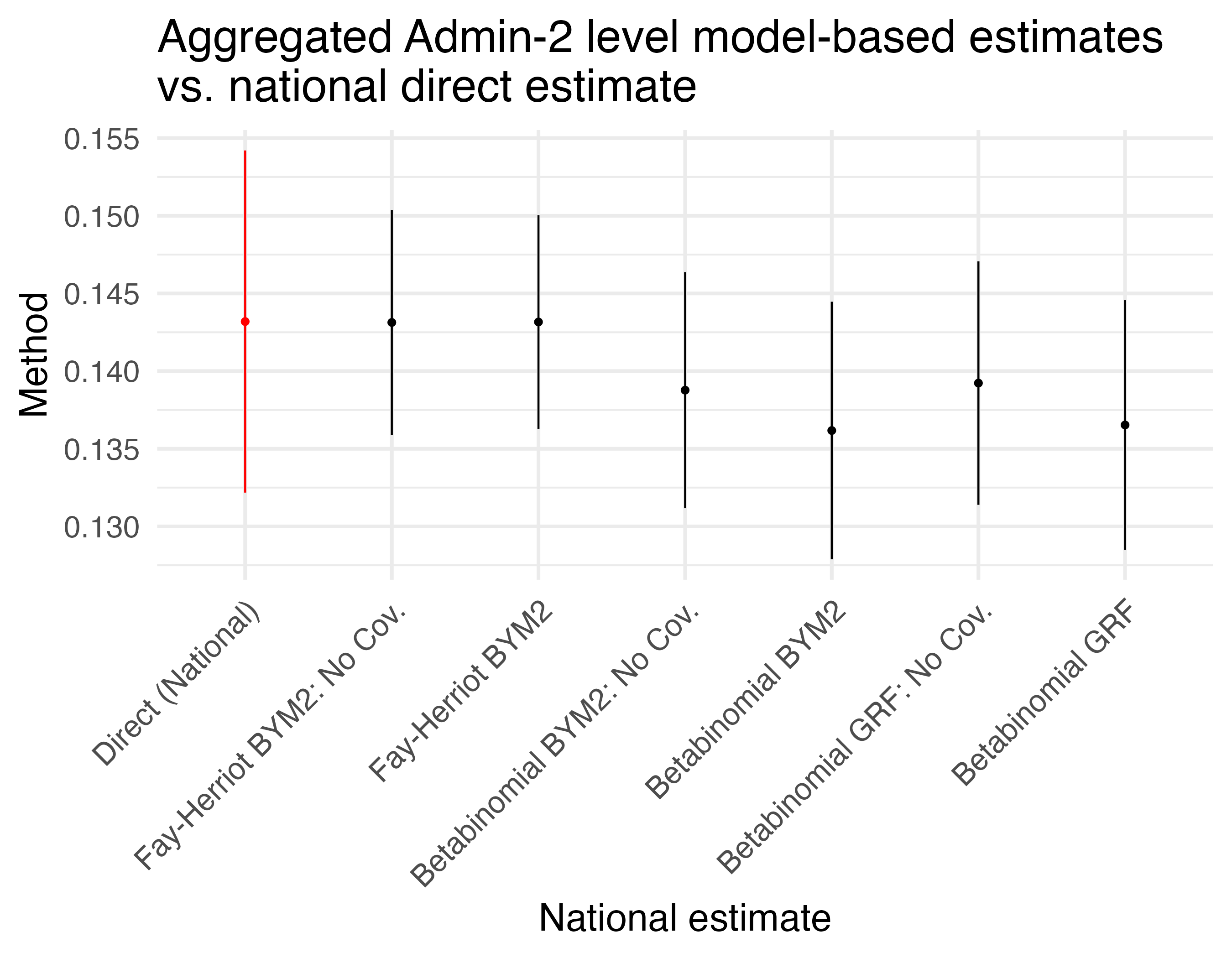}
\caption{Comparison of aggregated model-based Admin-2 level estimates and direct estimates at national level. The BYM2 models are specified at Admin-2.}
\label{fig:Zambia_aggregated_adm2_natl}
\end{figure}

In Figures  \ref{fig:Zambia_aggregated_adm1_natl} and \ref{fig:Zambia_aggregated_adm2_natl} we aggregate the model-based areal estimates to the national level. The GRF results are the same in both figures, while the BYM2 models are specified at Admin-1 in Figure \ref{fig:Zambia_aggregated_adm1_natl} and and at Admin-2 in \ref{fig:Zambia_aggregated_adm2_natl}. In both plots, we see reasonable agreement with the weighted estimates from the two Fay-Herriot models but downward bias for the unit-level models. The downward bias is greater in the covariate models, which suggests there may be problems with the aggregation step.

The comparisons between Admin-1 estimates to direct estimates is given in Section \ref{sec:scatterplots}.

\section{Comparison of Additional Models}

\begin{figure}[htbp]
\includegraphics[width=\columnwidth]{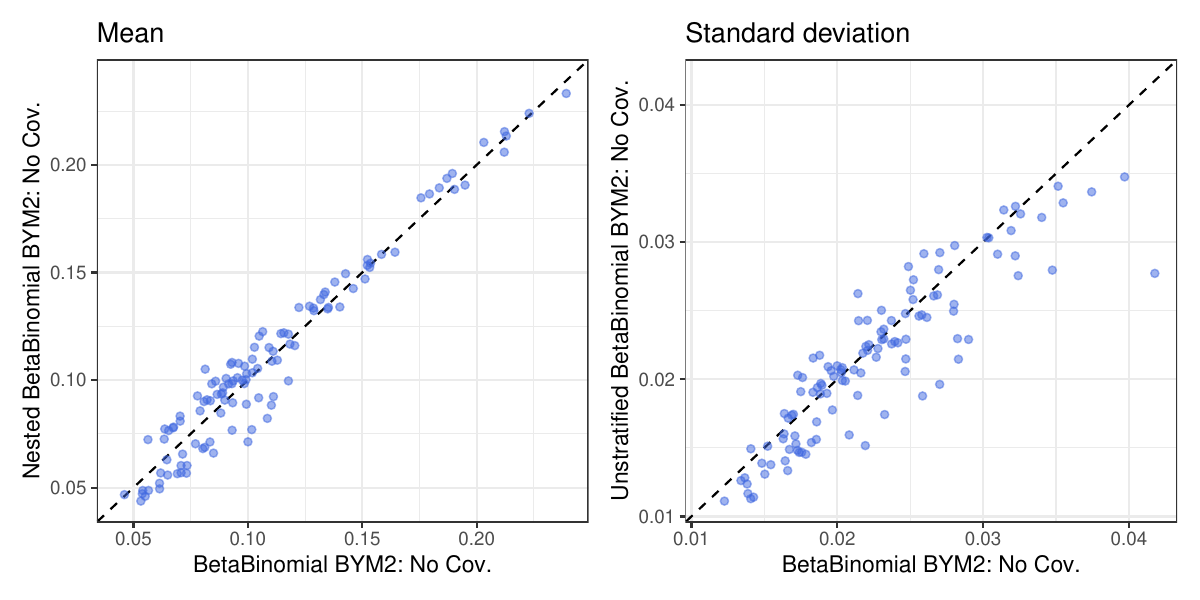}
\includegraphics[width=\columnwidth]{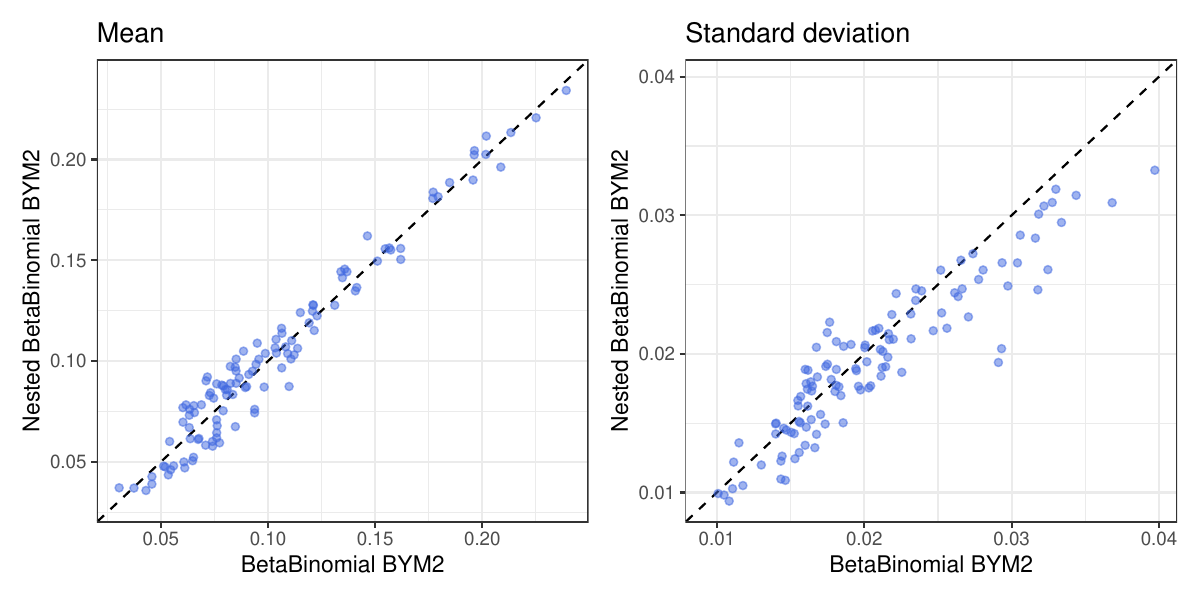}
\caption{Comparison of the betabinomial BYM2 model and the nested betabinomial BYM2 model with fixed effects at Admin-1. Top row: without covariates. Bottom row: with covariates}
\label{fig:Zambia_nest_or_no_nest}
\end{figure}

In Figure  \ref{fig:Zambia_nest_or_no_nest} we compare the betabinomial BYM2 model with and without Admin-1 area fixed effects.
Overall, there are not great differences in this example, because the data are quite abundant at Admin-1. In other examples, we have found nested models to be very useful.

\begin{figure}[htbp]
\includegraphics[width=\columnwidth]{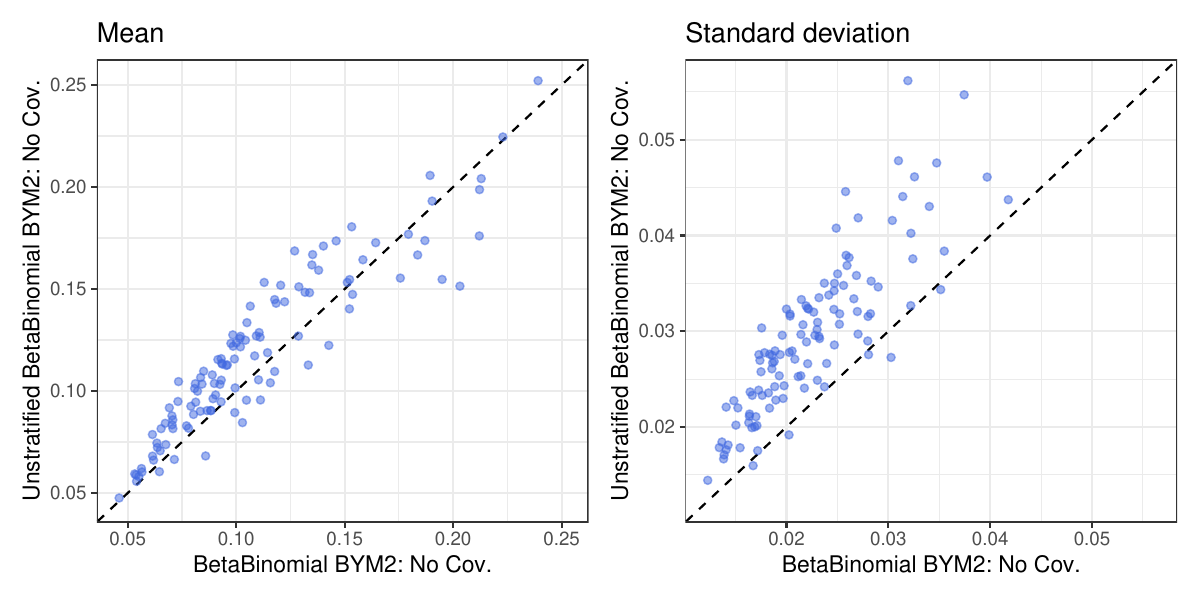}
\includegraphics[width=\columnwidth]{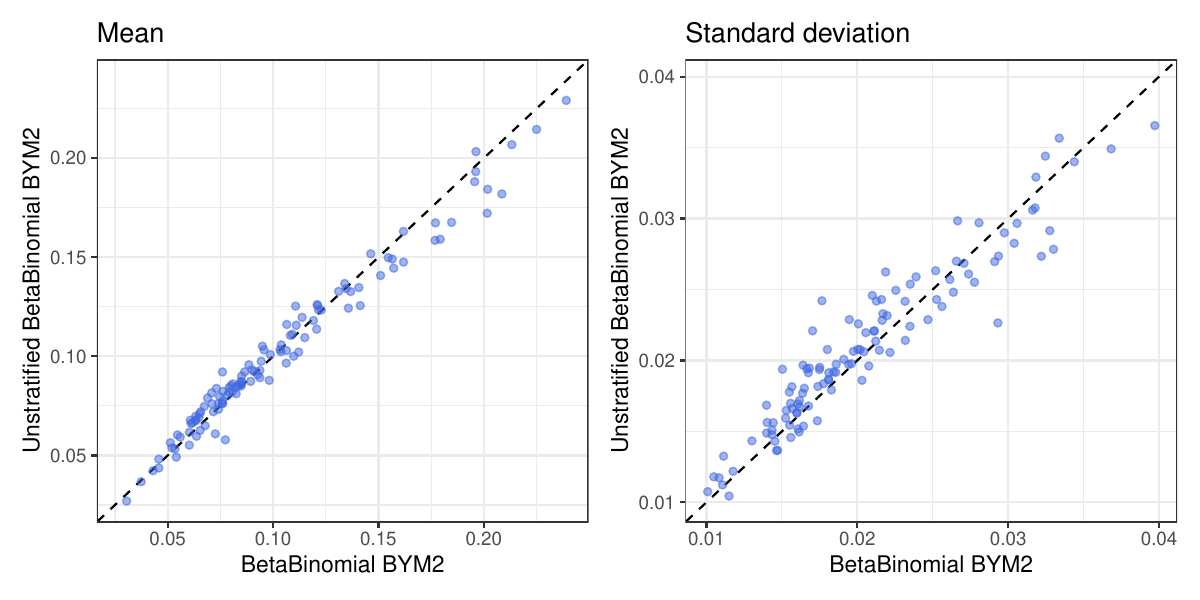}
\caption{Comparison of unstratified betabinomial BYM2 models and the stratified betabinomial BYM2 models which include an urban/rural effects. Top row: without covariates. Bottom row: with covariates}
\label{fig:Zambia_strat_or_no_strat}
\end{figure}

\begin{figure}[htbp]
\centering
\includegraphics[width=.9\columnwidth]{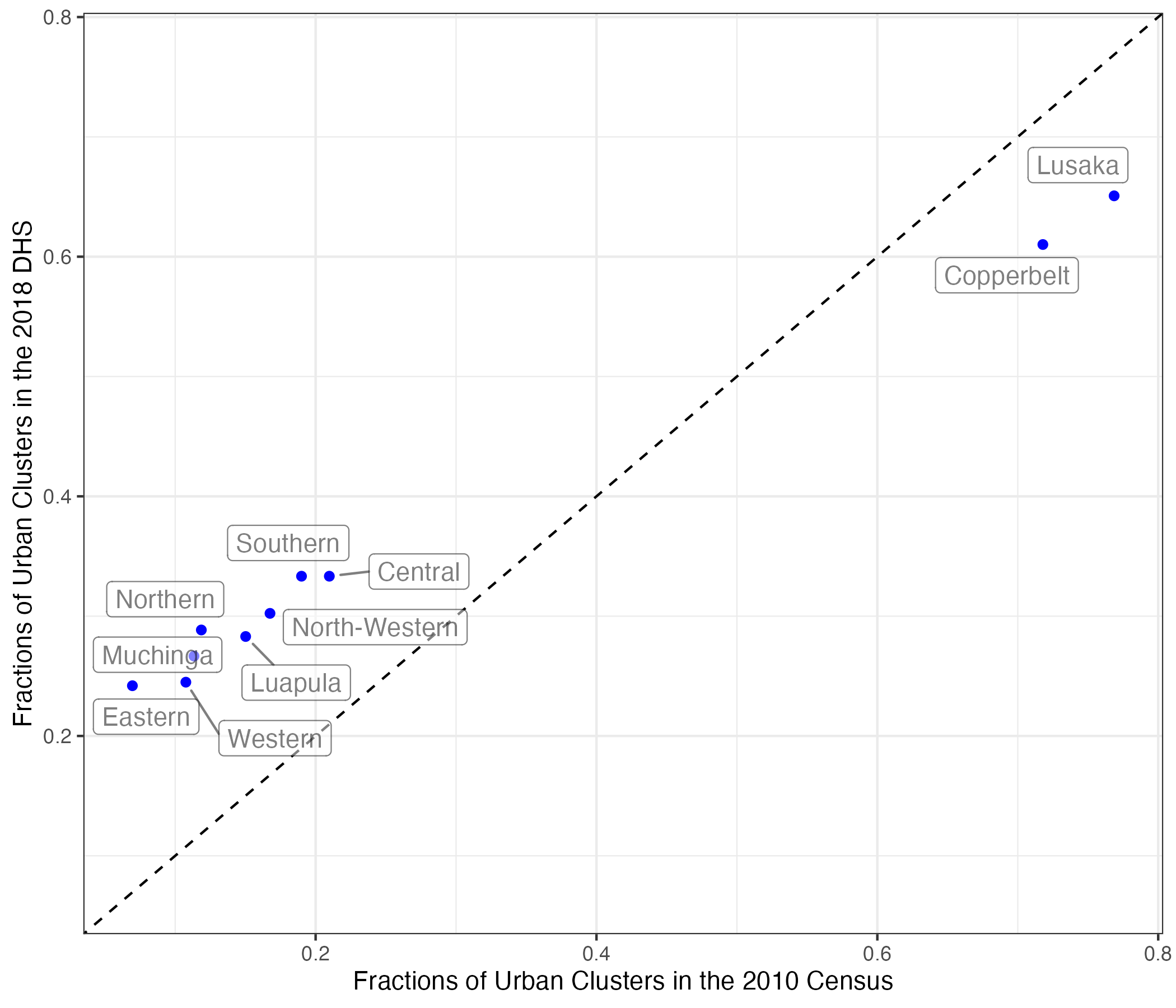}
\caption{Comparison of the fraction of urban clusters in the 2010 census and the 2018 DHS samples.}
\label{fig:Zambia_urban_frac}
\end{figure}

In Figure \ref{fig:Zambia_strat_or_no_strat}, we compare the betabinomial BYM2 model with and without the urban fixed effect. In the top row, the model without the urban fixed effect, i.e.,~the unstratified model with no covariates, leads to higher prevalence estimates when compared to the stratified model. This is due to the oversampling of urban clusters, where HIV prevalence is higher, in the survey. This is shown in Figure \ref{fig:Zambia_urban_frac} where urban clusters are oversampled in 8 out of 10 Admin-1 areas, with the two that are not, both being more ubanized. If we ignore the differential sampling aspect we overestimate HIV prevalence. In the bottom row, the presence of covariates in the unstratified model  helps alleviate the bias but, to avoid bias the safest option is to include an urban/rural fixed effect in the model.

\clearpage
\onecolumn
\section{Maps}

\begin{figure}[htbp]
\centering
\includegraphics[width=.8\textwidth]{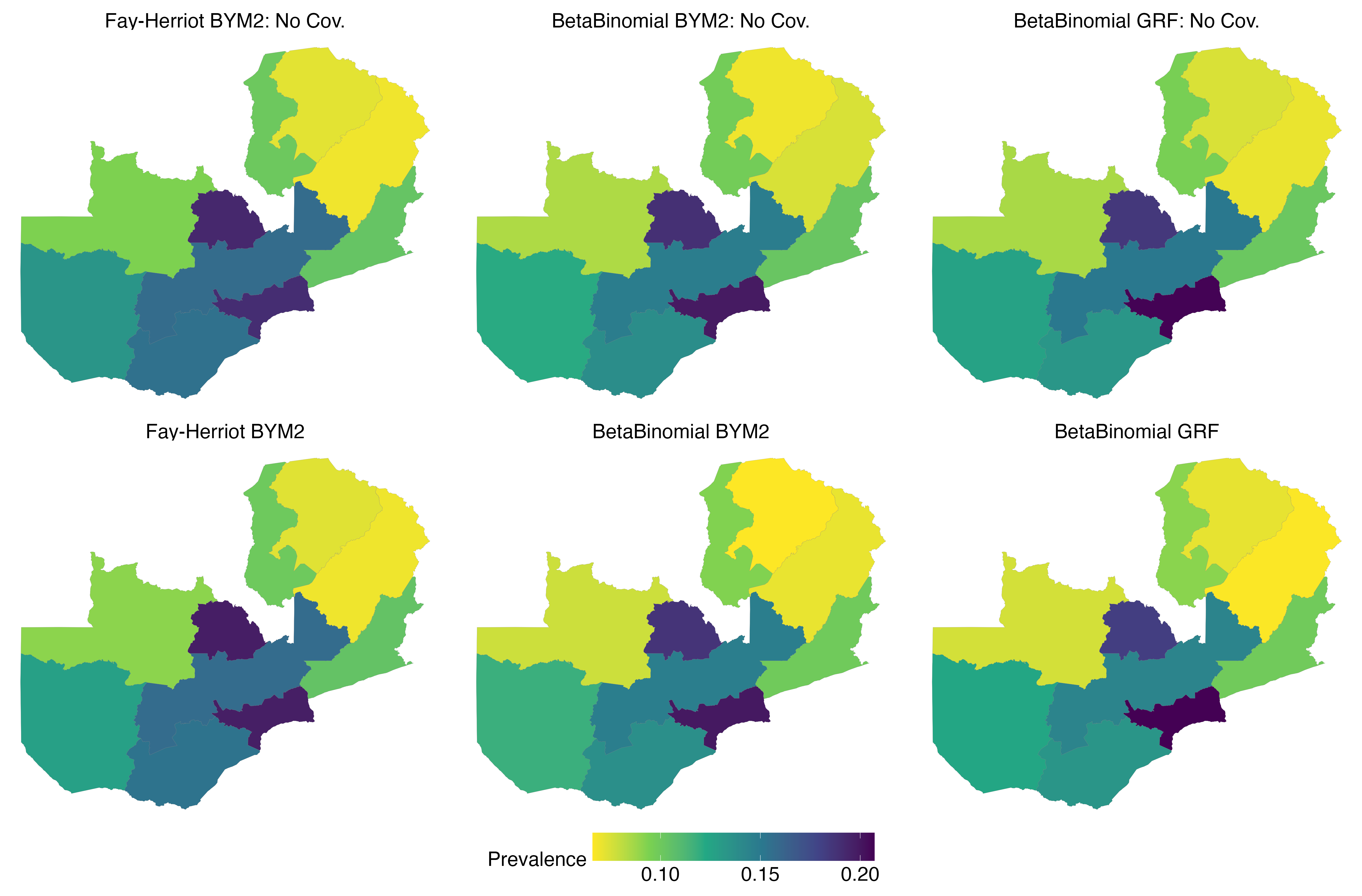}
\caption{Comparison of model-based Admin-1 level estimates from all six SAE models. Top row: no covariates (apart from urban/rural) in the model. Bottom row: covariates in the model.}
\label{fig:ZMB_map_admin1_est}
\end{figure}

\begin{figure}[htbp]
\centering
\includegraphics[width=.8\textwidth]{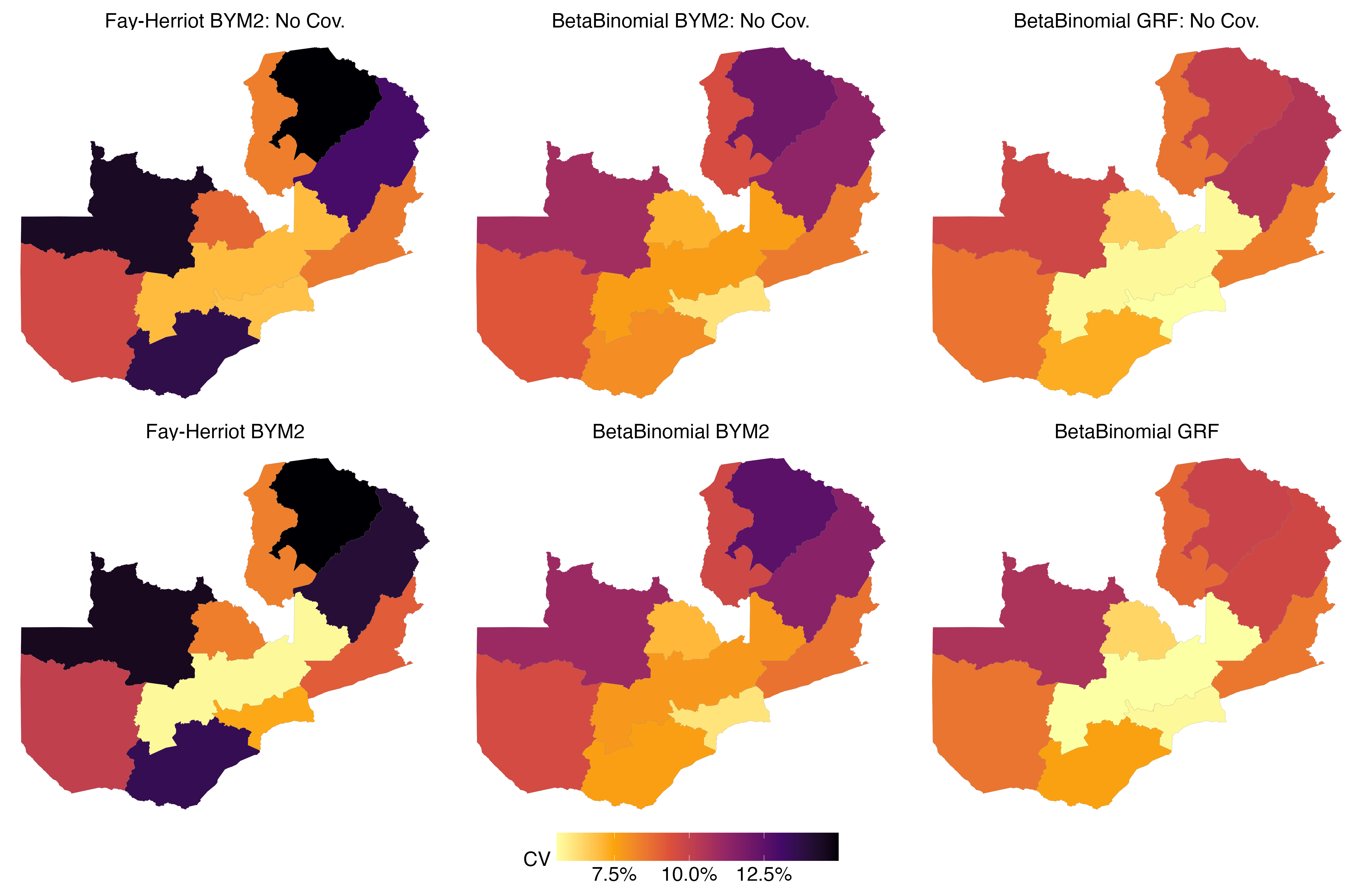}
\caption{Comparison of model-based Admin-1 level estimates of CV from all six SAE models. Top row: no covariates (apart from urban/rural) in the model. Bottom row: covariates in the model.}
\label{fig:ZMB_map_admin1_cv}
\end{figure}

\begin{figure}[htbp]
\centering
\includegraphics[width=.8\textwidth]{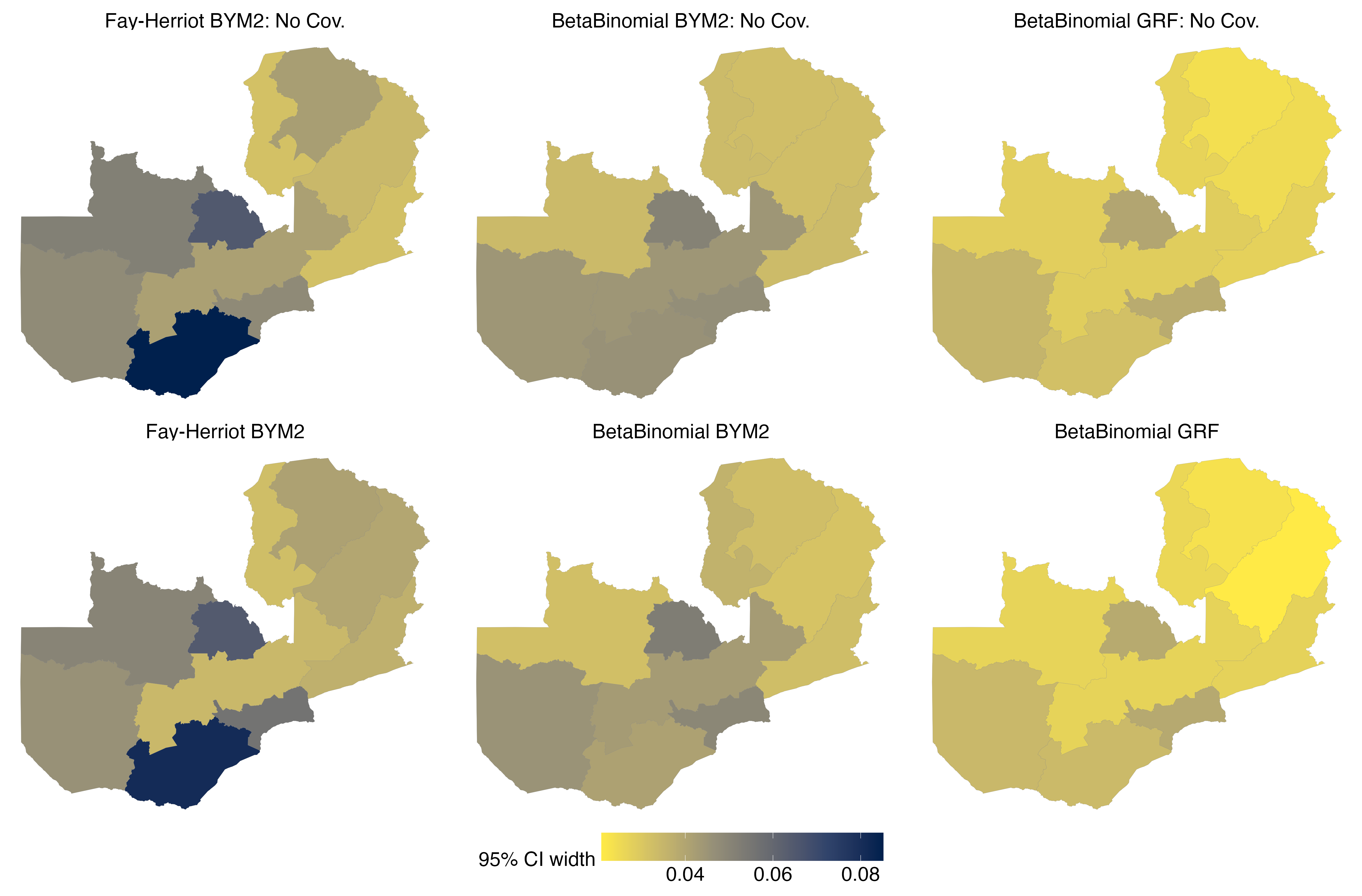}
\caption{Comparison of model-based Admin-1 level estimates of $95\%$ posterior interval width from all six SAE models. Top row: no covariates (apart from urban/rural) in the model. Bottom row: covariates in the model.}
\label{fig:ZMB_map_admin1_width}
\end{figure}

Figures \ref{fig:ZMB_map_admin1_est}, \ref{fig:ZMB_map_admin1_cv}, \ref{fig:ZMB_map_admin1_width} show the point estimates, coefficients of variation (CV), and $95\%$ interval widths for the six models at the Admin-1 level. Plots of the direct estimates are included in the main paper.

\begin{figure}[htbp]
\centering
\includegraphics[width=.8\textwidth]{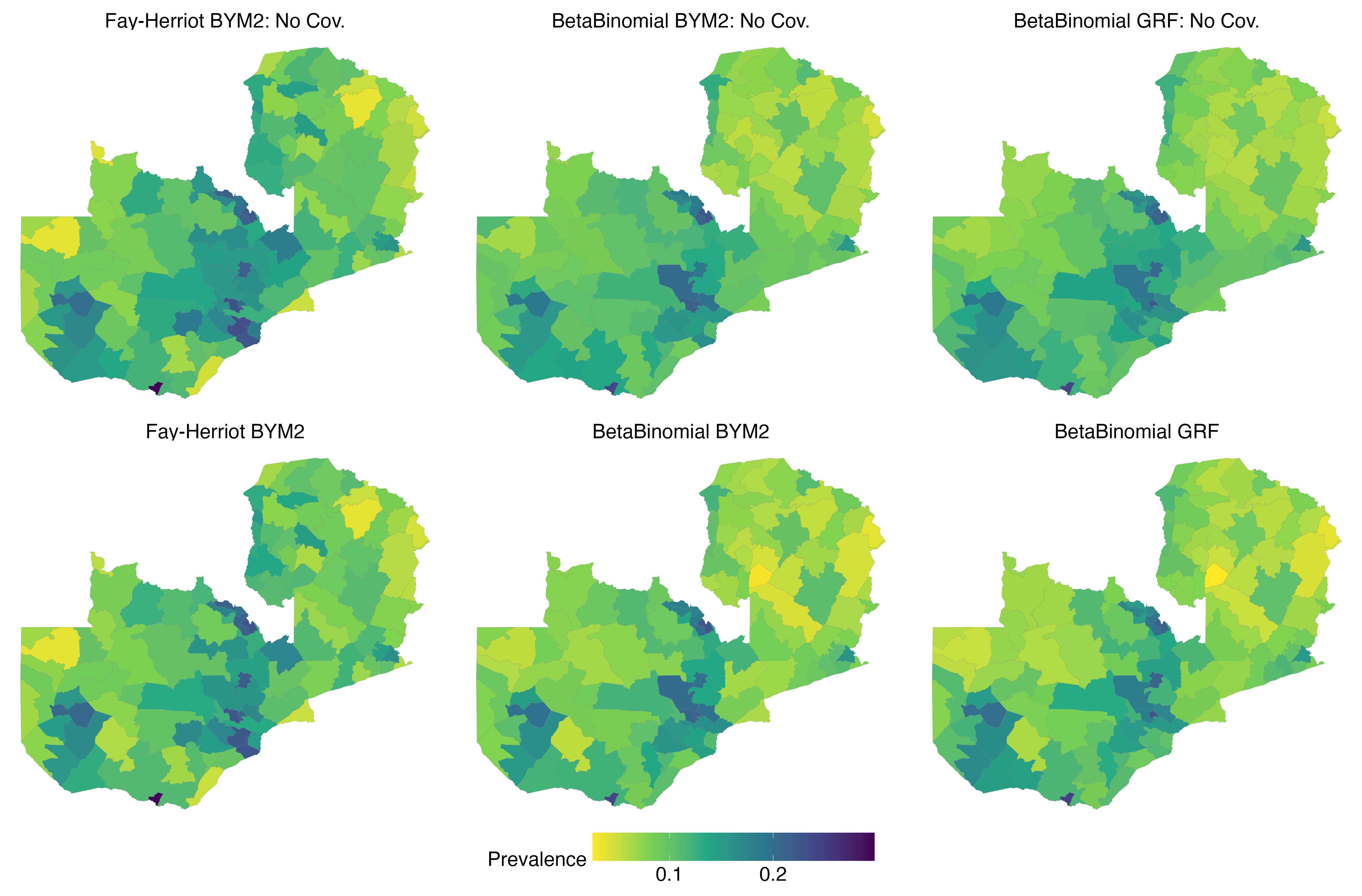}
\caption{Comparison of model-based Admin-2 level estimates from all six SAE models. Top row: no covariates (apart from urban/rural) in the model, bottom row, covariates in the model.}
\label{fig:ZMB_map_admin2_est}
\end{figure}

\begin{figure}[htbp]
\centering
\includegraphics[width=.8\textwidth]{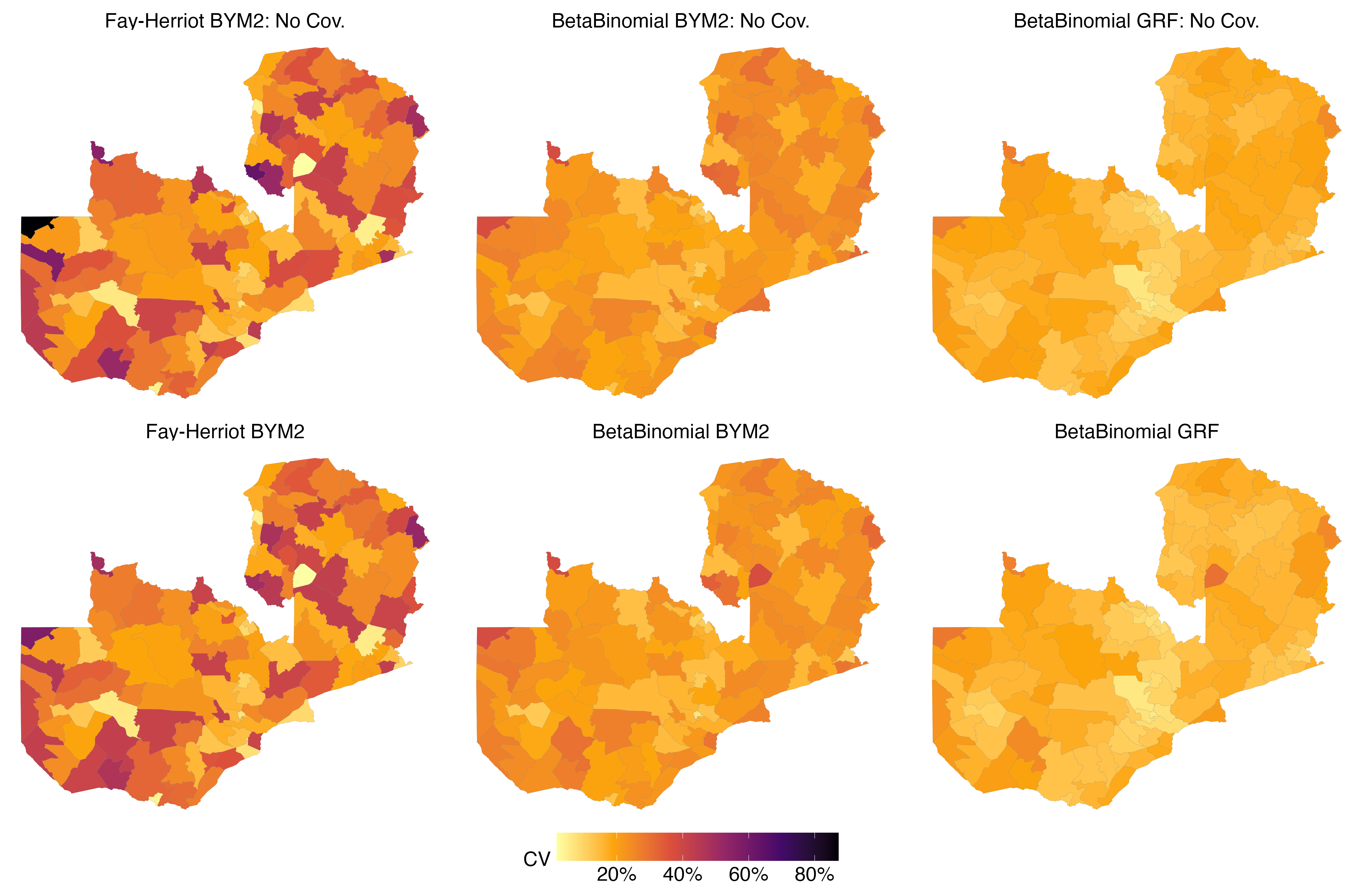}
\caption{Comparison of model-based Admin-2 level estimates of CV from all SAE six models. Top row: no covariates (apart from urban/rural) in the model. Bottom row: covariates in the model.}
\label{fig:ZMB_map_admin2_cv}
\end{figure}

\begin{figure}[htbp]
\centering
\includegraphics[width=.8\textwidth]{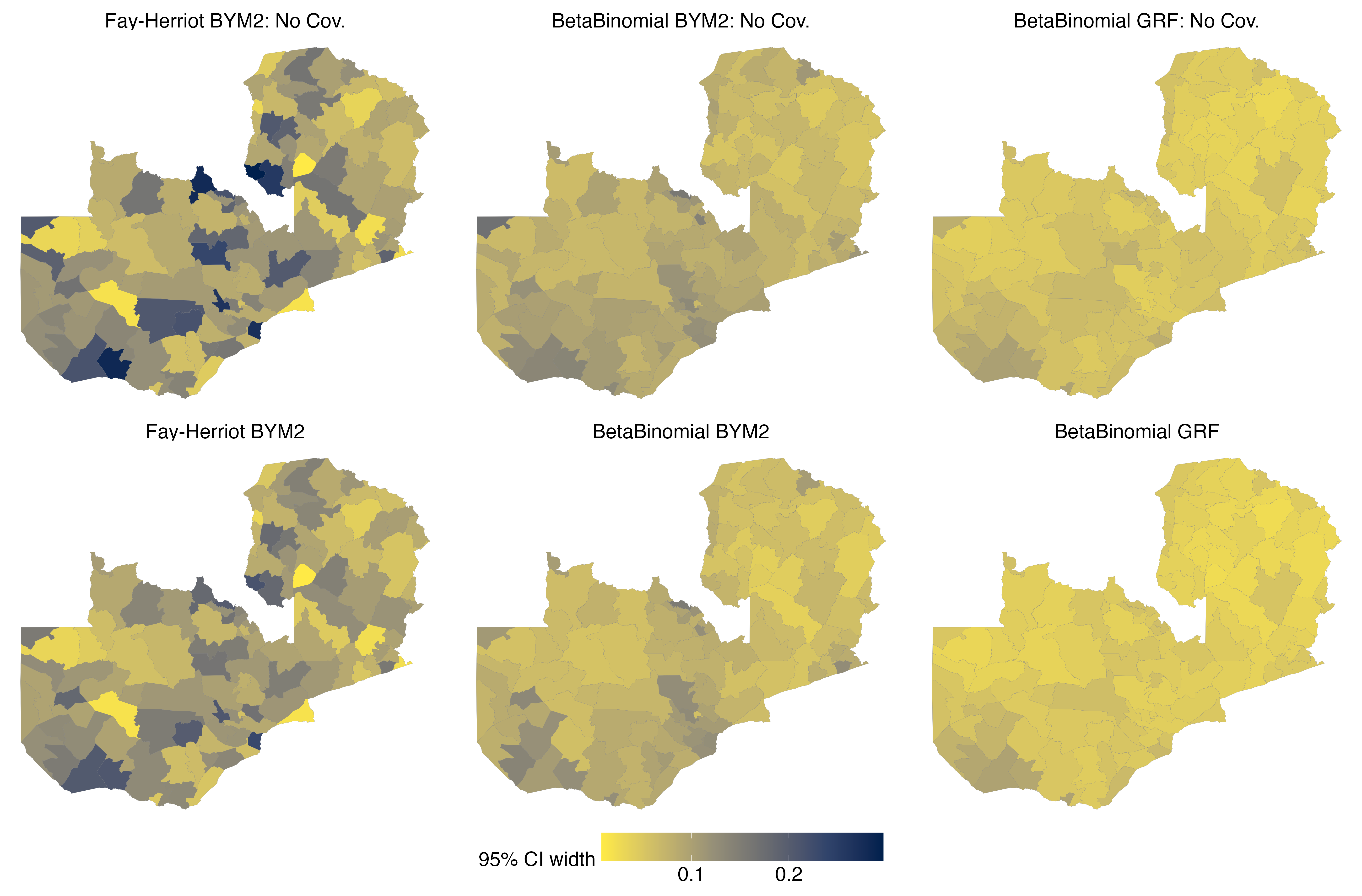}
\caption{Comparison of model-based Admin-2 level estimates of $95\%$ posterior interval width from all six SAE models. Top row: no covariates (apart from urban/rural) in the model. Bottom row: covariates in the model.}
\label{fig:ZMB_map_admin2_width}
\end{figure}

Figures \ref{fig:ZMB_map_admin2_est}, \ref{fig:ZMB_map_admin2_cv}, \ref{fig:ZMB_map_admin2_width} show the point estimates, CVs, and $95\%$ interval widths for the six models at the Admin-2 level. Plots of the direct estimates are included in the main paper.

\clearpage
\onecolumn
\section{Scatterplots}\label{sec:scatterplots}

\begin{figure}[htbp]
\includegraphics[width=\textwidth]{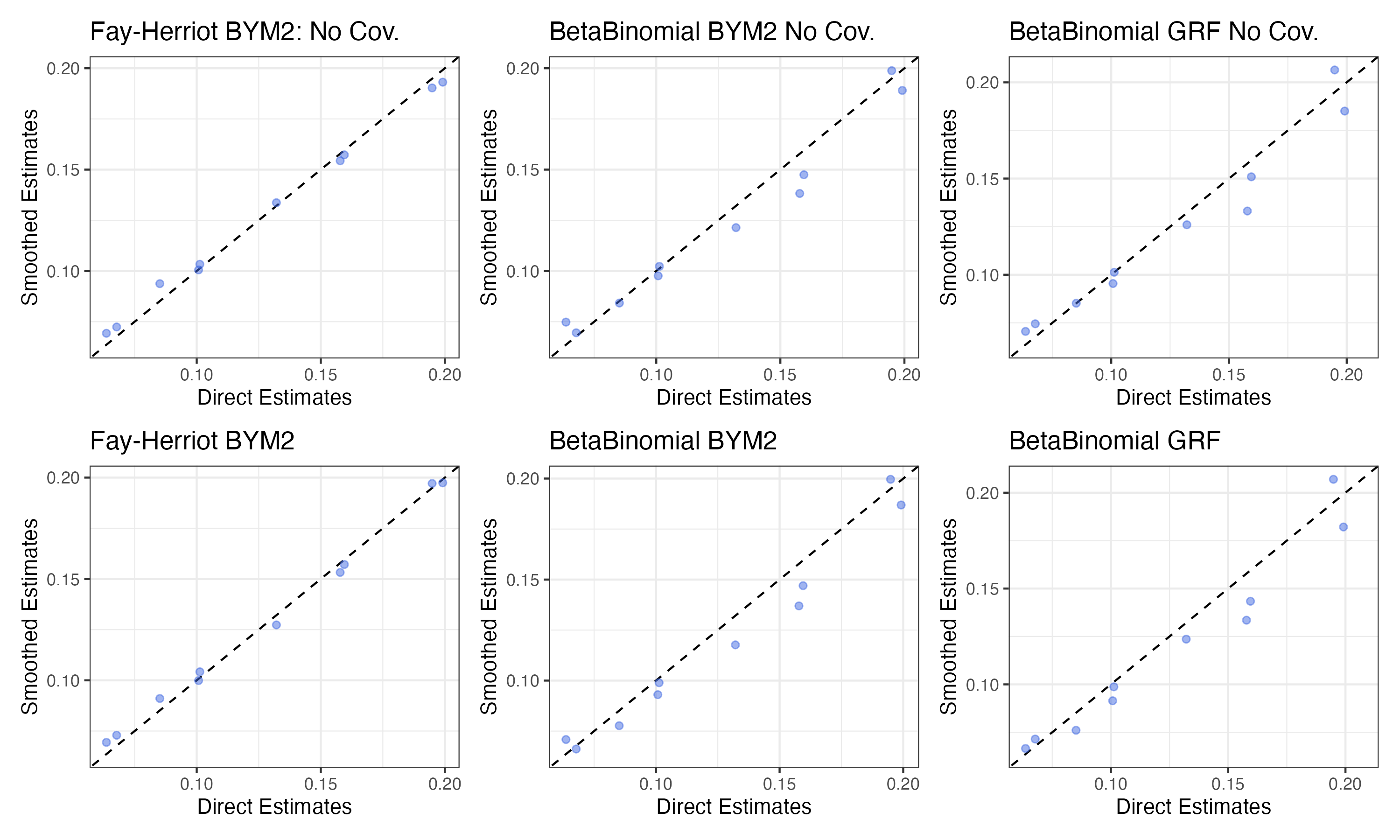}
\caption{Comparison of model-based Admin-1 level estimates and direct estimates. Top row: no covariates (apart from urban/rural) in the model. Bottom row: covariates in the model.}
\label{fig:ZMB_scatter_admin1}
\end{figure}

\begin{figure}[htbp]
\includegraphics[width=\textwidth]{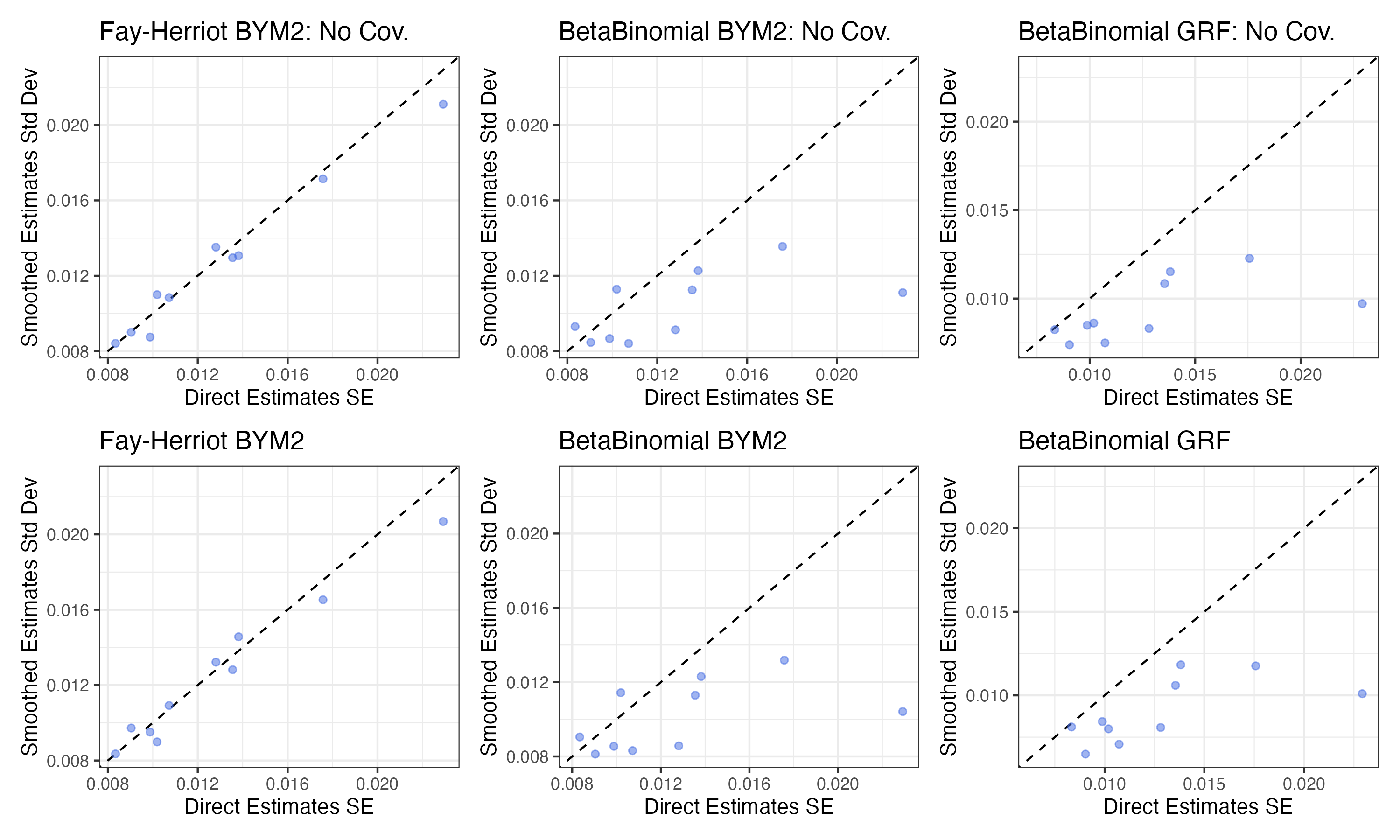}
\caption{Comparison of the model-based standard deviations at  Admin-1 level estimates and the standard errors of the direct estimates. Top row: no covariates (apart from urban/rural) in the model. Bottom row: covariates in the model.}
\label{fig:ZMB_scatter_admin1_sd}
\end{figure}

Figures \ref{fig:ZMB_scatter_admin1} and  \ref{fig:ZMB_scatter_admin1_sd} give, for Admin-1 areas, a scatter plot of the (aggregated) estimates and their associated standard errors from the six SAE models versus the direct estimates. There is little shrinkage at Admin-2 though the betabinomial models show underestimation, consistent with what we saw in Figure \ref{fig:Zambia_aggregated_adm1_natl}, when national estimates were compared.
Of course, for the unit-level non-linear models that we are using, we do not have design consistency.

\begin{figure}[htbp]
\includegraphics[width=\textwidth]{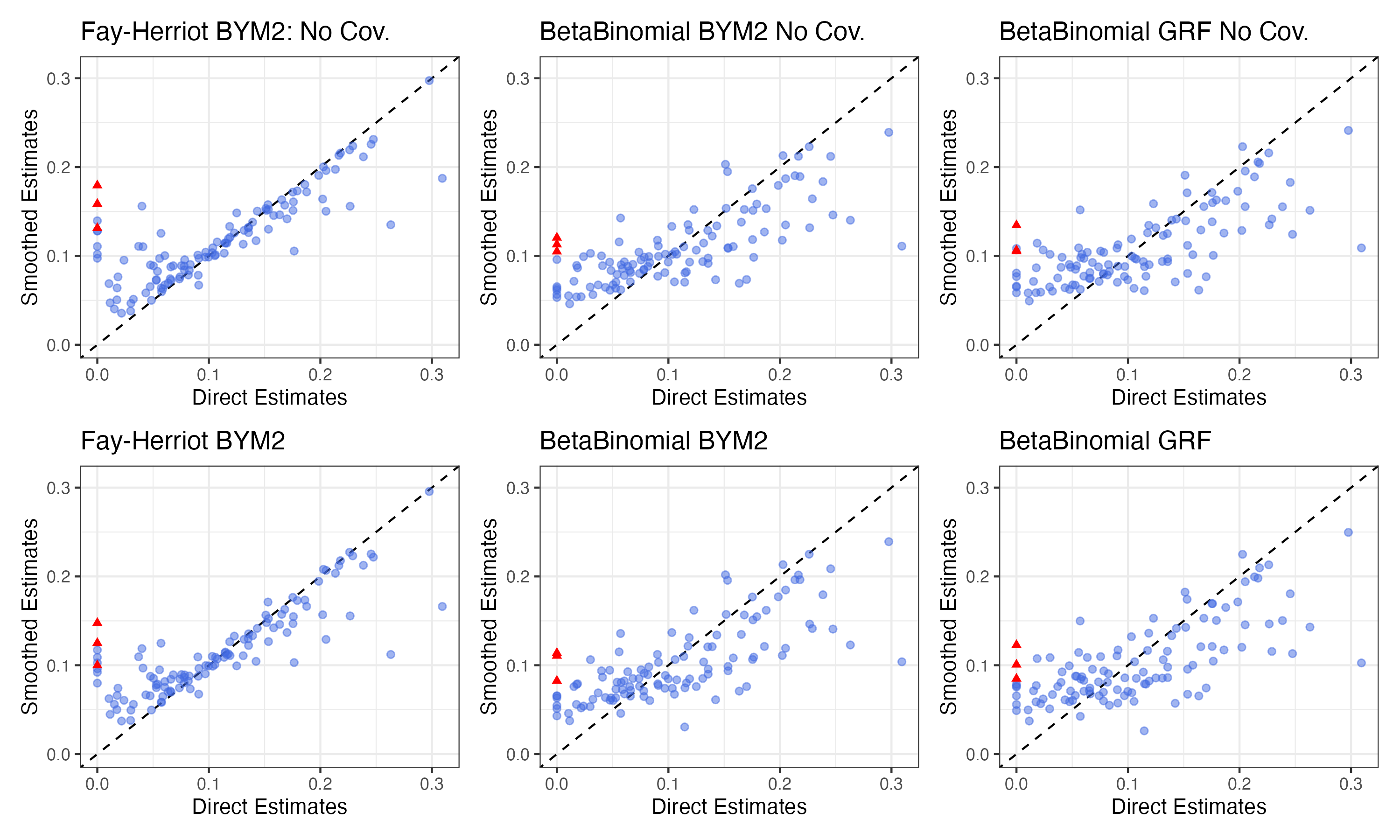}
\caption{Comparison of model-based Admin-2 level estimates and direct estimates. Top row: no covariates (apart from urban/rural). Bottom row: covariates in the model.}
\label{fig:ZMB_scatter_admin2}
\end{figure}

\begin{figure}[htbp]
\includegraphics[width=\textwidth]{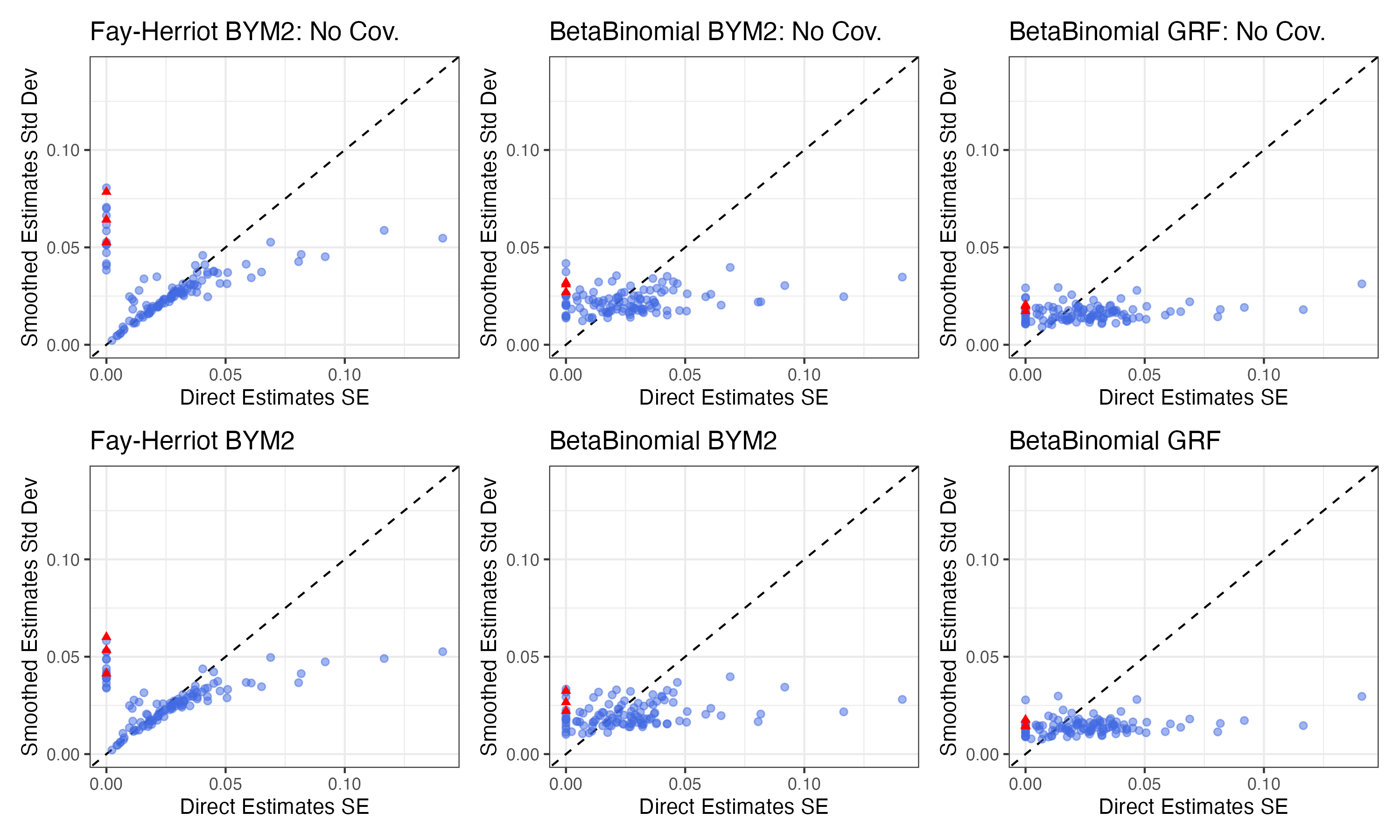}
\caption{Comparison of the model-based standard deviations at  Admin-2 level estimates and the standard errors of the direct estimates. Top row: no covariates (apart from urban/rural) in the model. Bottom row: covariates in the model.}
\label{fig:ZMB_scatter_admin2_sd}
\end{figure}

Figure \ref{fig:ZMB_scatter_admin2} and \ref{fig:ZMB_scatter_admin2_sd} show the area-level estimates versus the direct estimates, though now for Admin-2 areas. The Admin-2 areas without valid direct estimates are shown as red triangles. 
The shrinkage in all models is apparent, as is the reduction in the uncertainty, as compared to the direct estimates.




In Figure \ref{fig:ZMB_scatter_admin2_sd}, we plot the posterior standard deviations (SDs) against the standard errors of the direct estimates. The Fay-Herriot SDs show predictable behavior, as one would expect from a linear mixed effects model,
with greater precision increases for the larger prevalence estimates. In contrast, the unit-level SDs are in a horizontal band and appear unrelated to the direct prevalence estimates. These SDs are based on aggregation of estimates from a generalized linear mixed effects model (specifically, a logistic model), for which very little insight is currently available.

\onecolumn
\clearpage
\section{Ridge plots}

\begin{figure*}[htbp]
\includegraphics[width=\textwidth]{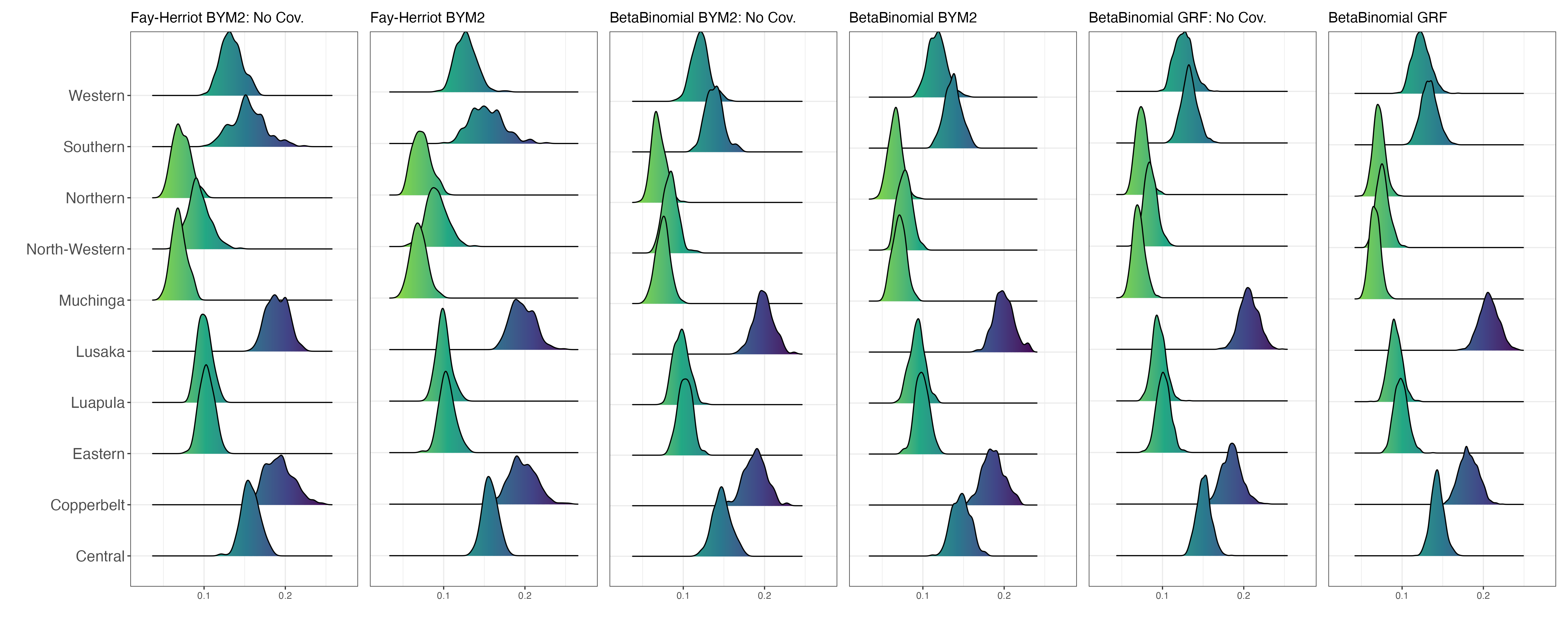}
\caption{Ridge plot of the posterior distributions of the Admin-1 HIV prevalence.}
\label{fig:ZMB_ridge_adm1}
\end{figure*}

\begin{figure*}[htbp]
\includegraphics[width=\textwidth]{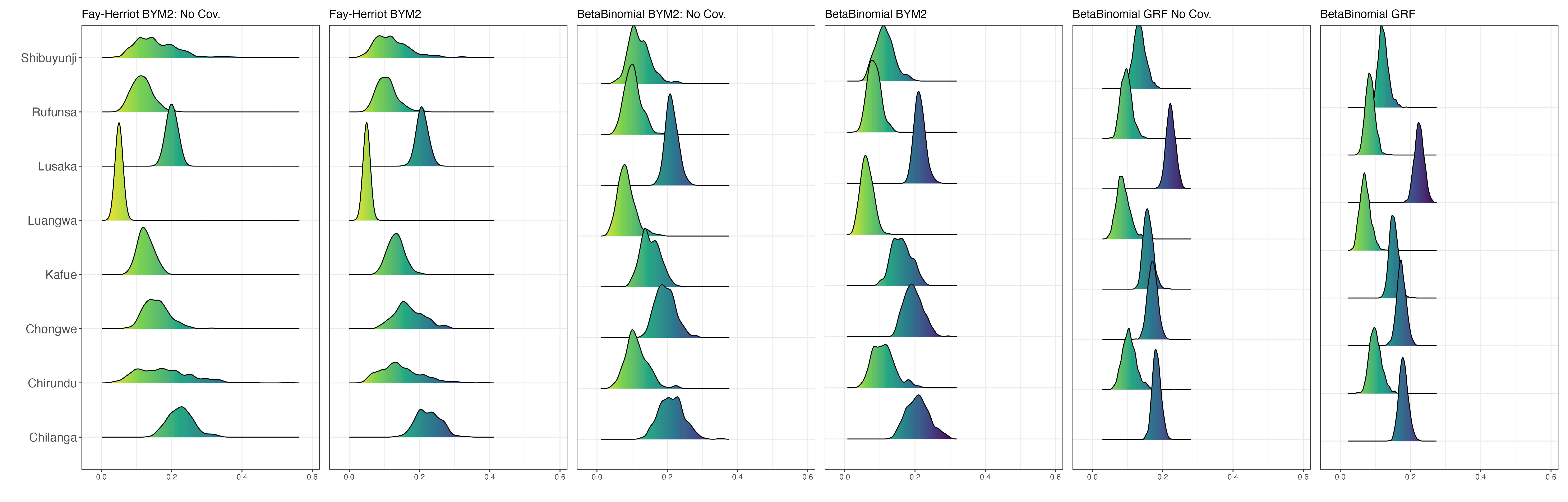}
\caption{Ridge plot of the posterior distributions of the Admin-2 HIV prevalence, for Admin-2 areas within Lusaka.}
\label{fig:ZMB_ridge_adm2_Lusaka}
\end{figure*}

\begin{figure*}[htbp]
\includegraphics[width=\textwidth]{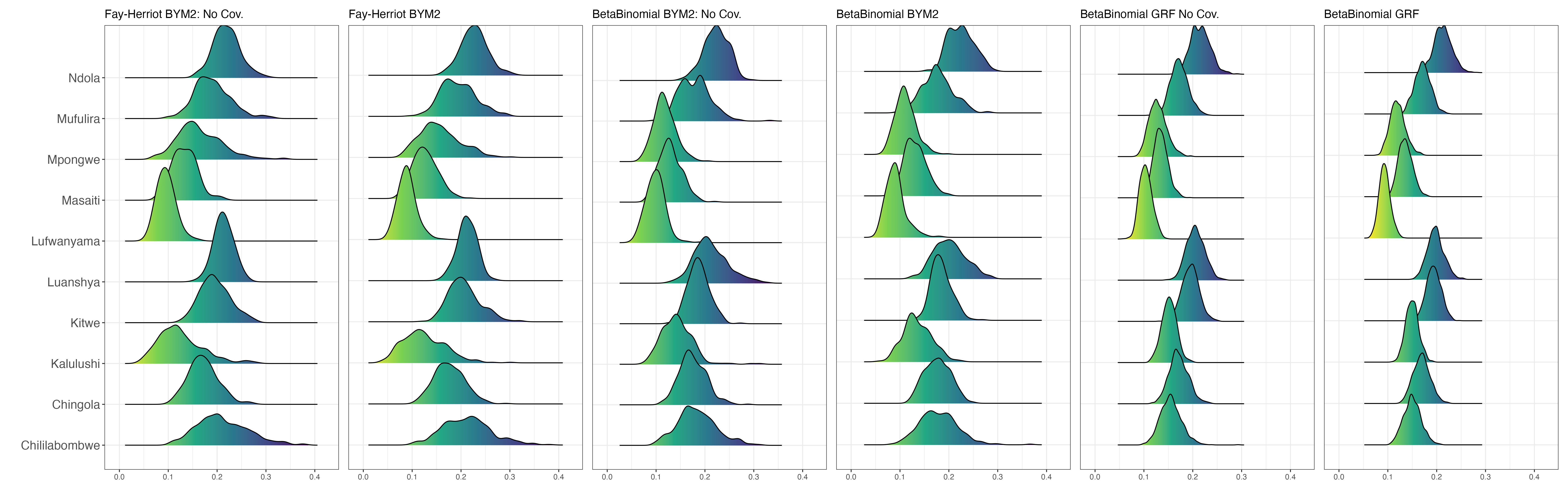}
\caption{Ridge plot of the posterior distributions of the Admin-2 HIV prevalence, for Admin-2 areas within Copperbelt.}
\label{fig:ZMB_ridge_adm2_Copperbelt}
\end{figure*}

Figure \ref{fig:ZMB_ridge_adm1} gives, for Admin-1 areas, ridge plots of the posterior marginal distribution of estimates from the six SAE models. Figures \ref{fig:ZMB_ridge_adm2_Lusaka} and \ref{fig:ZMB_ridge_adm2_Copperbelt} show the same plot for Admin-2 areas within the two Admin-1 areas with the highest prevalence, Lusaka and Copperbelt. 
Note that the scales are different on the three figures. As we would expect, the Admin-2 distributions are wider, reflecting the greater sparsity at Admin-2.

The Admin-1 distributions are relatively similar across models. he Admin-2 distributions show 

\clearpage
\section{Ranking plots}
\begin{figure*}[htbp]
\includegraphics[width=\textwidth]{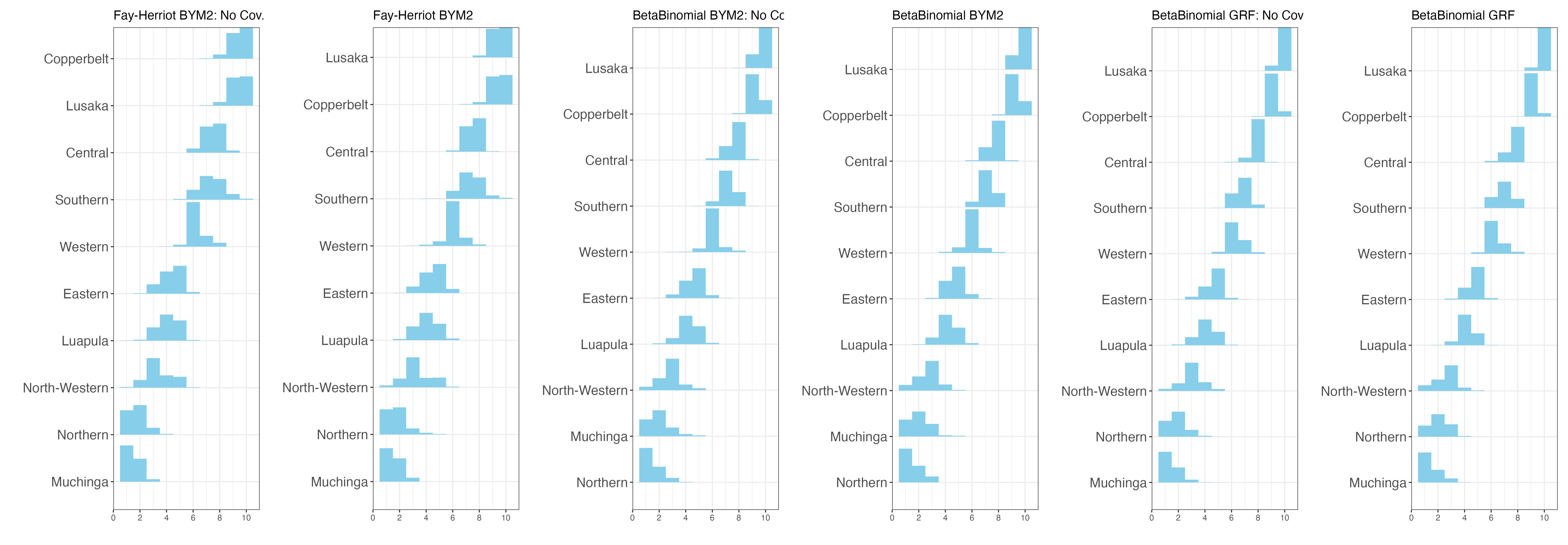}
\caption{Ranking plot of the posterior distributions of the Admin-1 HIV prevalence. The areas are ranked from low prevalence (bottom left) to high prevalence (top right).}
\label{fig:ZMB_rank_adm1}
\end{figure*}

Figure \ref{fig:ZMB_rank_adm1} gives, for Admin-1 areas, a plot of the posterior marginal distribution of the area rankings from the six SAE models, where high ranking correspond to high prevalence. 
With high probability, Copperbelt and Lusaka have relatively high HIV prevalence. Muchinga and Northern have relatively low prevalence. Across models, the results are relatively consistent.

\begin{figure*}[htbp]
\includegraphics[width=\textwidth]{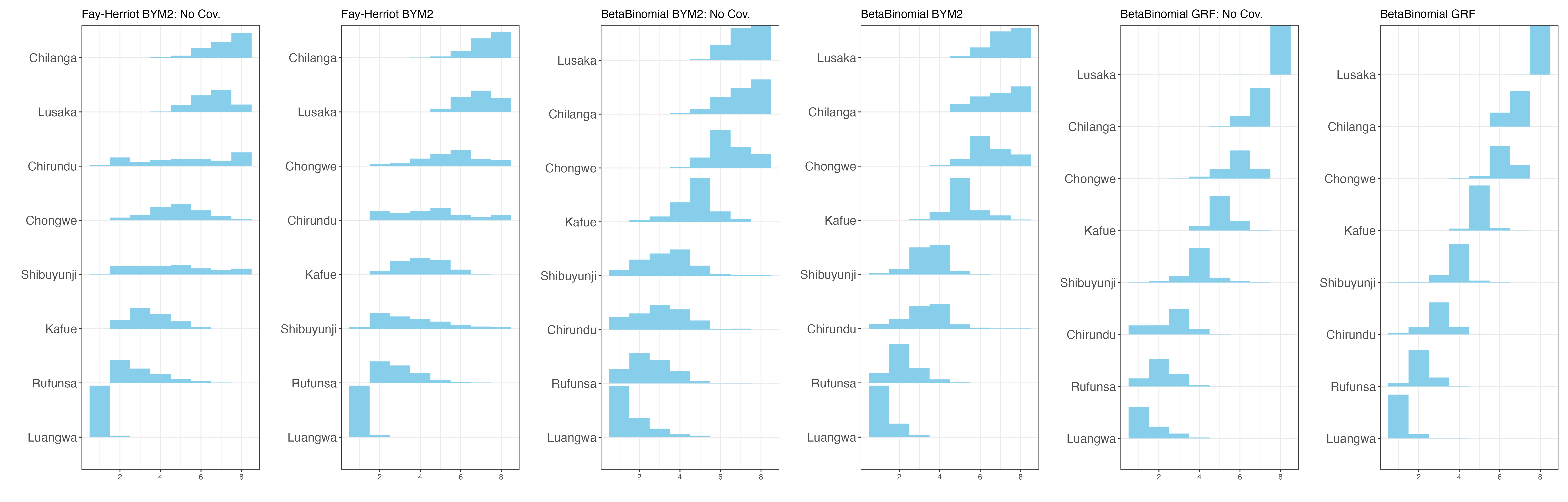}
\caption{Ranking plot of the posterior distributions of the Admin-2 HIV prevalence, for Admin-2 areas within Lusaka. The areas are ranked from low prevalence (bottom left) to high prevalence (top right).}
\label{fig:ZMB_rank_adm2_Lusaka}
\end{figure*}

\begin{figure*}[htbp]
\includegraphics[width=\textwidth]{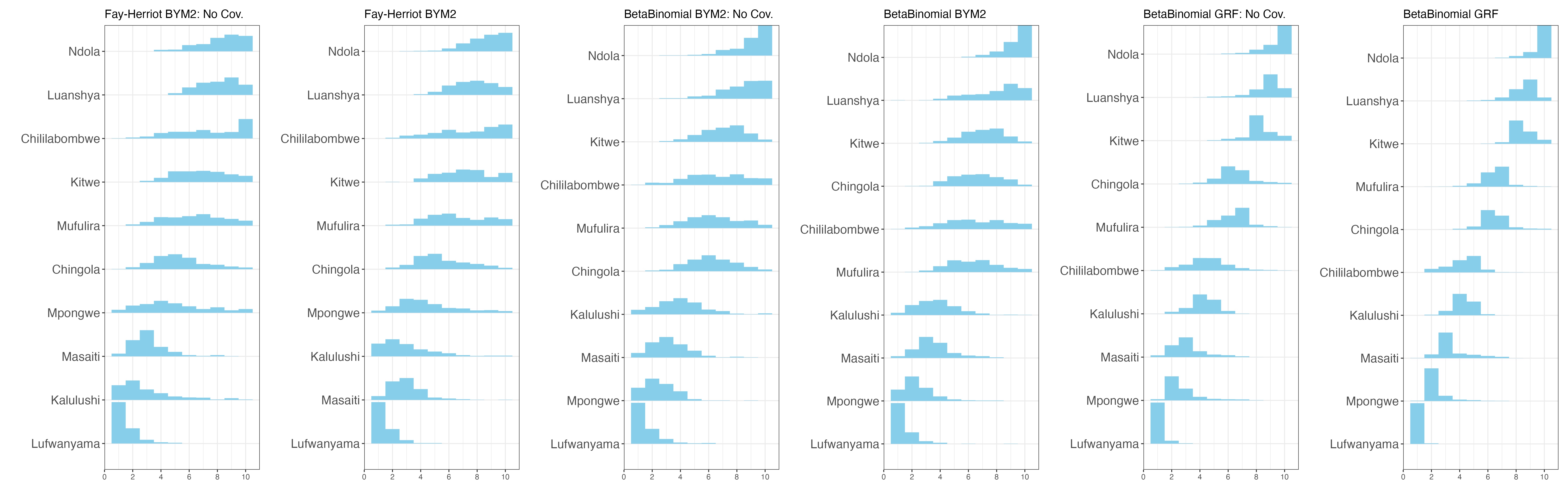}
\caption{Ranking plot of the posterior distributions of the Admin-2 HIV prevalence, for Admin-2 areas within Copperbelt. The areas are ranked from low prevalence (bottom left) to high prevalence (top right).}
\label{fig:ZMB_rank_adm2_Copperbelt}
\end{figure*}

Figures \ref{fig:ZMB_rank_adm2_Lusaka} and \ref{fig:ZMB_rank_adm2_Copperbelt} show the posterior marginal distributions of the rankings for Admin-2 areas within the two Admin-1 areas with the highest prevalence, Lusaka (8 Admin-2 areas) and Copperbelt (10 Admin-2 areas). These ranking plots have flatter distributions than we saw for Admin-1 areas, because the data are more sparse. We also see more variation in the results under different models, which also reflects the data sparseness.

\clearpage
\section{Exceedance probability plots}

\begin{figure}[htbp]
\includegraphics[width=\textwidth]{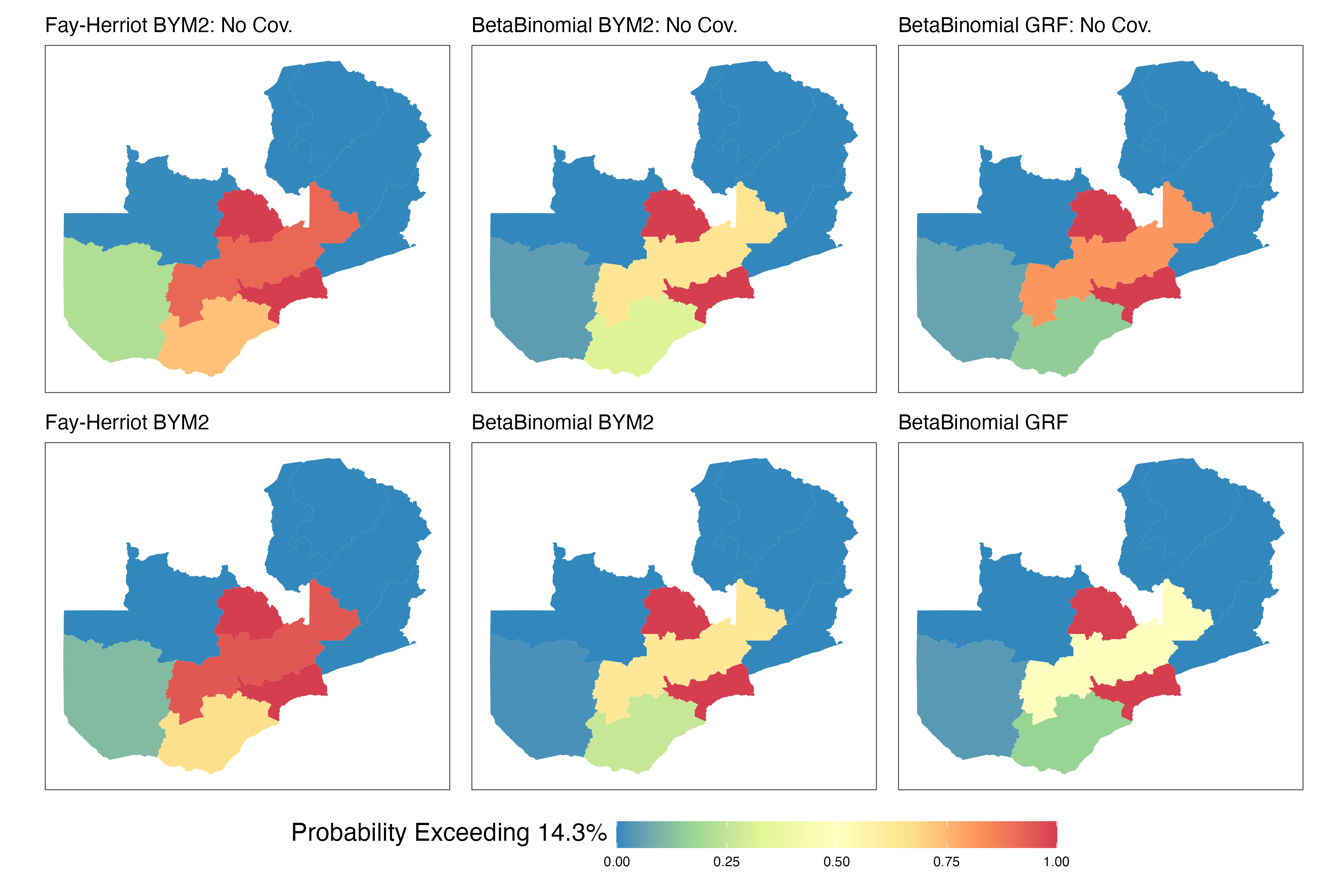}
\caption{Posterior probability of HIV prevalence exceeding $14.3\%$ (the national weighted estimate) for Admin-1 areas.}
\label{fig:ZMB_exceed_national_adm1}
\end{figure}

\begin{figure}[htbp]
\includegraphics[width=\textwidth]{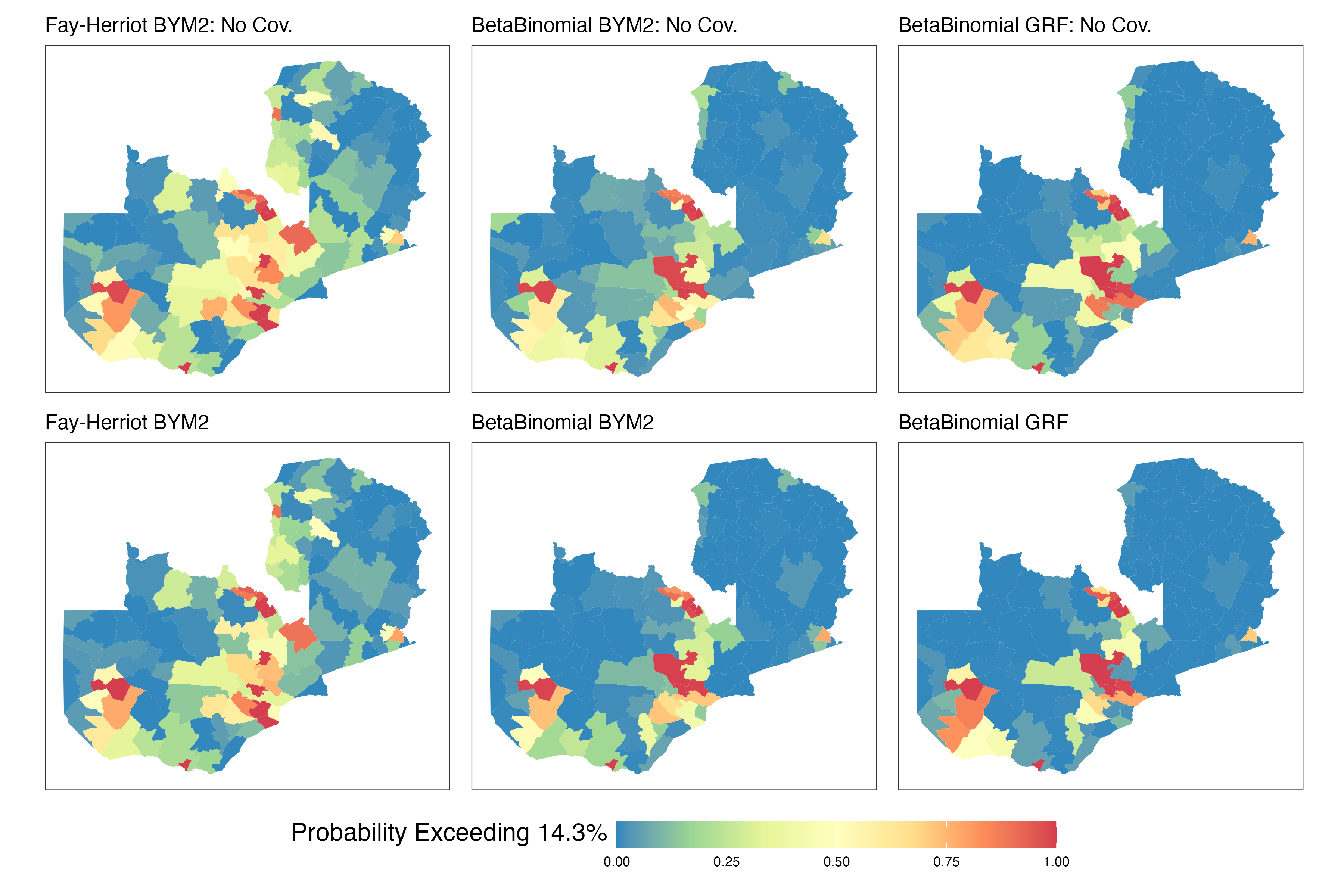}
\caption{Posterior probability of HIV prevalence exceeding $14.3\%$ (the national weighted estimate) for Admin-2 areas.}
\label{fig:ZMB_exceed_national_adm2}
\end{figure}

Prevalence estimates can also be used to compute other informative quantities of interest. For example, Figures \ref{fig:ZMB_exceed_national_adm1} and  \ref{fig:ZMB_exceed_national_adm2} show, respectively, the probabilities that the prevalence of each Admin-1 and Admin-2 area exceeds $14.3\%$ (the weighted national estimate of HIV prevalence). For Admin-1 areas, across all models, the highest posterior probabilities are in Copperbelt and Lusaka. For Admin-2 areas, we see there is large variation both across Zambia and also within Admin-2 areas.

\clearpage
\section{Model parameter estimates}

In Tables \ref{tab:fixed}--\ref{tab:hyper}, we provide parameter estimates with associated poster standard deviations for the models used in the Zambia application. The urban coefficient is highly significant in all models, which emphasizes the importance of its inclusion. When the three additional covariates are added to the model, the coefficient is attenuated a little, but it remains highly significant.
The overdispersion parameter in the betabinomial drops slightly in magnitude when covariates are added to the model.

\begin{table*}[ht]
\centering
\caption{Fixed effect posterior mean  estimates in all models with the associated posterior standard deviations.}
\begin{tabular}{lllllll}
  \hline
& Model & Intercept & urban & access & malaria & log(night time light + 1) \\
  \hline
Admin-1 & Fay-Herriot BYM2 No Cov.& -1.99 (0.109) &  &  &  &  \\
  &   Fay-Herriot BYM2 & -5.88 (4.64) &  & -7.15 (9.03) & -0.008 (0.158) & 0.037 (0.041) \\
 & BetaBinomial BYM2 No Cov. & -2.38 (0.091) & 0.897 (0.072) &  &  &  \\
  & BetaBinomial BYM2 & -5.91 (0.919) & 0.654 (0.106) & -6.74 (1.74) & 0.012 (0.065) & 0.008 (0.008) \\
  \hline
 Admin-2 & Fay-Herriot BYM2 No Cov.& -2.13 (0.052) &  &  &  &  \\
 & Fay-Herriot BYM2 & -4.34 (1.113) &  & -4.08 (2.19) & 0.058 (0.087) & 0.062 (0.023) \\
   & BetaBinomial BYM2 No Cov. & -2.39 (0.056) & 0.878 (0.079) &  &  &  \\
  & BetaBinomial BYM2 & -6.08 (0.975) & 0.641 (0.107) & -7.07 (1.86) & 0.028 (0.064) & 0.011 (0.009) \\
  \hline
 GRF & BetaBinomial GRF No Cov. & -2.44 (0.159) & 0.848 (0.073) &  &  &  \\
   & BetaBinomial GRF  & -6.33 (1.015) & 0.649 (0.105) & -7.52 (1.88) & 0.042 (0.092) & 0.005 (0.008) \\
   \hline
\end{tabular}
\label{tab:fixed}
\end{table*}

\begin{table*}[ht]
\centering
\caption{Hyperparameter posterior mean estimates in all models with the associated posterior standard deviations.}
\begin{tabular}{lllllll}
  \hline
& \multicolumn{1}{l}{\multirow{2}{*}{Model}}  & \multicolumn{1}{c}{Overdispersion}& BYM2 model & BYM2 model & GRF model  & GRF model  \\
&&\multicolumn{1}{c}{$1/(d+1)$}&\multicolumn{1}{c}{$\phi$}&\multicolumn{1}{c}{$\sigma_u$}&\multicolumn{1}{c}{range $r$}&\multicolumn{1}{c}{$\sigma_\omega$}\\
  \hline
Admin-1 &Fay-Herriot BYM2 No Cov.&  & 0.398 (0.254) & 7.202 (3.59) &  &  \\
 & Fay-Herriot BYM2 &  & 0.389 (0.254) & 10.85 (6.90) &  &  \\
  &BetaBinomial BYM2 No Cov. & 0.026 (0.004) & 0.357 (0.241) & 15.11 (8.66) &  &  \\
  &BetaBinomial BYM2 & 0.023 (0.004) & 0.383 (0.251) & 15.17 (8.67) &  &  \\
  \hline
Admin-2  &Fay-Herriot BYM2 No Cov.&  & 0.703 (0.183) & 2.424 (0.647) &  &  \\
& Fay-Herriot BYM2 &  & 0.434 (0.250) & 3.755 (1.112) &  &  \\
  &BetaBinomial BYM2 No Cov. & 0.021 (0.004) & 0.653 (0.210) & 8.336 (3.254) &  &  \\
  &BetaBinomial BYM2 & 0.019 (0.004) & 0.701 (0.204) & 9.184 (3.709) &  &  \\
  \hline
GRF   &BetaBinomial GRF No Cov. & 0.021 (0.004) &  &  & 3.57 (1.668) & 0.406 (0.083) \\
  &BetaBinomial GRF  & 0.020 (0.004) &  &  & 7.00 (5.16) & 0.413 (0.102) \\
   \hline
\end{tabular}
\label{tab:hyper}
\end{table*}

\onecolumn
\clearpage
\bibliographystyle{chicago}
\bibliography{spatepi2}

\end{document}